\documentclass[a4paper,12pt]{article}
\usepackage[latin1]{inputenc}
\usepackage{graphicx}
\usepackage{amsfonts}
\usepackage{amssymb}
\usepackage{graphics}
\begin{document}
\title{Gravitational Instability of Yang-Mills Cosmologies}
\author
{A. F\"uzfa$^{1,2}$}

\maketitle
\begin{small}   

\textit{
${}^{1}$ University of Namur, Unit\'e de Syst\`emes Dynamiques, Rue de Bruxelles, 61, B-5000 Namur, Belgium\\
${}^{2}$ F.N.R.S. Research Fellow}\\

E-mail: afu@math.fundp.ac.be  

\end{small}

\begin{abstract}
The gravitational instability of Yang-Mills cosmologies is numerically studied with the hamiltonian formulation of the spherically
symmetric Einstein-Yang-Mills equations with $SU(2)$ gauge group. On the short term,
the expansion dilutes the energy densities of the Yang-Mills fluctuations due to their conformal invariance. In this early regime,
the gauge potentials appear oscillating quietly in an interaction potential quite similar to the one of the homogeneous case.
However, on the long term, the expansion finally becomes significantly inhomogeneous and no more mimics a conformal transformation of the metric.
Thereafter, the Yang-Mills fluctuations enter a complex non-linear regime, accompanied by diffusion, while their associated energy
contrasts grow.
\end{abstract}
\pagebreak
\section{Introduction}
The spherically symmetric Einstein-Yang-Mills (EYM) equations with $SU(2)$ gauge group has been extensively studied in 
the past decades, from cosmological solutions (cf. \cite{galtsov,cervero,hosotani,henneaux,shchigolev})
to static configurations (cf. \cite{bartnik,volkov}). The homogeneous and anisotropic cosmological models with Yang-Mills (YM) fields
were considered in \cite{darian1,darian2,barrow} and exhibit chaotic features. For a detailed review of the litterature prior to 1999, see \cite{volkov} and references therein. More recently, some authors have investigated non-abelian Born-Infeld cosmology \cite{galtsov2}
in which the YM-Born-Infeld lagrangian is no more scale invariant and therefore yields to 
different behaviour from classical YM fields, especially near the singularity. \\
\\
However, most of the work done so far provides (semi-)analytical and/or numerical solutions in the case of only one independent variable:
time (cosmological solutions) or radius (gravitating non abelian solitons and black holes). Nevertheless, some authors \cite{shchigolev} have
proposed time-dependent inhomogeneous analytical solutions of EYM theory but at the cost of strong restrictions on the form
of the metric and the gauge potentials. As well, perturbation methods have been used several times
to study the instability of self-gravitating non-abelian solutions \cite{volkov}. Here, we use the hamiltonian formulation of spherically
symmetric EYM equations with $SU(2)$ gauge group to build a numerical method that allows us to find numerical time-dependent inhomogeneous
solutions under a few general assumptions.
In this paper, we apply it to the question of gravitational instability of YM cosmologies, by studying collapsing shells, which
was done for Minkowski space in \cite{shells}.\\
\\
In the case of homogeneous and isotropic spacetimes, the conformal invariance of YM fields allows to solve separately Einstein
and YM equations as the resulting solution is a radiation dominated universe. 
The general ansatz for open, flat and closed cosmologies filled with $su(2)-$valued YM fields was proposed in \cite{galtsov},
for which the solution in \cite{hosotani} is a special case. 
In this paper, we will investigate departures from this ansatz, consequent to some initial perturbations,
in order to characterize their gravitational instability. Although, our approach differs from usual perturbation theory in the sense
that we do not linearize the equations, we consider here time-dependent inhomogeneous numerical solutions \textit{around} the general
ansatz.
One will argue that the conformal invariance of YM fields implies that any primeval excitation of these fields
will be diluted as the universe expands. We will see that this is true only in the first times of the evolution of the fluctuations,
when the expansion is not too inhomogeneous and looks like a conformal transformation of the metric. Thereafter, interesting phenomena
such as diffusion and self-interaction arise which results in growing instabilities.\\
\\
In section 2, we remind the reader about the hamiltonian formulation of the spherically symmetric EYM system
with $SU(2)$ gauge group, firstly introduced in \cite{cordero}, then we recall the Gal'tsov-Volkov ansatz that we will use as a background solution.
Thereafter, the numerical method used to integrate the hamiltonian formulation is introduced in section 3, a test on the homogeneous
ansatz being provided. The section 4 is devoted to a qualitative discussion of the gravitational instability from the numerical results
presented in the figures at the end of the paper. Finally, we conclude by some comments on the application of our results to cosmology
as well as some perspectives.
\section{Theoretical Framework}
\subsection{Hamiltonian Formulation}
The hamiltonian approach of the gravitational field equations is the well-known \textit{Arnowitt-Deser-Misner} formalism \cite{ADM} which
can be completed by a hamiltonian treatment of YM equations describing a gauge interaction \cite{cordero,sundermeyer}. 
The spherically symmetric inhomogeneous vacuum space-times were studied in \cite{berger,moussiaux} while the spherically symmetric EYM 
equations with gauge group $SU(2)$ were first discussed in \cite{cordero}. We briefly recall here the hamiltonian equations of the EYM system
and fix the gauges in which we will work for the rest of this paper.\\
The coupled EYM action can be written as
\begin{equation}
\label{action1}
S=\int\sqrt{-g}\left\{-\frac{1}{2\kappa}\mathcal{R}-\frac{1}{4}F_{\mu\nu}^{\bf{a}}F^{\mu\nu}_{\bf{a}}\right\}d^4 x 
\end{equation}
where $\kappa=8\pi G$ ($c=1$), $g$ is the determinant of the 4-metric, $\mathcal{R}$ is the scalar curvature, $F_{\mu\nu}^{\bf{a}}$ are the components of the YM
field strength tensor\footnote{Gauge indices will be indicated by bold latin letters or bold numbers while
latin indices are only spatial and greek indices are for spacetime.}. 
For a particular gauge group $G$ locally specified by its Lie algebra 
$\left[T_{\bf{a}},\;T_{\bf{b}}\right]=if_{\bf{abc}}T_{\bf{c}}$, the field strength tensor is given through its components by
$ F_{\mu\nu}^{\bf{c}}=\partial_\mu A_{\nu}^{\bf{c}}-\partial_\nu A_{\mu}^{\bf{c}}+f_{\bf{abc}}A_{\mu}^{\bf{a}}A_{\nu}^{\bf{b}}$ where the $A_{\mu}^{\bf{a}}$'s
are the components of the one-form connection $\mathbf{A}=A_{\mu}dx^\mu=A_{\mu}^{\bf{a}}T_{\bf{a}}dx^\mu\cdot$ 
Here, we will focus on the gauge group $SU(2)$
with generators $T_{\bf{a}}=\frac{1}{2}\tau_{\bf{a}}$ where the $\tau_{\bf{a}}$ 
are the usual Pauli matrices and the structure constants are given by the completely antisymmetric tensor
$f_{\bf{abc}}=\epsilon_{\bf{abc}}$. The normalization is 
$Tr(T_{\bf{a}}T_{\bf{b}})=\frac{1}{2}\delta_{\bf{ab}}\cdot$\\
Assuming spherical symmetry of the spatial slices of the universe, we write the metric in a \textbf{3+1} decomposition as follows :
\begin{equation}
\label{metric}
ds^2=\left(-N^2+N_\chi N^\chi\right)dt^2+2N_\chi d\chi dt+e^{2\mu}d\chi^2+e^{2\lambda}d\Omega^2,
\end{equation}
where $d\Omega^2=d\theta^2+\sin^2\theta\; d\varphi^2 $ is the solid angle element, $N$ and $N_\chi$ are the lapse and shift function respectively\footnote{The angular components of the shift vector $N_\theta =g_{02}$ and $N_\varphi =g_{03}$ vanish exactly due to spherical symmetry.},
all these components $N,\;N_\chi,\;\mu,\;\lambda$ being functions of the coordinates $\chi$ and $t$.\\
If the gauge fields $A_\mu$ exhibit spherical symmetry up to a gauge
(see \cite{cordero,forgacs,henneaux}), they can be written thanks to the so-called
$Witten\; ansatz$ \cite{witten} (with the same conventions as in \cite{bartnik}):
\begin{eqnarray}
A_{\mu}dx^{\mu}&=&a T_{\bf{3}} dt + b T_{\bf{3}} d\chi+ \left(c T_{\bf{1}}+d T_{\bf{2}}\right)d\theta+ \nonumber\\
&&\left(\cot \theta\; T_{\bf{3}}+c T_{\bf{2}}-d T_{\bf{1}}\right)\sin\theta d\varphi,\label{connection}
\end{eqnarray}
where $a,\; b,\; c,\; d$ are all functions of both coordinates $\chi$ and $t$. In Eq.(\ref{connection}), only 
three components are of physical relevance since the potential $d$ can be gauged away by a transformation 
that preserves the $SO(3)$ symmetry (see \cite{galtsov,galtsov2} and references therein for a detailed discussion).\\
If we now let the spatial components of the metric $g_{ij}$ and the connection $A_{i}^{\bf{a}}$ ($i=1,\;2,\;3$) be canonical
variables and if we consider the lapse and shift functions $N,\;N_\chi$ as well as the electric components $A_0^{\bf{a}}$ of the YM connection as being 
Lagrange multipliers, it is possible to define some momenta conjugated to these variables and then to rewrite the coupled EYM action 
Eq.(\ref{action1}) under constrained hamiltonian form (cf. \cite{cordero,ADM,sundermeyer} and references therein):
\begin{eqnarray}
\label{action2}
S=\int\int\left\{\pi^{ij}\;\partial_0 g_{ij}+\pi_{A}^{i\;\bf{a}}\;\partial_0 A_{i\;\bf{a}}-N\mathcal{H}_{\perp}-N^i \mathcal{H}_i
-A_{0}^{\bf{a}}\mathcal{G}_{\bf{a}}\right\}dt\; d^3 x,
\end{eqnarray}
where $\pi^{ij}=\sqrt{^{(3)}g}\left(g^{ij}K_l^l-K^{ij}\right)$ ($K^{ij}$ being the extrinsic curvature tensor),  
$\pi_{A}^{j\;\bf{a}}=\frac{\sqrt{^{(3)}g}}{N}\left(g^{ij}F_{0i}^{\bf{a}}-N^k g^{ij}F_{ki}^{\bf{a}}\right)$. In Eq.(\ref{action2}),
the generators of normal and tangential deformations ($\mathcal{H}_\perp$ and $\mathcal{H}_i$, respectively) and those
of internal gauge transformations $\mathcal{G}_{\bf{a}}$ are given by\footnote{The gauge coupling constant has been set to one.}
\begin{eqnarray}
\mathcal{H}_\perp&=&-\sqrt{-g}\left[R+g^{-1}\left(\frac{1}{2}\pi^2-\pi^{ij}\pi_{ij}\right)\right]\nonumber\\
&&+\frac{\sqrt{-g}}{2}g_{ij}\left(\pi_{A}^{j\;\bf{a}}\pi_{A\;\;\bf{a}}^{j}+\mathcal{B}^{i\;\bf{a}}\mathcal{B}^{i}_{\;\bf{a}}\right)\approx 0\\
\mathcal{H}_i&=&-2\pi_{i\;\;|j}^{\; j}+\epsilon_{ijk}\pi_{A\;\bf{a}}^{j}\mathcal{B}^{k\;\bf{a}}\approx 0\\
\mathcal{G}_{\bf{a}}&=&-\pi_{A\;{\bf{a}},i}^{i}+f_{\bf{ab}}^{\;\;\;\bf{c}}\pi_{A\;\bf{c}}^i\;A_i^{\bf{b}}\approx 0
\end{eqnarray}
with $R$ the scalar curvature of the 3-metric and $\pi=\pi_i^i$, $|$ denoting the covariant derivative w.r.t. the 3-metric
and the vector $\mathcal{B}_{\bf{a}}^i$ reading
\begin{equation}
\mathcal{B}_{\bf{a}}^i=\frac{1}{2}\epsilon^{ijk}\left(A_{k{\bf{a}},j}-A_{j{\bf{a}},k}-f_{{\bf{cba}}}A^{\bf{c}}_{j}A^{\bf{b}}_{k}\right)\cdot
\end{equation}
Let us now focus on the spherically symmetric hamiltonian formalism.\\
For the gravitational part of the EYM system, the canonical variables are chosen as\footnote{We work in the following non coordinate basis : $(\partial_\chi, \partial_\theta, \frac{1}{\sin\theta}\partial_\phi)$}
$g_{ij}=diag\left(e^{2\mu},e^{2\lambda},e^{2\lambda}\right)$
while their conjugate momenta are given by $\pi_{ij}=diag\left(e^{-2\mu}\displaystyle{\frac{\pi_{\mu}}{2}},e^{-2\lambda} \displaystyle{\frac{\pi_{\lambda}}{4}},e^{-2\lambda} \displaystyle{\frac{\pi_{\lambda}}{4}}\right)$.
Variation of Eq.(\ref{action2}) w.r.t. the lapse and shift functions $N$ and $N_\chi$ will provide the hamiltonian and super-momentum constraints
while the variation w.r.t. the canonical variables $\mu$, $\lambda$ and their conjugate momenta $\pi_\mu$ and $\pi_\lambda$ will lead
to the Hamilton equations. As we are interested in perturbations of homogeneous solutions and as the YM lagrangian in Eq.(\ref{action1})
is conformally invariant, we will work in the conformal gauge for the 
gravitational field: $N(\chi,t)=R(t)$ where $R(t)$ is the scale factor given by the Friedmann equation and $N_\chi(\chi,t)=0$. 
After the variation and the choice of gauge, we now have for the dynamics of the gravitational field the following constraints:
\begin{eqnarray}
\mathcal{H}_{1}=0&\equiv&-\mu'\pi_{\mu}-\lambda' \pi_{\lambda}+\pi_{\mu}'-16\pi N e^{2\lambda+\mu}\;T^{0}_{1}=0\label{h1=0}\\
\mathcal{H}=0&\equiv&
\frac{1}{8}e^{-\mu-2\lambda}\left\{8e^{4\lambda}(-4\mu'\lambda'+6\lambda^{'2}+4\lambda'')-16e^{2\mu+2\lambda}+\pi_{\mu}^{2}-2\pi_{\mu}
\pi_\lambda\right\}\nonumber\\
&&+16\pi e^{2\lambda+\mu}\;T^0_0=0,\label{h0=0}
\end{eqnarray}
and the following Hamilton equations,
\begin{eqnarray}
\dot{\mu}&=& \frac{1}{4}e^{-\mu-2\lambda}N(\pi_\mu-\pi_\lambda)\nonumber\\
\dot{\lambda}&=& -\;\frac{1}{4}e^{-\mu-2\lambda}N\pi_\mu\label{einstein1}\\
\dot{\pi_\mu}&=& \frac{1}{8}e^{-\mu-2\lambda}N
\left\{16e^{2\lambda}(e^{2\mu}-\lambda^{'2}e^{2\lambda})+\pi_\mu(\pi_\mu-2\pi_\lambda)\right\}\nonumber\\
&&-16\pi\;N\; e^{2\lambda+\mu}\;T_1^1\nonumber\\
\dot{\pi_\lambda}&=& \frac{1}{4}e^{-\mu-2\lambda}N
\left\{16e^{4\lambda}(\mu'\lambda'-\lambda^{'2}-\lambda'')+\pi_\mu(\pi_\mu-2\pi_\lambda)\right\}\nonumber\\
& & -32\pi\;N\;e^{2\lambda+\mu}\;T_2^2\nonumber
\end{eqnarray}
where $T_{\mu}^\nu$ are the components of the YM stress-energy tensor (see further), and where a prime and a dot denote
derivations w.r.t. position $\chi$ and time $t$ respectively.\\
On the YM side, we have, according to Eq.(\ref{connection}),
only one electrical component $A_{0}^{\bf{3}}=a$ and three couples of variables and momenta 
$(A_{1}^{\bf{3}},\;\pi_{A}^{1,\bf{3}})=(b,\;\pi_b)$, $(A_{2}^{\bf{1}},\;\pi_{A}^{2,\bf{1}})=(A_{3}^{\bf{2}},\;\pi_{A}^{3,\bf{2}})=(c,\;\pi_c)$, 
$(A_{2}^{\bf{2}},\;\pi_{A}^{2,\bf{2}})=-(A_{3}^{\bf{1}},\;\pi_{A}^{3,\bf{1}})=(d,\;\pi_d)\cdot$   
Varying the YM action w.r.t. these variables and fixing the gauge $d(\chi,\;t)=0$ so that $\pi_d=\displaystyle{\frac{e^\mu}{N}ac}$
leads to the following constraint
\begin{equation}
\label{ym1}
\mathcal{G}^{\bf{3}}=0\equiv\pi_b '+2\frac{e^\mu}{N}ac^2=0,
\end{equation}
and to the following Hamilton equations
\begin{eqnarray}
\dot{b}&=&a'+Ne^{\mu-2\lambda}\pi_b\nonumber\\
\dot{c}&=&Ne^{-\mu}\label{ym3}\pi_c\nonumber\\
\dot{\pi_b}&=&-2Ne^{-\mu}bc^2\label{ym2}\\
\dot{\pi_c}&=&\frac{e^\mu}{N}a^2 c+\left(Ne^{-\mu}c'\right)'-Ne^{-\mu}b^2c-Ne^{\mu-2\lambda}c(c^2-1)\nonumber\\
\dot{a}&=&N^2\; e^{-2\mu}\left(b'+2\frac{bc'}{c}-\mu'\; b\right)-2N\;e^{-\mu}\frac{a\pi_c}{c}-\frac{N}{4}\;e^{-\mu-2\lambda}(\pi_\mu-\pi_\lambda)a\nonumber\\
&&+\frac{\dot{N}}{N}a\cdot\nonumber
\end{eqnarray}
With these variables, we can now write the components of the YM stress-energy tensor 
$T_{\mu\nu}=-2F^{\bf{a}}_{\mu\alpha}F^{\;\;\alpha \bf{a}}_{\nu}+\frac{1}{2}g_{\mu\nu}F^{\bf{a}}_{\alpha\beta}F^{\alpha\beta \bf{a}}$
\begin{eqnarray}
T^{0}_{0}&=&e^{-4\lambda}\frac{\pi_{b}^{2}}{2}+e^{-2\lambda-2\mu}\pi_{c}^{2}+\frac{c^{'2}N^2+e^{2\mu}a^2 c^2+N^2 b^2 c^2}{N^2}e^{-2\lambda-2\mu}\nonumber\\
&&+e^{-4\lambda}\frac{\left(c^2 -1\right)^2}{2}\label{t00}\\
T^{1}_{1}&=&e^{-4\lambda}\frac{\pi_{b}^{2}}{2}-e^{-2\lambda-2\mu}\pi_{c}^{2}-\frac{c^{'2}N^2+e^{2\mu}a^2 c^2+N^2 b^2 c^2}{N^2}e^{-2\lambda-2\mu}\nonumber\\
&&+e^{-4\lambda}\frac{\left(c^2 -1\right)^2}{2}\label{t11}\\
T^{2}_{2}&=&-\frac{e^{-4\lambda}}{2}\left(\pi_{b}^{2}+(c^2 -1)^2\right)\label{t22}\\
T^{0}_{1}&=&2\frac{e^{-2\lambda}}{N^2}\left(Ne^{-\mu}\pi_c c'+abc^2\right)=-N^2 e^{-2\mu}T^{1}_{0},\label{t01}
\end{eqnarray}
in order to achieve the coupling between gauge and gravitation sectors. Eqs.(\ref{h1=0}-\ref{t01}) are the hamiltonian formulation of the well-known
spherically symmetric EYM equations with gauge group $SU(2)$. In the next paragraph, we recall the solution for those fields
in a Friedmann-Lemaître universe before describing the method used to study gravitational instability of such solutions.
\subsection{Yang-Mills Cosmologies}
$Non-abelian$ YM fields can fill a non trivial, radiation dominated, Friedmann-Lemaître universe due to their increased number
of degrees of freedom, and the couplings between them, compared to the abelian Maxwell theory. In particular, the case of gauge group $SU(2)$ has been extensively studied,
so we refer the reader to \cite{galtsov,hosotani,cervero,henneaux,galtsov2,volkov} and others references therein for a review.\\
With the definitions of previous paragraph, we can write the homogeneous solutions we will use for the background in which the fluctuations
will evolve. First, the geometry is given by (in the conformal gauge $N=R(t)$, upper script $B$ stands for $background$):
\begin{eqnarray}
\mu^B&=&\log R\nonumber\\
\lambda^B&=&\log R+\log\Sigma\nonumber\\
\pi_\mu^B&=&-4\dot{R}R\Sigma^2\label{geombkg}\\
\pi_\lambda^B&=&-8\dot{R}R\Sigma^2\nonumber
\end{eqnarray}
where $\Sigma=\sin\chi,\;\chi,\;\sinh\chi$ for closed, flat and open cosmologies respectively.
The conformal invariance and the consequent vanishing trace of the YM stress-energy tensor result in a radiation-dominated universe
for which the Friedmann equation Eq.(\ref{h0=0}) reads (using Eqs.(\ref{geombkg})):
\begin{equation}
H^2=\left(\frac{\dot{R}}{R}\right)^2=\frac{\kappa\mathcal{E}^2}{2R^2}-k,
\end{equation}
with $\mathcal{E}$ some constant to be precised in one moment.
This allows us to write the following solutions
for the scale factor $R(t)$ (see also \cite{landau}).
\begin{eqnarray}
R(t)&=&\sqrt{\frac{\kappa}{3}\mathcal{C}}\;t\;(k=0)\nonumber\\
R(t)&=&\sqrt{\frac{\kappa}{3}\mathcal{C}}\;\cos(t)\;(k=+1)\label{rbkg}\\
R(t)&=&\sqrt{\frac{\kappa}{3}\mathcal{C}}\;\sinh (t)\;(k=-1)\nonumber
\end{eqnarray}
for flat, elliptic and hyperbolic universes, respectively. The constant $\mathcal{C}=\rho R^4$ is a first integral of the conservation equations $\nabla_{\mu}T_{0}^{\mu}=0$. \\
The gauge fields and their conjugate momenta are given by the following ansatz \cite{galtsov,galtsov2}:
\begin{eqnarray}
c^B(\chi,t)&=&\sqrt{1+\Sigma^2\left(\sigma^2-k\right)}\nonumber\\
a^B(\chi,t)&=&-\dot{\sigma}\Sigma\frac{\sqrt{1-k\Sigma^2}}{1+\Sigma^2\left(\sigma^2-k\right)}\nonumber\\
b^B(\chi,t)&=&\sigma\Sigma^2\frac{\left(\sigma^2-k\right)}{1+\Sigma^2\left(\sigma^2-k\right)}\label{gaugebkg}\\
\pi_b^B(\chi,t)&=&\dot{\sigma}\Sigma^2\nonumber\\
\pi_c^B(\chi,t)&=&\frac{\dot{\sigma}\sigma\Sigma^2}{\sqrt{1+\Sigma^2\left(\sigma^2-k\right)}}\nonumber
\end{eqnarray}
where $k=+1,\;0,\;-1$ for closed, flat and open cosmologies respectively, and
$\sigma$ is a function of time. By substituting the ansatz for the gauge fields Eqs.(\ref{gaugebkg}) into Eqs.(\ref{ym1}-\ref{ym2}),
the YM equations now reduce to
\begin{equation}
\label{sigma1}
\ddot{\sigma}+2\sigma\left(\sigma^2-k\right)=0,
\end{equation}
or, expressed as a first integral
\begin{equation}
\label{sigma2}
\dot{\sigma}^2+\left(\sigma^2-k\right)^2=\mathcal{E}^2,
\end{equation}
where $\mathcal{E}^2=\frac{2}{3}\mathcal{C}=\frac{2}{3}\rho R^4$. The conformal invariance
of YM fields is now manifest as the scale factor $R(t)$
does not appear in Eqs.(\ref{sigma1},\ref{sigma2}). 
Therefore, in the conformal gauge, the time part of the gauge fields $\sigma(t)$ oscillates in the potential well $\left(\sigma^2-k\right)^2$.
In the synchronous gauge $N=1$, it can be shown that the frequency of the oscillation decreases with expansion.\\ 
In the rest of the paper, we will focus on the case of flat space as the observed features of gravitational instability can be
deduced from the general properties of the homogeneous solution presented in this paragraph and can be therefore extrapolated to curved
cosmologies.
\section{Overview of the Method}
In order to integrate the hamiltonian EYM equations, we will use a two-step numerical procedure often referred
as \textit{free evolution} method. First, we solve the three constraints
Eqs.(\ref{h1=0}-\ref{h0=0}) and Eq.(\ref{ym1}) -- this consists of the \textit{initial value problem} (IVP) -- then we propagate the initial data
of the fields imposed or found 
by the IVP with the Hamilton equations Eqs.(\ref{einstein1}-\ref{ym2}) and check the accuracy and stability of evolution by evaluating the error made on the constraints.
The growing violation of the constraints in such procedures and the development of formalisms reducing it are very hot topics in numerical 
relativity (see for example \cite{yoneda}, and \cite{seidel} for an introduction to numerical relativity). 
Here, we choose the canonical approach to the EYM system (original ADM
formalism \cite{ADM} for gravitation and classical treatment for YM \cite{cordero}) which we will
rewrite in section 3.2 in order to improve stability and accuracy.\\
\\
But before going further, let us give some comments on the particular choice of gauges we made for both the gravitational and YM fields.
For the first one, we chose a slightly modificated version of the famous \textit{geodesic slicing}: $N(\chi,t)=1$, $N_i(\chi,t)=0$,
which we called the conformal gauge: $N(\chi,t)=R(t)$ and $N_i(\chi,t)=0$. Even though the geodesic condition is well-known
to produce coordinate singularities (cf. \cite{york}), it is the most efficient for weak gravitational fields such as instabilities
in a homogeneous background. The reason why we chose a conformal gauge with $N=R(t)$ rather than a synchronous one $N=1$ 
is because the YM fields oscillate
at a fixed frequency in the first while 
their period is stretched with time in the second.
Therefore, using a synchronous gauge would have increased the time over which the fluctuations around the homogeneous solution
significantly change. And, as we used an explicit scheme, it was crucial to reduce this time in order to avoid numerical discrepancies.\\
Furthermore, our choice of a gauge dependent formalism ($d$ was set to $0$) was prefered for two reasons. The first is that it will sweep away
one degree of freedom in the hamiltonian formalism, which simplifies the equations and therefore lowers truncation errors as
well as avoiding those related to the couple $(d,\;\pi_d)$ of unknowns.
The second is because it simplifies the algorithm. Indeed, with a gauge-invariant formalism, the electric potential $A_0^{\bf{3}}=a$
which represents the freedom of making arbitrary gauge transformations in the YM fields during the evolution of the system
has to be determined by an additionnal gauge condition. By fixing the gauge,
we are left with a Hamilton equation for this component.
However, by reducing in this way the number of degrees of freedom in the analysis, we cancel an additional scalar field in the theory,
in the context of the two-dimensional abelian Higgs mechanism that consitutes the spherically symmetric EYM system with gauge group $SU(2)$
(cf. \cite{forgacs,volkov}).
But now, we have learned from a more deep analysis of our results that these scalar fields and the interactions related to it bring
a good part of the energy of the fluctuations. Part of our future work will be to examine how energy spread out into the different modes
of the fully gauge invariant YM equations and how it affects the results presented here.

\subsection{Initial Value Problem (IVP)}
In this problem, we have to determine the initial distribution of nine fields 
$(\mu_0,\;\lambda_0,\;\pi_{\mu,0},\;\pi_{\lambda,0},\;a_0,\;b_0,\;c_0,\;\pi_{b,0},\;\pi_{c,0})\footnote{Lower script zero means at the time $t=0$, for example $\mu_0=\mu(\chi,0)\cdot$}$
that verifies the three constraints (two for the gravitational field and one from the gauge sector) of the EYM hamiltonian system.
Here, we will only write the formulas for a perturbation of a flat Friedmann-Lemaître universe as they can easily be generalized to closed and 
hyperbolic cosmologies.
In order to do this, we fix six fields as follows:
\begin{itemize}
\item[$\bullet$] We start working some time after the Big Bang to avoid primeval singularities : $R(0)=1$. 
Please note that the formula given in
section 2.2 for the scale factors should be translated accordingly.
\item[$\bullet$] We set $\lambda_0=\ln \chi$ so that the initial circonferential distances are taken as references for the coordinate $\chi$.
\item[$\bullet$] We perturb the following fields around their homogeneous values :
\begin{eqnarray}
\dot{\lambda}_0&=&\sqrt{\frac{\kappa}{3}\rho_0^{B}}+\epsilon_h(\chi)\nonumber\\
a_0&=&a_0^B+\epsilon_a(\chi)\nonumber\\
b_0&=&b_0^B+\epsilon_b(\chi)\label{pert}\\
c_0&=&c_0^B+\epsilon_c(\chi)\nonumber\\
\dot{c}_0&=&\pi_{c,0}^{B}+\epsilon_{\dot{c}}(\chi),\nonumber
\end{eqnarray}
where $\rho_0^{B}$ is the initial background density, taken as a parameter. The functions $\epsilon_i \;(\chi)$ are also provided and will be
taken so as to have convenient boundary conditions for the fields. Typically, we will study the gravitational instability of shells in a Friedmann-Lemaître background, i.e. we will impose that the $\epsilon_i$'s vanish at the boundaries of the $\chi$ coordinate interval. For example,
we can take for the $\epsilon_i$'s some gaussian functions.
\end{itemize}
From there, we solve the super-momentum constraint Eq.(\ref{h1=0}) w.r.t. to $\pi_{\lambda,0}$ and the YM constraint Eq.(\ref{ym1}) w.r.t. $\mu_0$
and we substitute the results into the hamiltonian constraint to get the following differential equation for $\pi_{b,0}$:
\begin{equation}
\label{ivp1}
A\phi''+B\phi'+C\phi^{'3}-4\pi\phi^2 \phi^{'3}=0
\end{equation}
where we put $\phi=\pi_{b,0}(\chi)=\pi_b(\chi,0)$ and where the coefficients are
\begin{eqnarray}
A&=&8a_0^2\;c_0^4\;\chi^3\nonumber\\
B&=&-4\chi^2\left(2a_0'a_0c_0^4\chi+8\pi c_0^{'2}a_0^2 c_0^4 +4c_0'a_0^2c_0^3\chi+8\pi a_0^2 b_0^2 c_0^6+a_0^2c_0^4\right)\label{ivp2}\\
C&=&2\dot{\lambda}_0'\dot{\lambda}_0\chi^5-8\pi a_0^2c_0^2 \chi^2+16\pi \chi^3 a_0 b_0 c_0^2 \dot{\lambda}_0-4\pi c_0^4+ 8\pi c_0^2
+3\dot{\lambda}_0^2\chi^4-4\pi\nonumber\\
&&+\chi^2+16\pi\chi^3 c_0'\dot{\lambda}_0\dot{c}_0-8\pi \chi^2\dot{c}_0^2\cdot\nonumber
\end{eqnarray}
The fields with lower script $0$ mentionned in Eq.(\ref{ivp2}) are those given by hypothesis in Eq.(\ref{pert}). From the solution $\phi=\pi_{b,0}$ of 
Eq.(\ref{ivp1}), one can complete the set of initial data with the following relations
\begin{eqnarray}
\mu_0&=&\ln\left(-\frac{\phi'}{2a_0^2 c_0^2}\right)\nonumber\\
\pi_{\mu,0}&=&-4\dot{\lambda}_0 e^{\mu_0}\chi^2\nonumber\\
\pi_{\lambda,0}&=&-32\pi e^{\mu_0}c_0'\dot{c}_0\chi-4e^{\mu_0}\chi^3\epsilon_h'-32\pi e^{\mu_0}a_0b_0c_0^2\chi+2\pi_{\mu,0}\label{ivp3}\\
\pi_{c,0}&=&e^{\mu_0}\dot{c}_0\cdot\nonumber
\end{eqnarray}
We used a finite difference scheme to solve Eq.(\ref{ivp1}) as follows: defining $\chi_i=\chi_{MIN}+\Delta\chi(i-1)$ ($i=1,\;\cdots,\;N_\chi+1$),
$\phi_i=\phi(\chi_i)$ and analogous definition for other fields, Eq.(\ref{ivp1}) becomes a non-linear system of $N_\chi-2$ algebraic equations
for determining the $\phi_i$'s:
\begin{eqnarray}
&&A_i \frac{\phi_{i+1}-2\phi_i+\phi_{i-1}}{\Delta\chi^2}+B_i \frac{\phi_{i+1}-\phi_{i-1}}{2\Delta\chi}+C_i \left(\frac{\phi_{i+1}-\phi_{i-1}}{2\Delta\chi}\right)^3\nonumber\\
&&-4\pi \phi_i^2\left(\frac{\phi_{i+1}-\phi_{i-1}}{2\Delta\chi}\right)^3=0 \label{ivp4}
\end{eqnarray} 
with the boundary conditions
\begin{eqnarray}
\phi_1&=&\pi_{b,0}^B(\chi_1)=\dot{\sigma}_0\chi_{MIN}^2\\
\phi_{N_\chi+1}&=&\pi_{b,0}^B(\chi_{N_\chi+1})=\dot{\sigma}_0\chi_{MAX}^2,\nonumber
\end{eqnarray}
where $\dot{\sigma}_0=\sqrt{\frac{2}{3}\rho_0^B-\sigma_0^4}$, $\sigma_0$ becoming another parameter setting the initial position in phase space
for the time part of the gauge fields filling the flat background. We choose a Newton-Raphson method (cf. \cite{recipe}) to solve Eq.(\ref{ivp4}) with the initial
guess : $\phi_i=\pi_{b,0}^B(\chi_i)=\dot{\sigma}_0\chi_i^2$ and stop the relaxation when the error on the equation is saturated. 
\subsection{Propagation of Hamilton Equations}
In this paragraph, we first rewrite the EYM constraints and Hamilton equations in order to render the numerical integration more 
accurate and stable.
Indeed, a direct integration of Eqs.(\ref{einstein1}) and Eqs.(\ref{ym2}) will have to recover first
the homogeneous solution Eqs.(\ref{geombkg}-\ref{gaugebkg}) before focusing to its perturbation. After having first directly integrated 
Eqs.(\ref{einstein1}) and Eqs.(\ref{ym2}), we found it more interesting to rewrite
the equations so that they would be more adapted to the study of the gravitational instability and less dependent on the homogeneous solution. 
The change of variables explained below results in a faster and more accurate method.\\
\\
Let us again assume a flat Friedmann-Lemaître background and proceed to the following change of variables:
\begin{eqnarray}
e^\mu&=&m(\chi,t)\;R(t)\nonumber\\
e^{2\lambda}&=&l(\chi,t)\;R^2(t)\chi^2\nonumber\\
\pi_\mu&=&-4\;R(t)\chi^2\;\pi_m(\chi,t)\nonumber\\
\pi_\lambda&=&-8\;R(t)\chi^2\;\pi_l(\chi,t)\nonumber\\
a(\chi,t)&=&-\frac{\alpha(\chi,t)\chi}{1+\chi^2\gamma(\chi,t)}\\
b(\chi,t)&=&\frac{\beta(\chi,t)\chi^2}{1+\chi^2\gamma(\chi,t)}\nonumber\\
c(\chi,t)&=&\sqrt{1+\chi^2\gamma(\chi,t)}\nonumber\\
\pi_b(\chi,t)&=&\pi_\beta(\chi,t)\chi^2\nonumber\\
\pi_c(\chi,t)&=&\frac{\pi_\gamma\chi^2}{\sqrt{1+\chi^2\gamma(\chi,t)}}\nonumber
\end{eqnarray}
so that the homogeneous solutions $(m^B,\;l^B,\;\pi_m^B,\;\pi_l^B,\;\alpha^B,\;\beta^B,\;\gamma^B,\;\pi_\beta^B,\;\pi_\gamma^B)$ are now a set of
constants w.r.t. $\chi$ : $(1,\;1,\;\dot{R},\;\dot{R},\;\dot{\sigma},\;\sigma^3,\;\sigma^2,\;\dot{\sigma},\;\sigma\dot{\sigma})\cdot$
In terms of these new variables, the hamiltonian formulation of EYM system Eqs.(\ref{h1=0}-\ref{t01}) now become for the constraints
\begin{eqnarray}
\mathcal{H}_{1}=0&\equiv&\frac{m'}{m}\pi_m\chi+\frac{l'\chi+2l}{l}\pi_l-2\pi_m-\chi\pi_m'-4\pi\;R^3\;m\;l\;\chi\;T_1^0=0\nonumber\\
\mathcal{H}=0&\equiv&
-\frac{m'R\left(l'\chi+2l\right)}{m^2\chi}-\frac{R\left(l'\chi+2l\right)^2}{4ml\chi^2}+\frac{R\left(l''\chi^2+4\chi l' +2l\right)}{m\chi^2}-\frac{mR}{\chi^2}\nonumber\\
&&+\frac{\pi_m}{mlR}\left(\pi_m-4\pi_l\right)+8\pi\;R^3\;m\;l\;T_0^0=0\label{constraints}\\
\mathcal{G}^{\bf{3}}=0&\equiv&\pi_\beta'\chi+2\pi_\beta-2m\alpha=0\nonumber
\end{eqnarray}
and for the Hamilton equations
\begin{eqnarray}
\dot{m}&=&-\frac{\dot{R}}{R}m-\frac{\pi_m-2\pi_l}{Rl}\nonumber\\
\dot{l}&=&-2\frac{\dot{R}}{R}l+2\frac{\pi_m}{mR}\label{einstein2}\\
\dot{\pi_m}&=&-\frac{\dot{R}}{R}\pi_m-\frac{mR}{2\chi^2}+\frac{\left(l'\chi+2l\right)^2R}{8ml\chi^2}-\frac{\pi_m}{2mlR}\left(\pi_m-4\pi_l\right)\nonumber\\
&& +4\pi\;R^3\;m\;l\;T_1^1\nonumber\\
\dot{\pi_l}&=&-\frac{\dot{R}}{R}\pi_l-\frac{m'R\left(l'\chi+2l\right)}{4m^2\chi}+\frac{R}{4m\chi^2}\left(4l'\chi+l''\chi^2+2l-\frac{\left(l'\chi+2l\right)^2}{2l}\right)\nonumber\\
&& -\frac{\pi_m}{2mlR}\left(\pi_m-4\pi_l\right)+4\pi\;R^3\;m\;l\;T_2^2\nonumber
\end{eqnarray}
with the following definitions for the components of the stress-energy tensor
\begin{eqnarray}
T_0^0&=&\frac{1}{2l^2\;R^4}\left(\pi_\beta^2+\gamma^2\right)+\frac{\left(\pi_\gamma^2\chi^2+\left(\gamma+\frac{\chi}{2}\gamma'
\right)^2+\beta^2\chi^2\right)}{m^2\;l\;R^4\left(1+\chi^2\gamma\right)}\nonumber\\
&&+\frac{\alpha^2}{l\;R^4\left(1+\chi^2\gamma\right)}\label{tmunu}\\
T_2^2&=&\frac{1}{2l^2\;R^4}\left(\pi_\beta^2+\gamma^2\right)\nonumber\\
T^0_1&=&\frac{2\chi}{l\;R^4\chi^2\left(1+\chi^2\gamma\right)}\left(\frac{\pi_\gamma}{m}\left(\gamma+\frac{\chi}{2}\gamma'
\right)-\alpha\beta\right)\nonumber
\end{eqnarray}
($T_1^1$ is the same expression as $T_0^0$ except for the sign of the last two terms). Finally, the YM equations can be written
\begin{eqnarray}
\dot{\beta}&=&\frac{2\beta\pi_\gamma \chi^2}{m\left(1+\chi^2\gamma\right)}-\frac{\alpha'}{\chi}-\frac{\alpha}{\chi^2}+\frac{m\pi_\beta}{l\;\chi^2}
+\frac{\alpha\left(2\gamma+\chi\gamma'\right)}{\left(1+\chi^2\gamma\right)}+\frac{m}{l}\pi_\beta \gamma\nonumber\\
\dot{\gamma}&=&2\frac{\pi_\gamma}{m}\nonumber\\
\dot{\pi_\beta}&=&-2\frac{\beta}{m}\label{ymii}\\
\dot{\pi_\gamma}&=&\frac{\left(\pi_\gamma^2-\beta^2\right)\chi^2}{m\left(1+\chi^2\gamma\right)}+\frac{m\alpha^2}{\left(1+\chi^2\gamma\right)}
-\frac{m'\left(2\gamma+\chi\gamma'\right)}{2m^2\;\chi}+\frac{2\gamma+4\chi\gamma'+\gamma''\chi^2}{2m\chi^2}\nonumber\\
&&-\frac{\left(2\gamma+\chi\gamma'\right)^2}{4m\left(1+\chi^2\gamma\right)}-\frac{m}{l\;\chi^2}\;\gamma-\frac{m}{l}\;\gamma^2\nonumber\\
\dot{\alpha}&=&-\frac{2\beta+\beta'\chi}{m^2}+\frac{m'}{m^3}\;\beta\chi+\alpha\left(\frac{\pi_m-2\pi_l}{mlR}+\frac{\dot{R}}{R}\right)\cdot\nonumber
\end{eqnarray}
In order to integrate Eqs.(\ref{einstein2}-\ref{ymii}), we used a second order difference scheme in both $\chi$ and $t$ (leapfrog-like) 
and, for the first step only, a forward time centered space scheme ($FTCS$, cf. \cite{recipe}). This explicit scheme is easy to implement
but yields a
very strong Courant condition in this case. Anyway, for the study of gravitational instability  during a few oscillations of $\sigma$ in Eq.(\ref{sigma1}), it
will be sufficient.  \\
Another crucial point is the question of boundary conditions, already mentionned in the previous paragraph. As we focus on shells
and all our initial perturbations have been chosen to vanish rapidly on the edges of the considered interval in $\chi$, we impose
that the boundary conditions follow the homogeneous evolution. For example, a field $f(\chi,t)\equiv f_{i,j}$ will have as
boundary conditions
\begin{eqnarray}
f_{1,j}&=&f^B_{1,j}\\
f_{N_\chi+1,j}&=&f^B_{N_\chi+1,j}\nonumber
\end{eqnarray}
where the upper script $B$ stands for the $background$ solution Eqs.(\ref{geombkg}-\ref{gaugebkg}).
\subsection{Test on the homogeneous solution}
\begin{figure}
\centerline{%
\begin{tabular}{c@{\hspace{5mm}}c@{\hspace{5mm}}c}
\includegraphics[scale=0.25,angle=0]{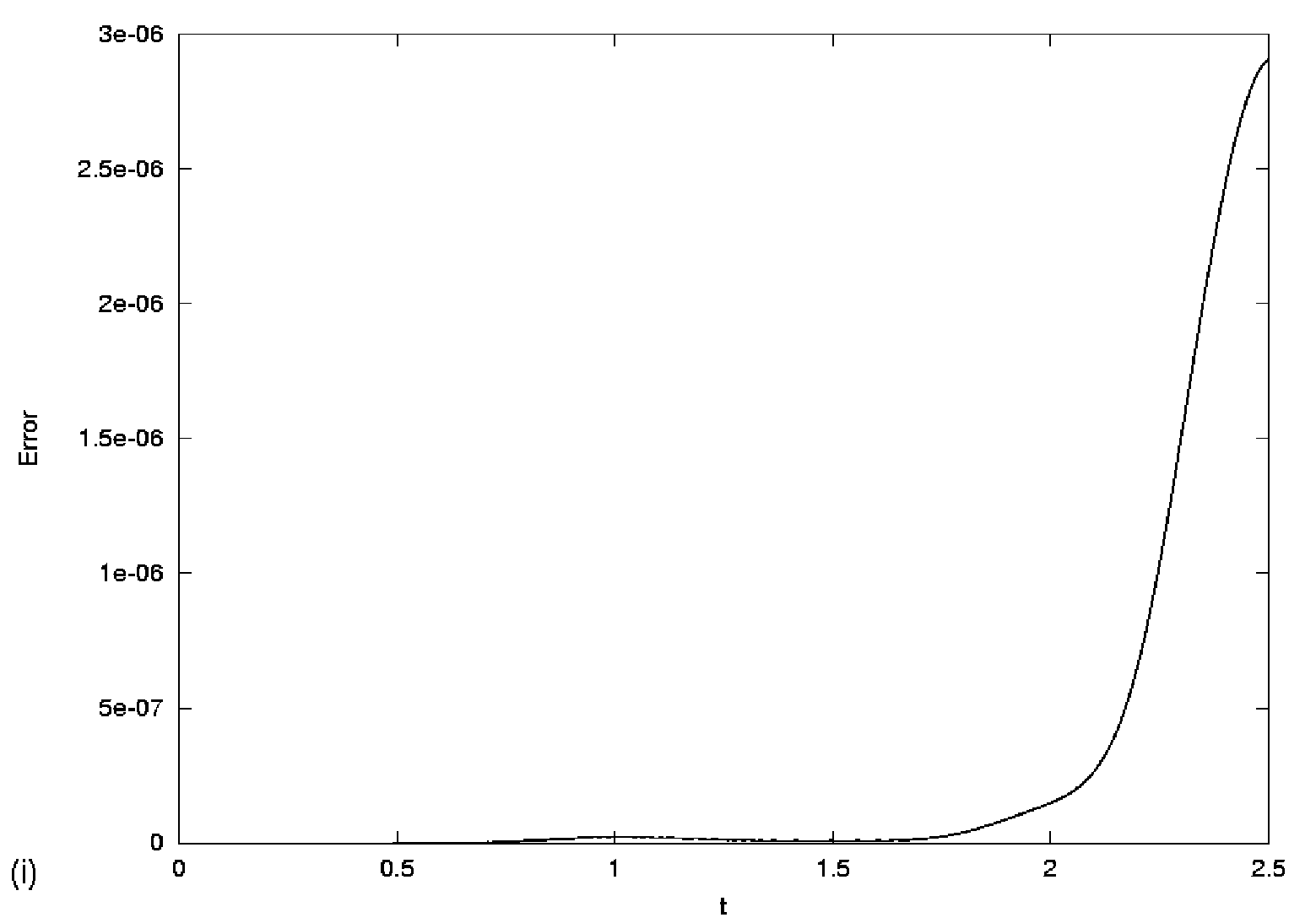} & 
\includegraphics[scale=0.25,angle=0]{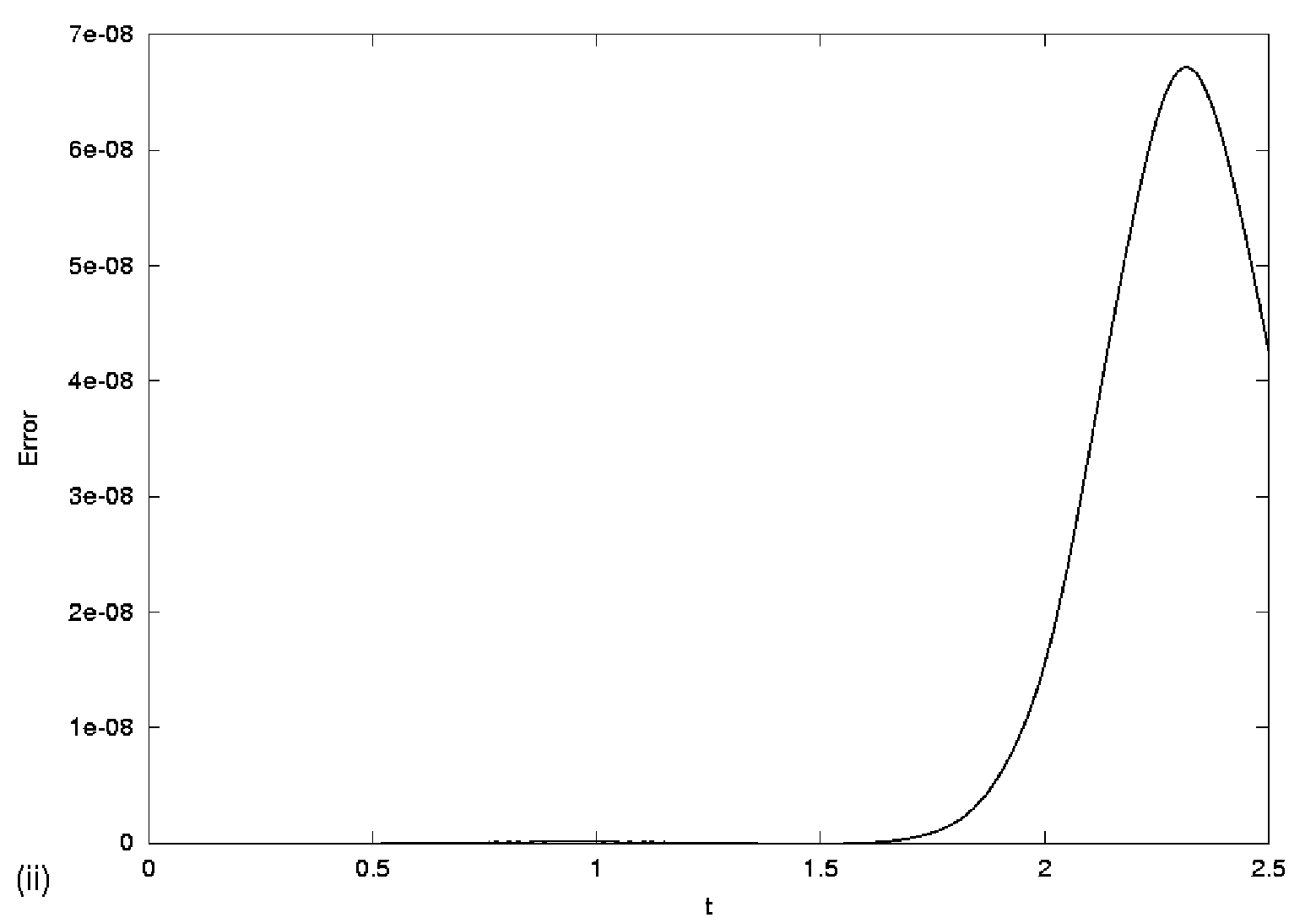} &
\includegraphics[scale=0.25,angle=0]{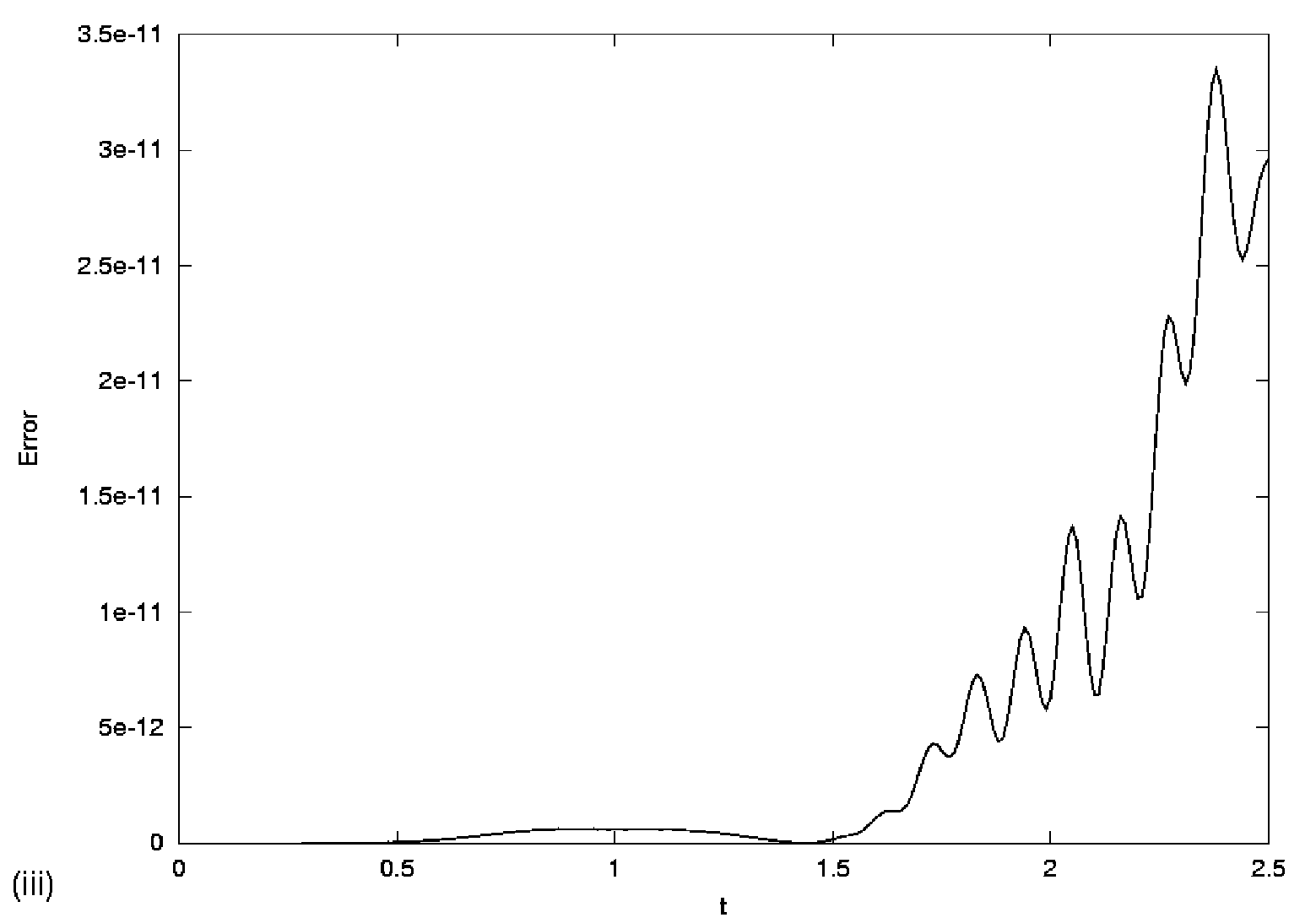} 
\end{tabular}}
\caption{\scriptsize {Mean quadratic errors on the homogeneous solution for the gauge variables.
\textsl{(i) $E_\alpha$ (ii) $E_\beta$ (iii) $E_\gamma$ ($\rho^B_0=15$, $\sigma_0=0$, $\Delta\chi=7\times 10^{-2}$, $\Delta t=10^{-7}$). }
} \normalsize}\label{errgauge}
\end{figure}

\begin{figure}
\centerline{%
\begin{tabular}{c@{\hspace{1cm}}c}
\includegraphics[scale=0.3,angle=0]{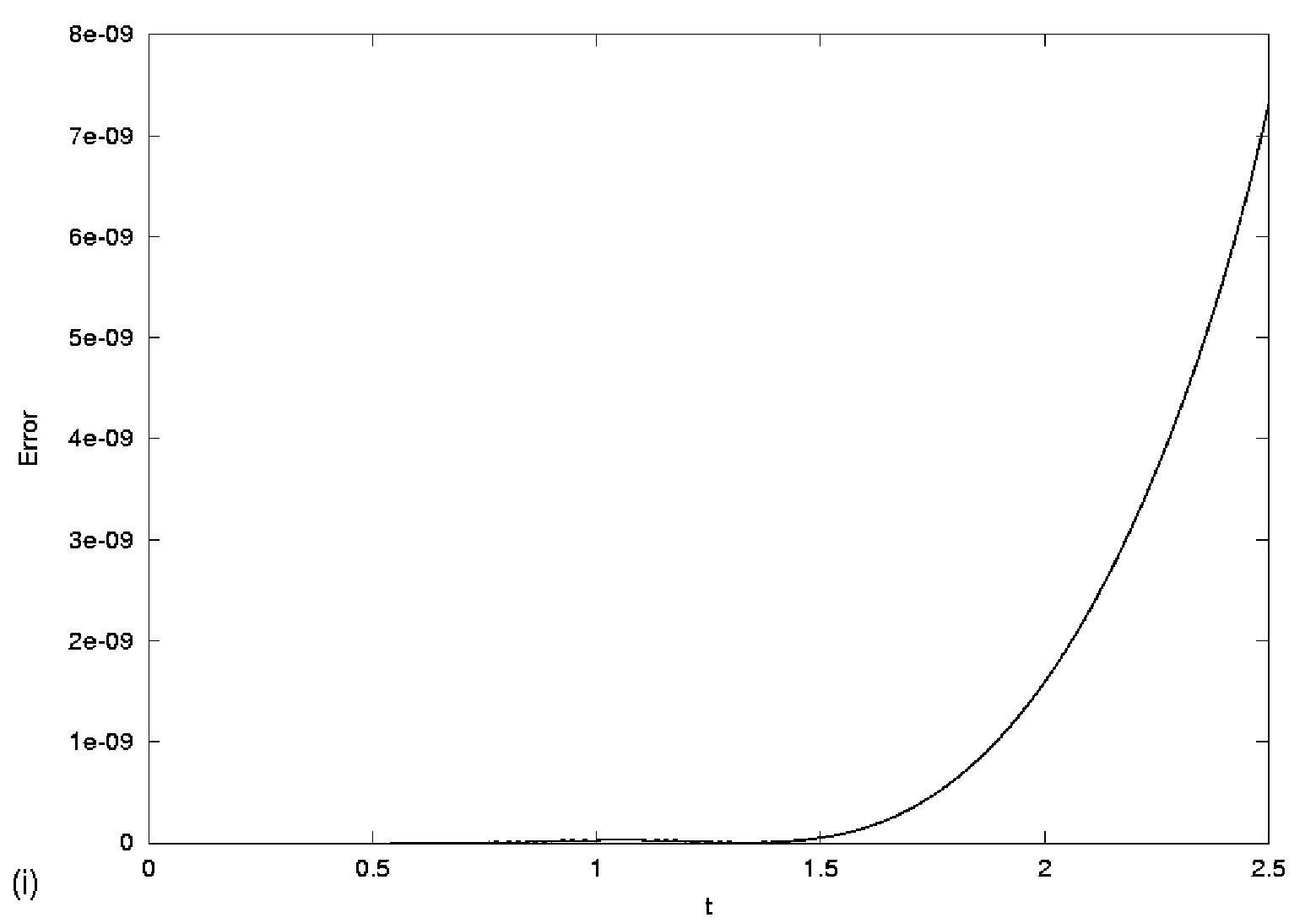} & 
\includegraphics[scale=0.3,angle=0]{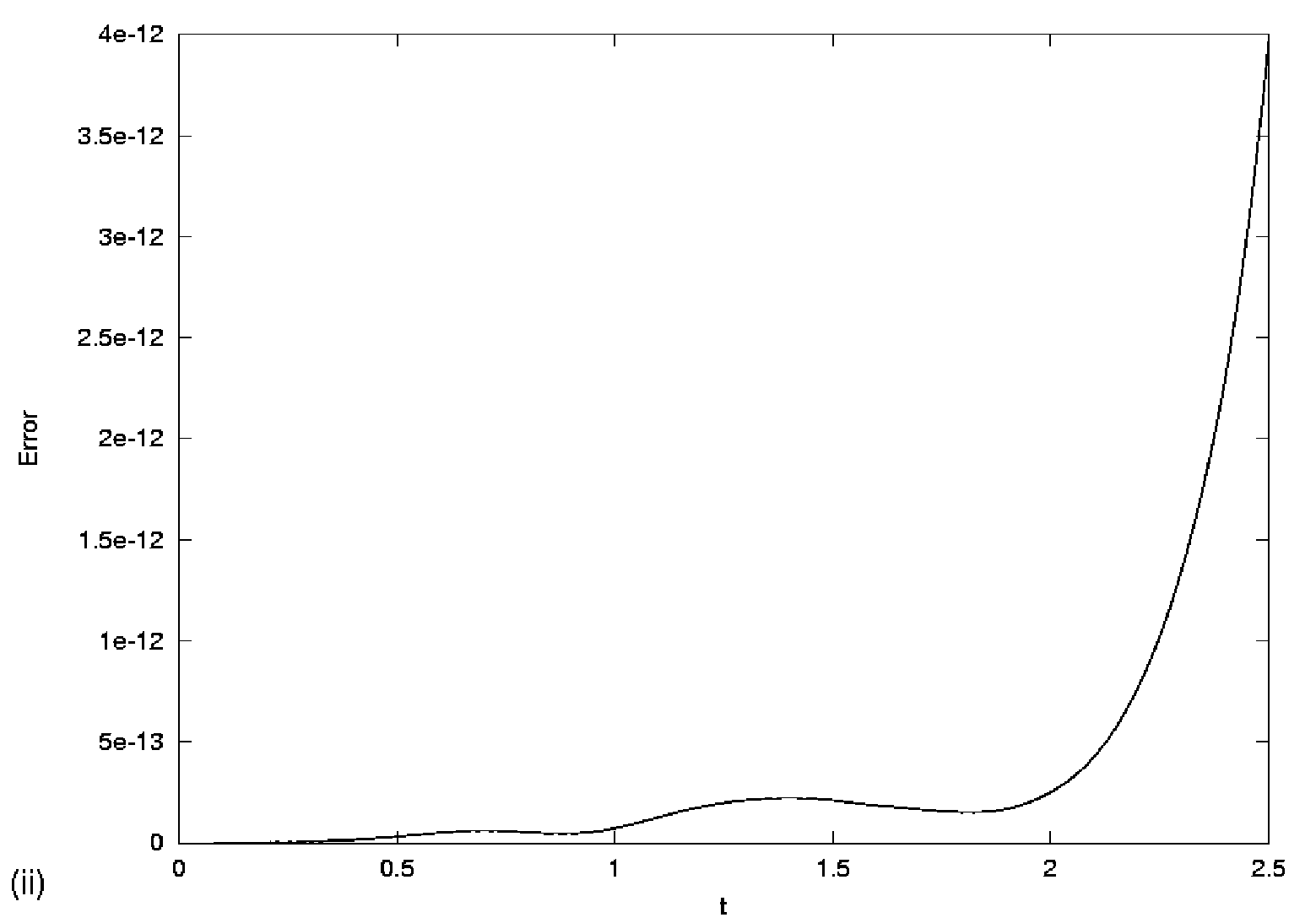} 
\end{tabular}}
\caption{\scriptsize {Mean quadratic errors on the homogeneous solution for the metric variables.
\textsl{(i) $E_m$ (ii) $E_l$ (same parameters as in Figure \ref{errgauge}).}
} \normalsize}\label{errgeom}
\end{figure}

\begin{figure}
\centerline{%
\begin{tabular}{c@{\hspace{5mm}}c@{\hspace{5mm}}c}
\includegraphics[scale=0.25,angle=0]{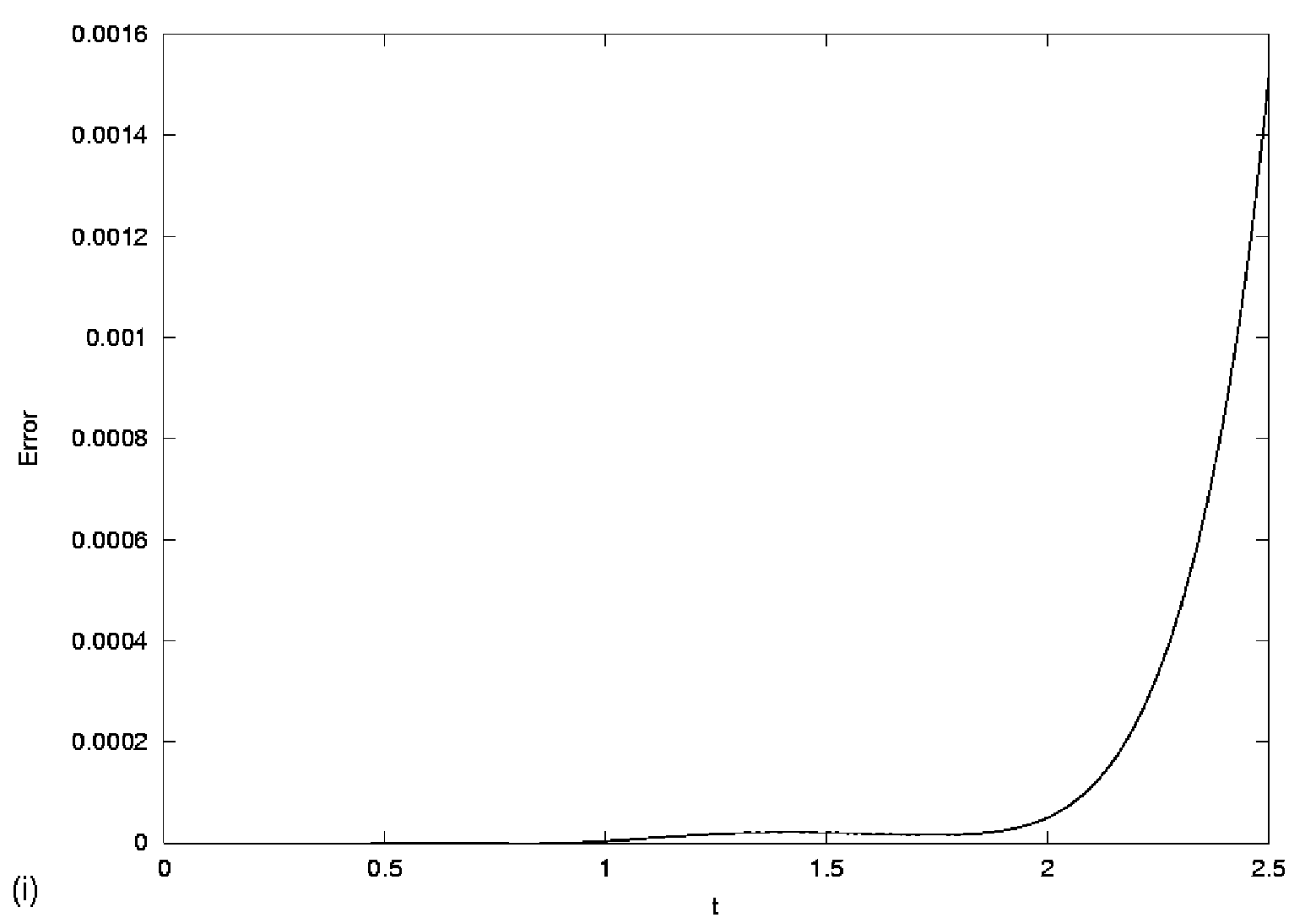} & 
\includegraphics[scale=0.25,angle=0]{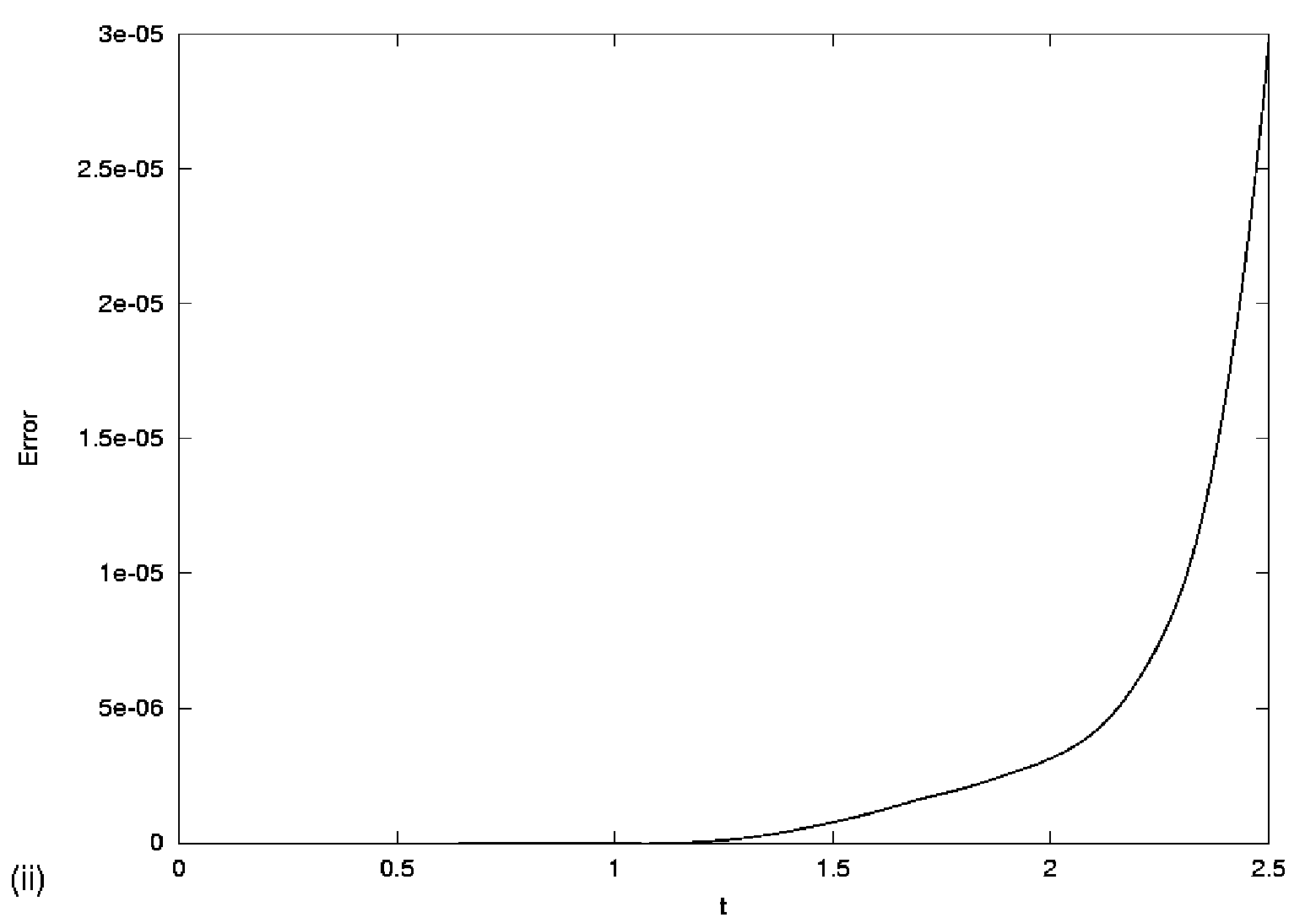} &
\includegraphics[scale=0.25,angle=0]{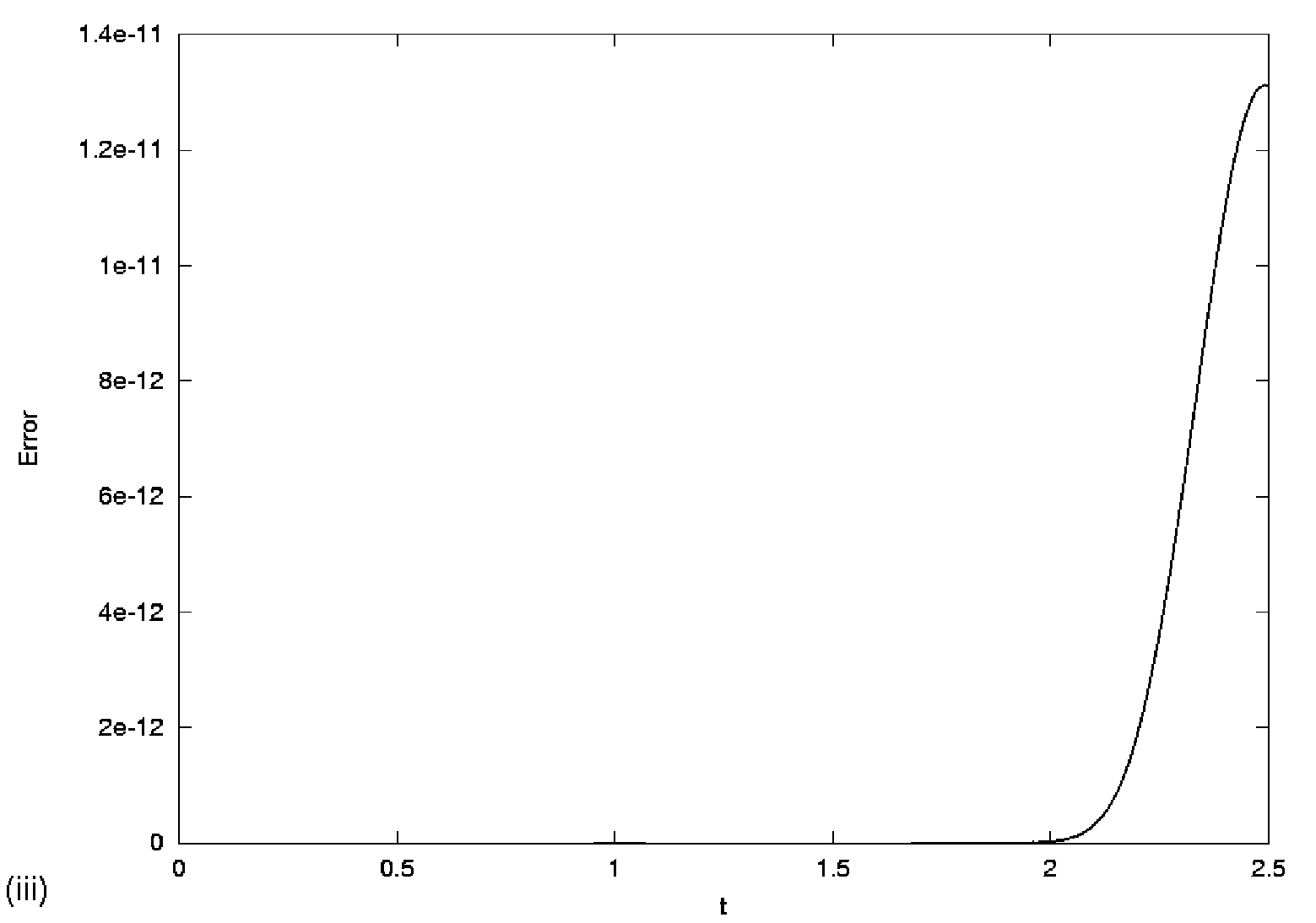} 
\end{tabular}}
\caption{\scriptsize {Mean quadratic errors on the homogeneous solution for the constraints.
\textsl{(i) $E_{\mathcal{H}}$ (ii) $E_{\mathcal{H}^{1}}$ (iii) $E_{\mathcal{G}^{\bf{3}}}$ (same parameters as in Figure \ref{errgauge}).}
} \normalsize}\label{errcons}
\end{figure}

In this paragraph, we briefly discuss the precision of the algorithm presented before. Figures \ref{errgauge} to \ref{errcons}
show the evolution of the mean quadratic error 
$$
E_f=\frac{1}{N_\chi}\sum_{i=1}^{N_\chi+1}\left(f_i-f^B_i\right)^2
$$
on the various fields used in the computation. The general divergent behaviour is due to the simple explicit scheme we have used. Therefore,
the scheme is unstable, partially due to the violation of the boundary conditions that introduce truncation errors on the edges of the grid (see next section). 
We found a very strong Courant condition, for example $\Delta\chi\approx 10^{-1}\; ; \; \Delta t \approx 10^{-6}$ to have an acceptable precision on the time scale of an oscillation period of the function $\sigma^B$ (measuring
the period of the gauge fields). For the gauge variables $\alpha,\;\beta$ and $\gamma$, we see in Figure \ref{errgauge} that
$E_\alpha >E_\beta >E_\gamma$, while for the metric variables $E_m > E_l$ (Figure \ref{errgeom}). For the constraints (Figure \ref{errcons}),
the boundary effects dominate the quadratic error on the hamiltonian and super-momentum constraints $\mathcal{H}$ and $\mathcal{H}_1$
though the errors are much lower away from the edges of the $\chi-$interval (see also Figure \ref{cons}). These boundary effects are present in
all the fields
in the computation (see also other 3D plots). In Figure \ref{errcons} $(iii)$, we see 
that the error on the gauge constraint $\mathcal{G}^{\bf{3}}$ is much lower than those of Einstein's part.
\section{Discussion}
\begin{figure}
\centerline{%
\begin{tabular}{c@{\hspace{1cm}}c}
\includegraphics[scale=0.3,angle=0]{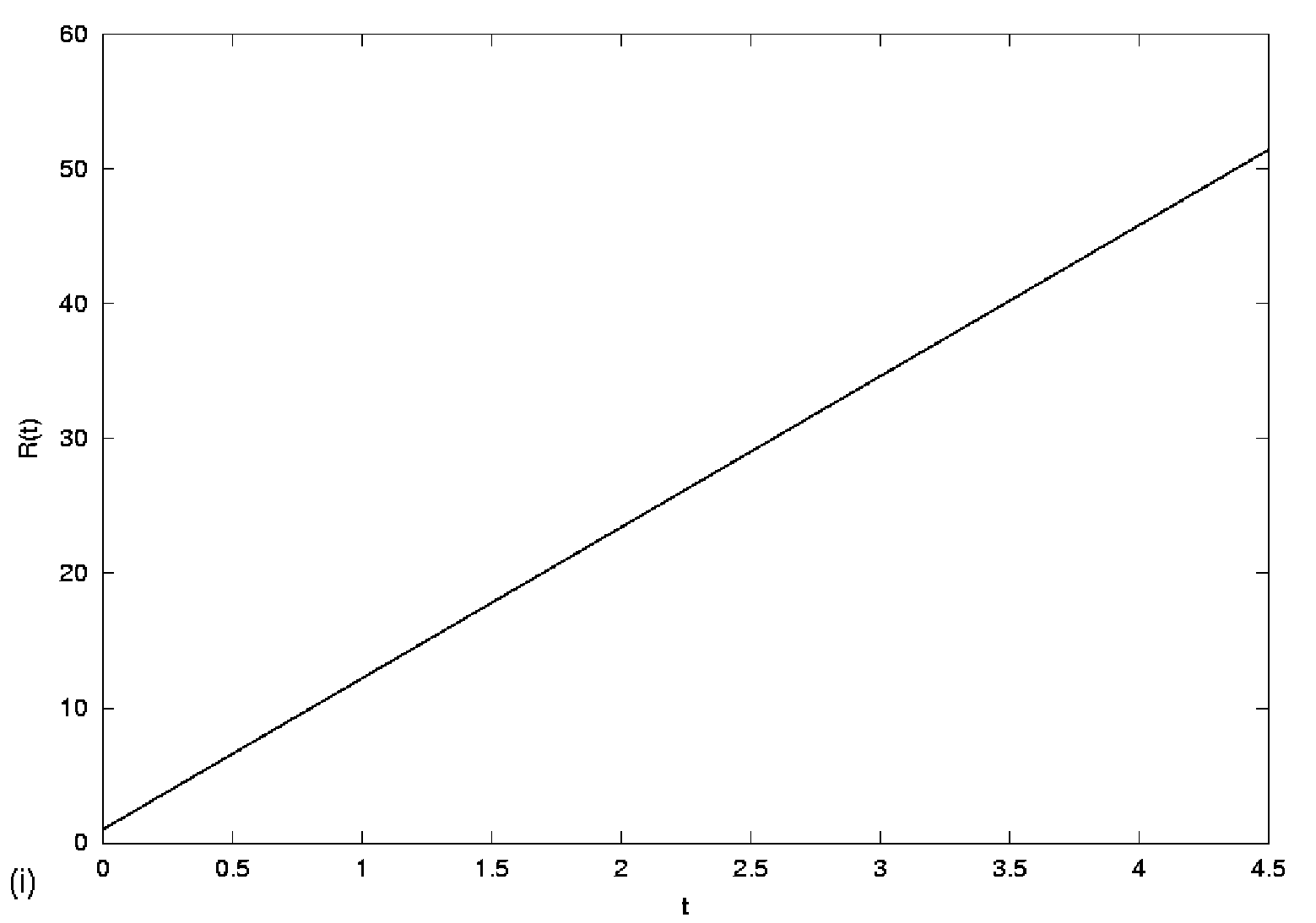} & 
\includegraphics[scale=0.3,angle=0]{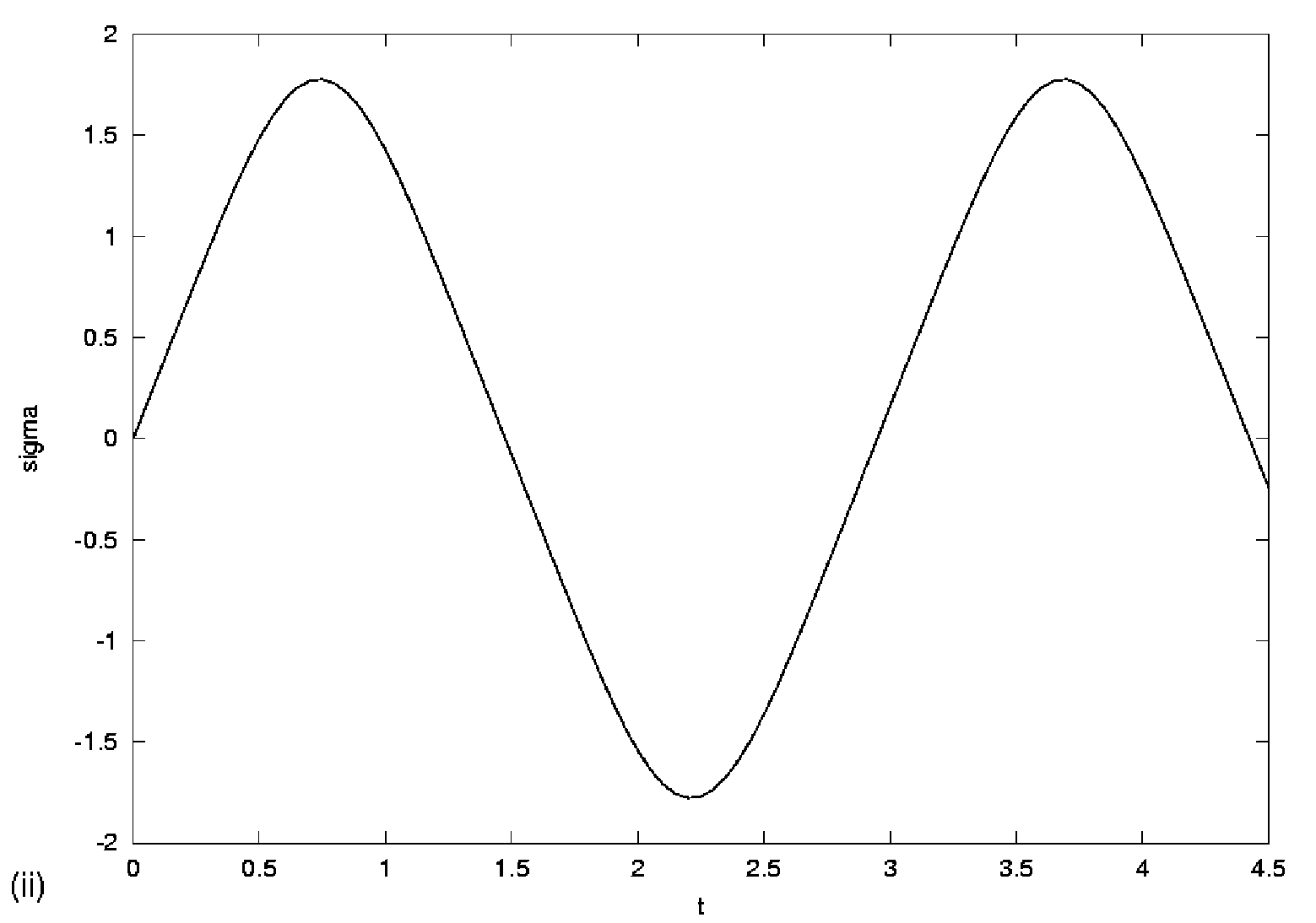} 
\end{tabular}}
\caption{\scriptsize {\textsl{Homogeneous solutions (i) $R(t)$ (ii) $\sigma(t)$ ($\rho^B_0=15$, $\epsilon_c=10^{-4}$, $\sigma_0=0$, $w=0.5$,
$L_H^0\approx 0.09$,
$\Delta\chi=9.3\times 10^{-2}$, $\Delta t=10^{-7}$). }
} \normalsize}\label{rsig}
\end{figure}

\begin{figure}
\centerline{%
\begin{tabular}{c@{\hspace{5mm}}c}
\includegraphics[scale=0.3,angle=0]{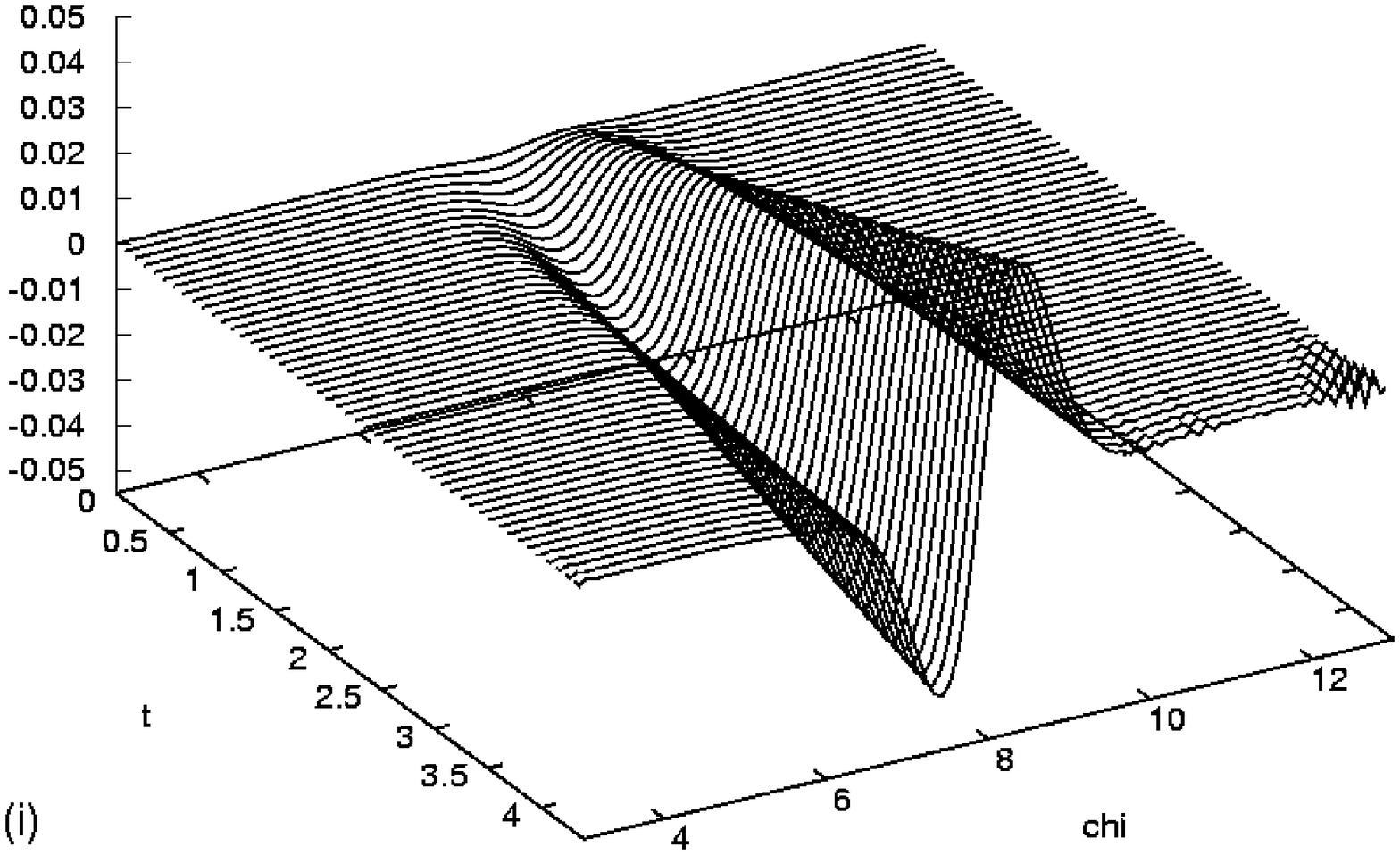} & 
\includegraphics[scale=0.3,angle=0]{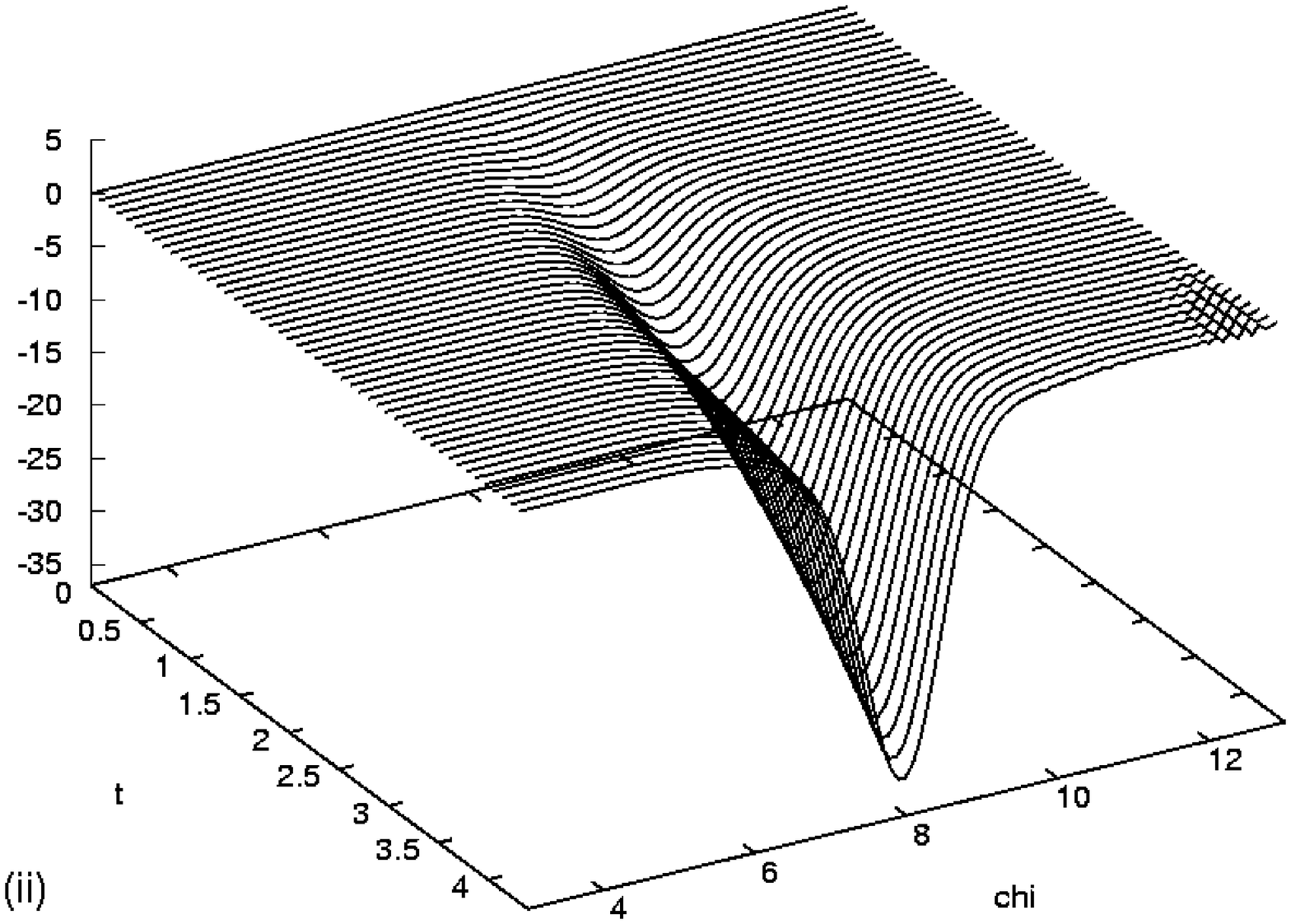} 
\end{tabular}}
\caption{\scriptsize {\textsl{Perturbations of the metric coefficients. (i) $(m-1)\;R(t)$ (ii) $(l-1)\;R^2\chi^2$ (same parameters as in Figure \ref{rsig})}
} \normalsize}
\label{ml}
\end{figure}

\begin{figure}
\centerline{%
\begin{tabular}{c@{\hspace{5mm}}c}
\includegraphics[scale=0.3,angle=0]{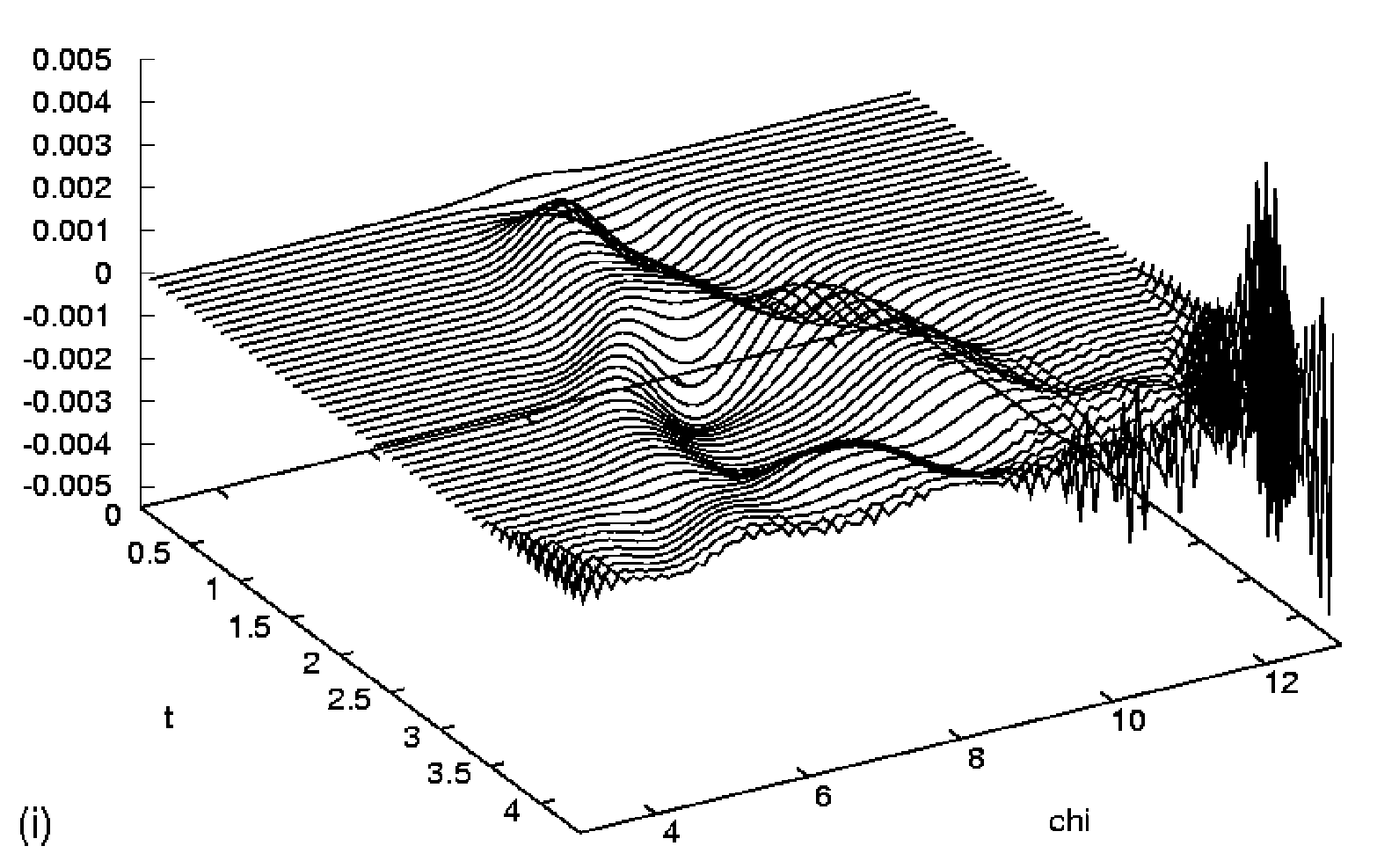} &
\includegraphics[scale=0.3,angle=0]{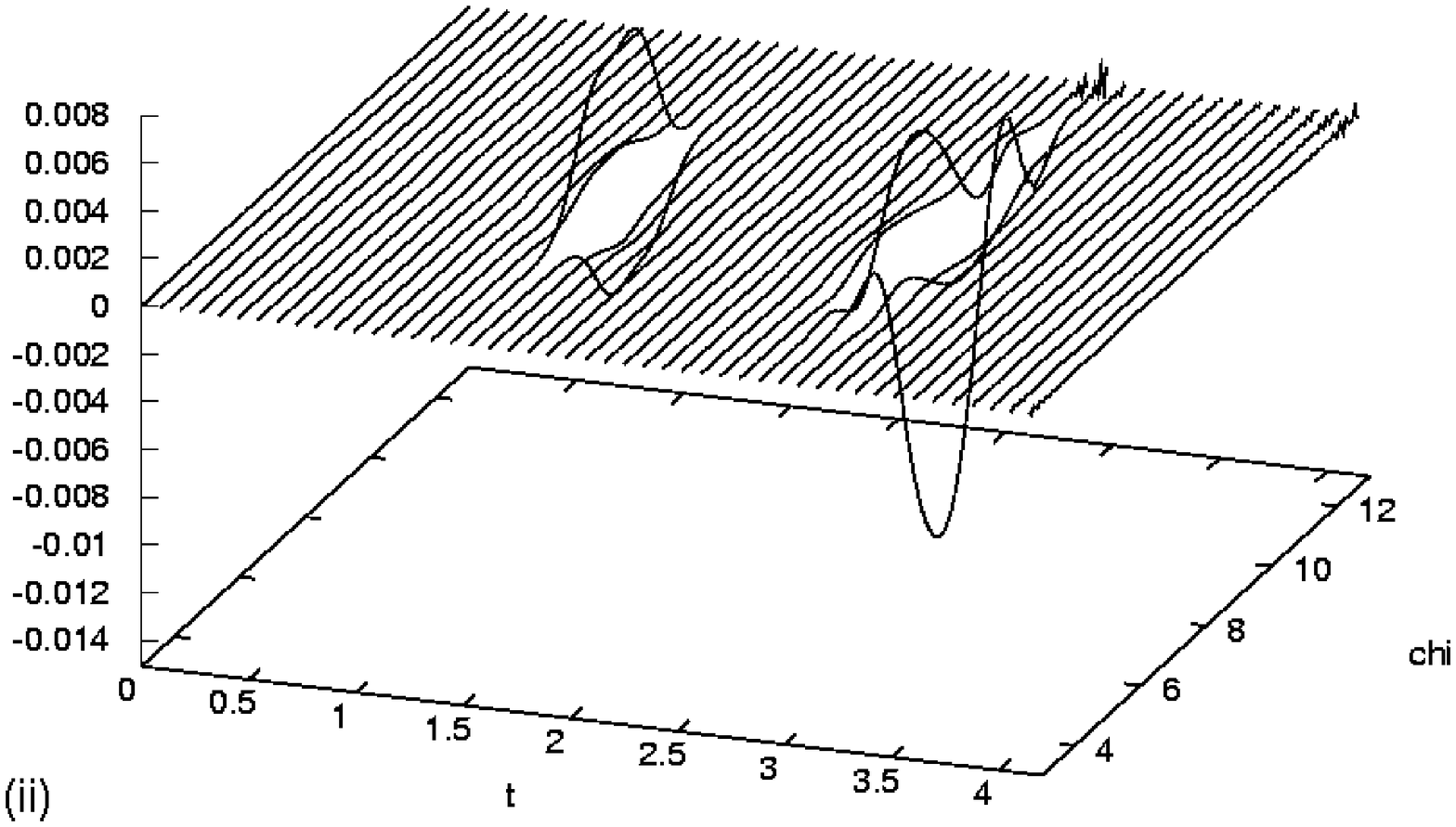} 
\end{tabular}}
\caption{\scriptsize {\textsl{Perturbation of the first gauge variable and the electric gauge potential. (i) $(\alpha-\dot{\sigma})$ (ii) $(a-a_{bkg})$ (same parameters as in Figure \ref{rsig}).}
} \normalsize}\label{a}
\end{figure}

\begin{figure}
\centerline{%
\begin{tabular}{c@{\hspace{5mm}}c}
\includegraphics[scale=0.3,angle=0]{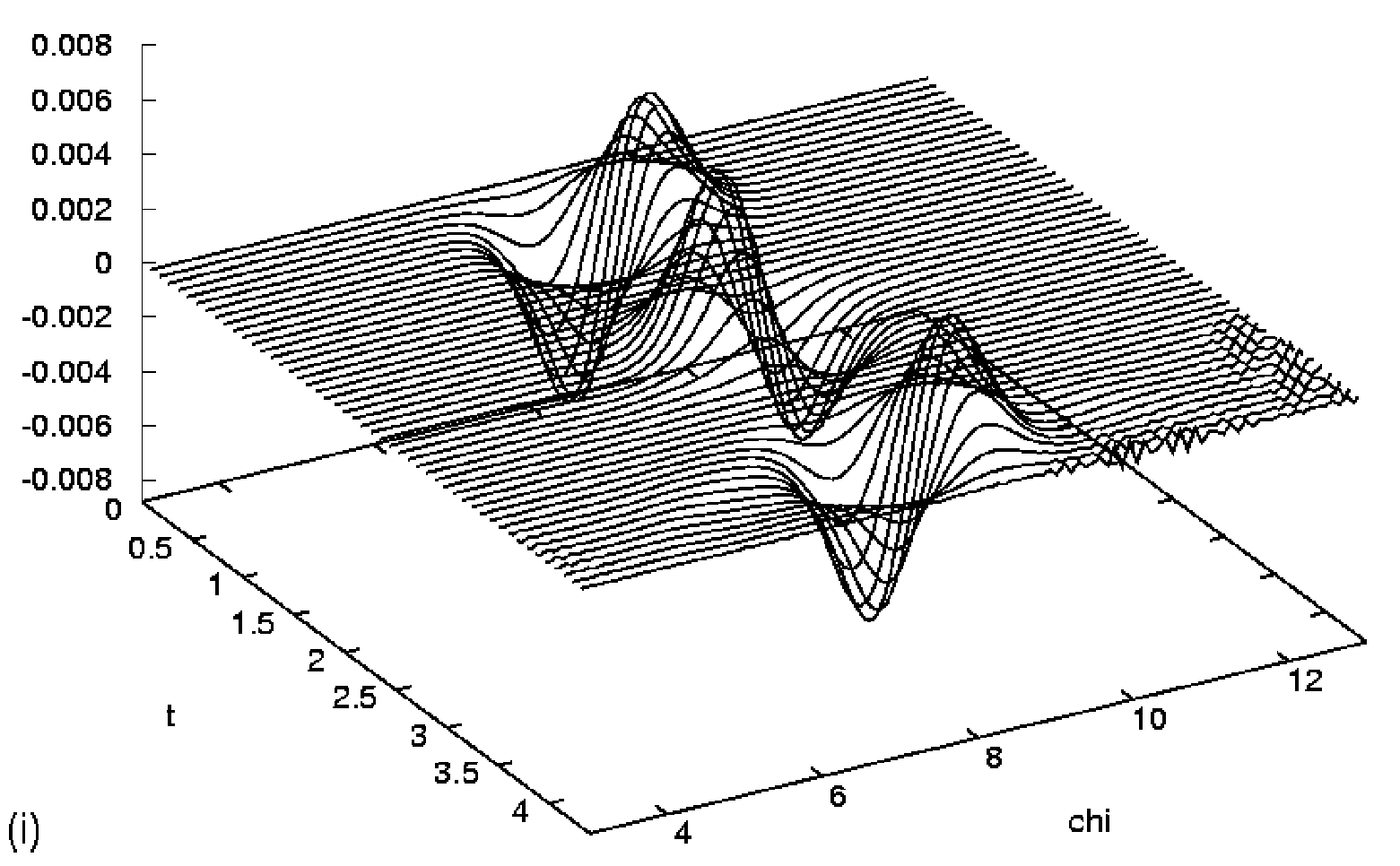} & 
\includegraphics[scale=0.3,angle=0]{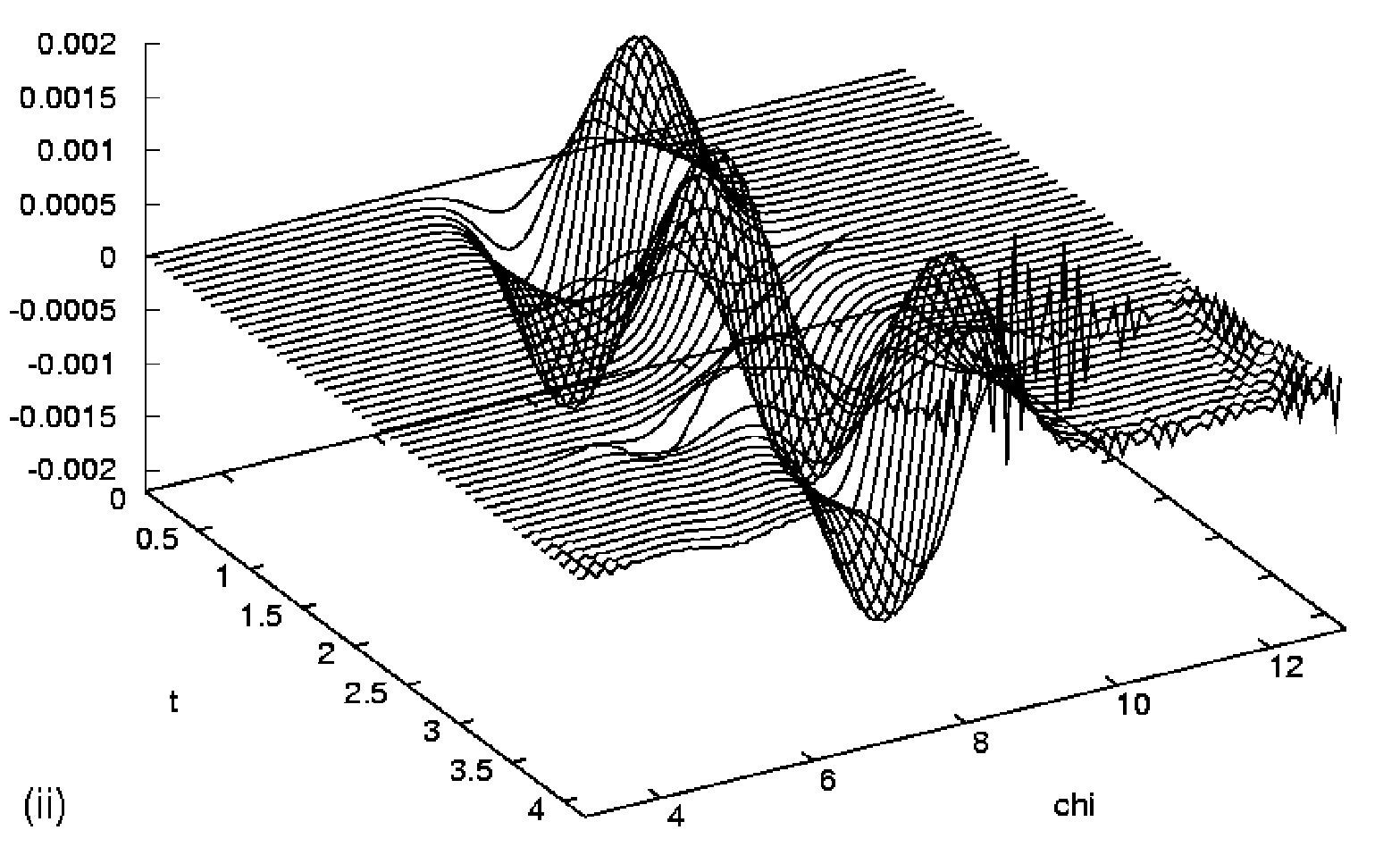} 
\end{tabular}}
\caption{\scriptsize {\textsl{Perturbation of the second gauge variable and the first magnetic gauge potential. (i) $\left(\beta-\sigma^3\right)$ (ii) $(b-b_{bkg})$ (same parameters as in Figure \ref{rsig}).}
} \normalsize}\label{b}
\end{figure}

\begin{figure}
\centerline{%
\begin{tabular}{c@{\hspace{5mm}}c}
\includegraphics[scale=0.3,angle=0]{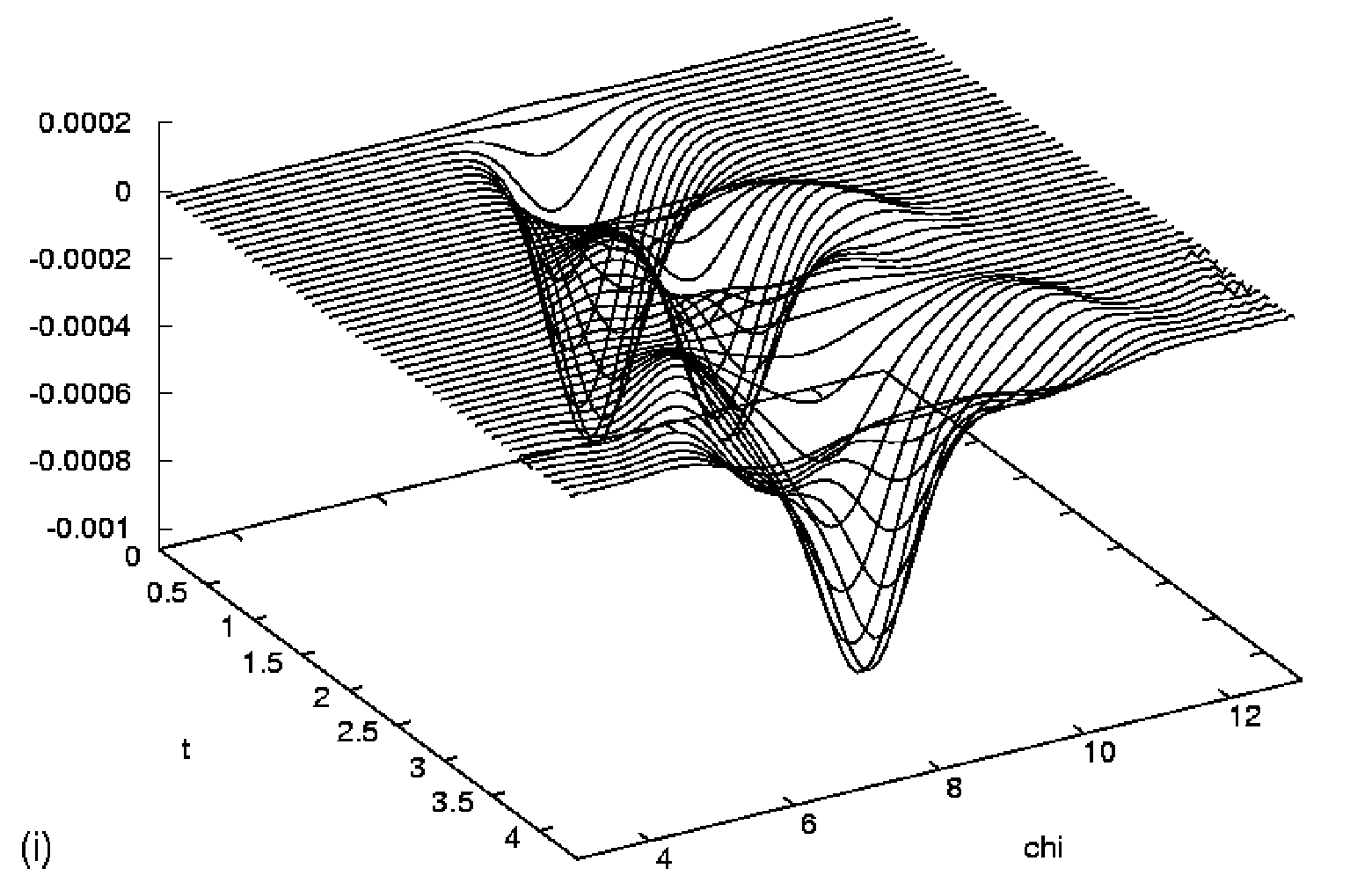} & 
\includegraphics[scale=0.3,angle=0]{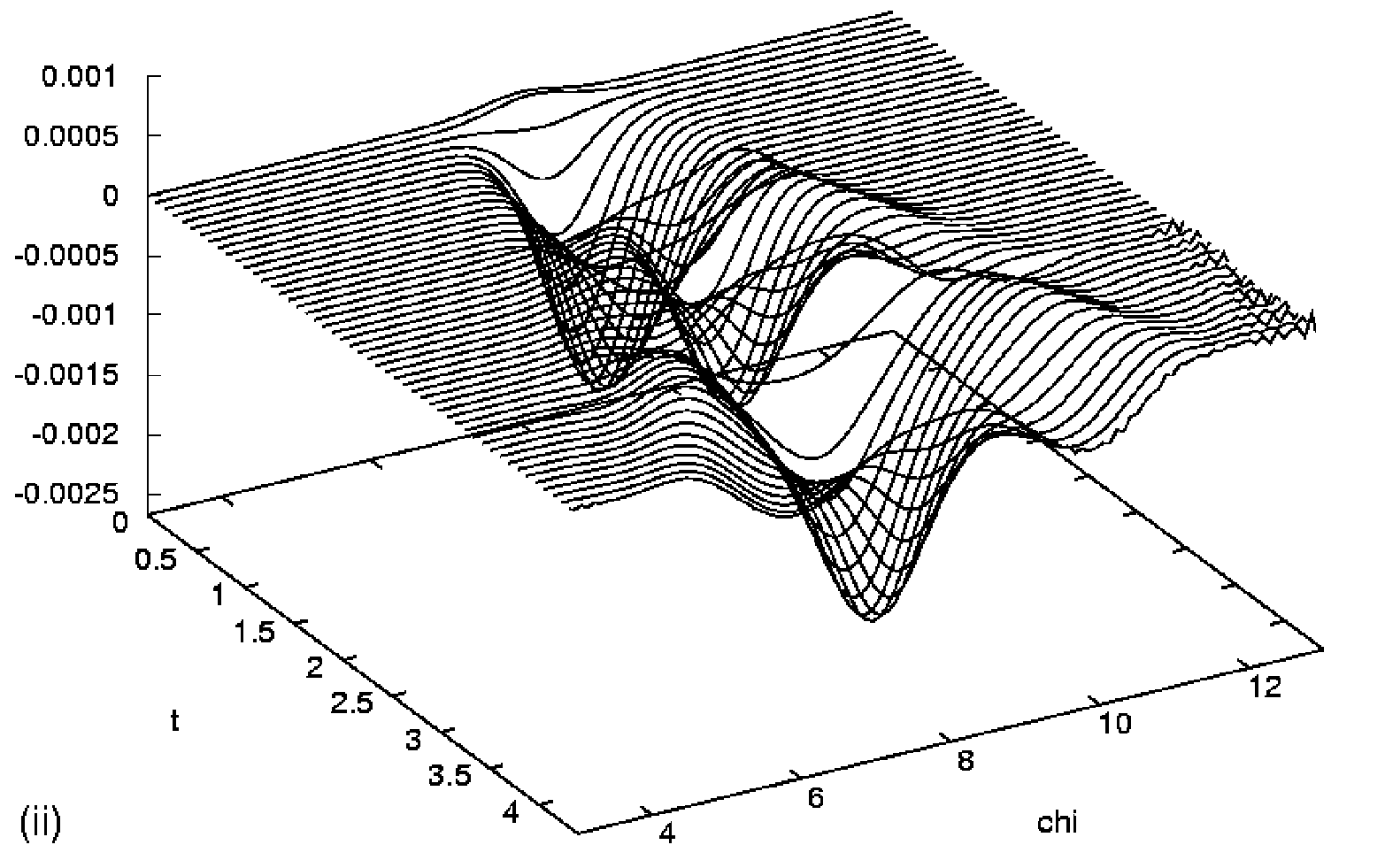} 
\end{tabular}}
\caption{\scriptsize {\textsl{Perturbation of the third gauge variable and the second magnetic gauge potential. (i) $\left(\gamma-\sigma^2\right)$ (ii) $(c-c_{bkg})$ (same parameters as in Figure \ref{rsig}).}
} \normalsize}\label{c}
\end{figure}

\begin{figure}
\centerline{%
\begin{tabular}{c@{\hspace{1mm}}c}
\includegraphics[scale=0.3,angle=0]{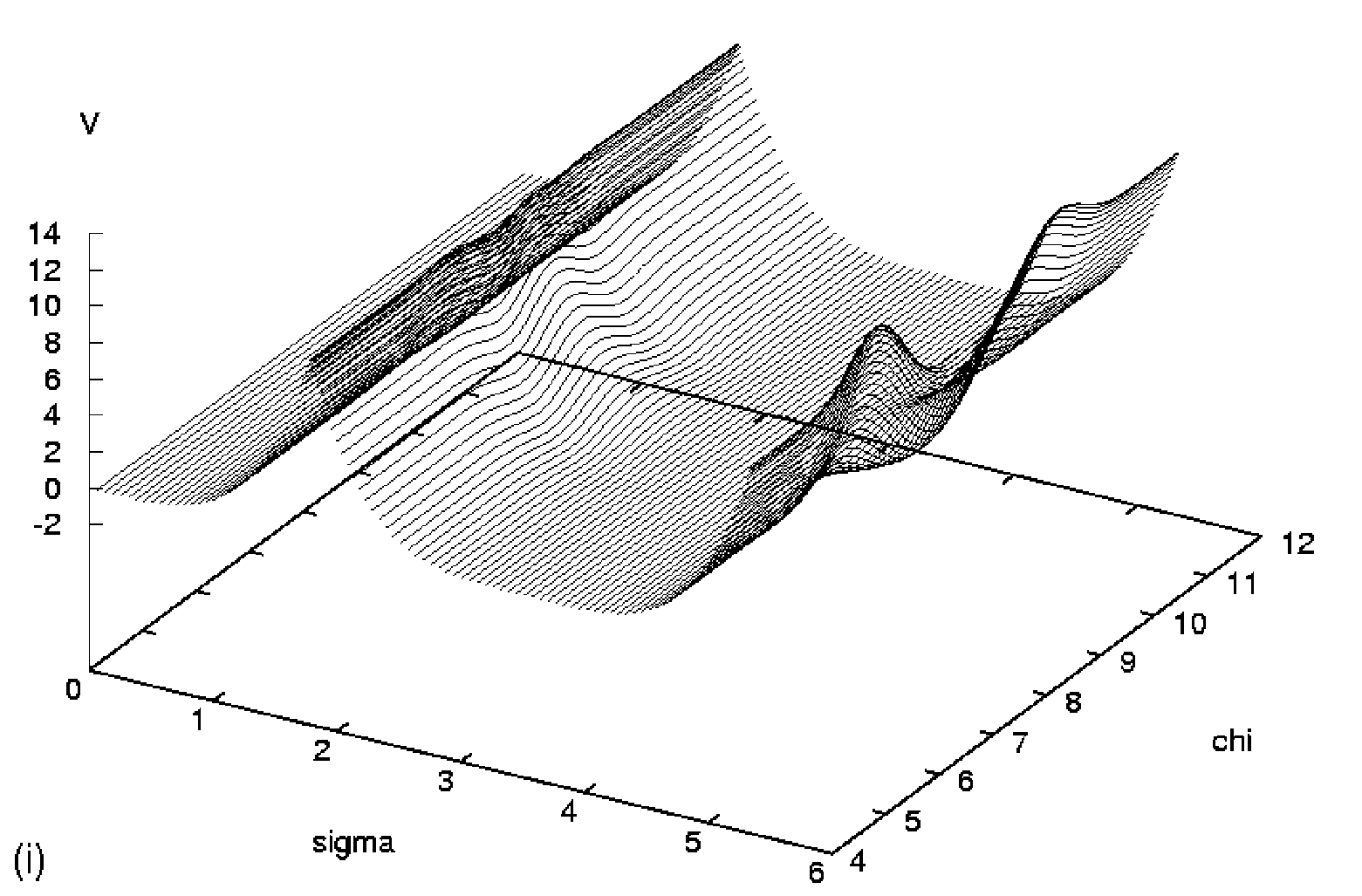} & 
\includegraphics[scale=0.3,angle=0]{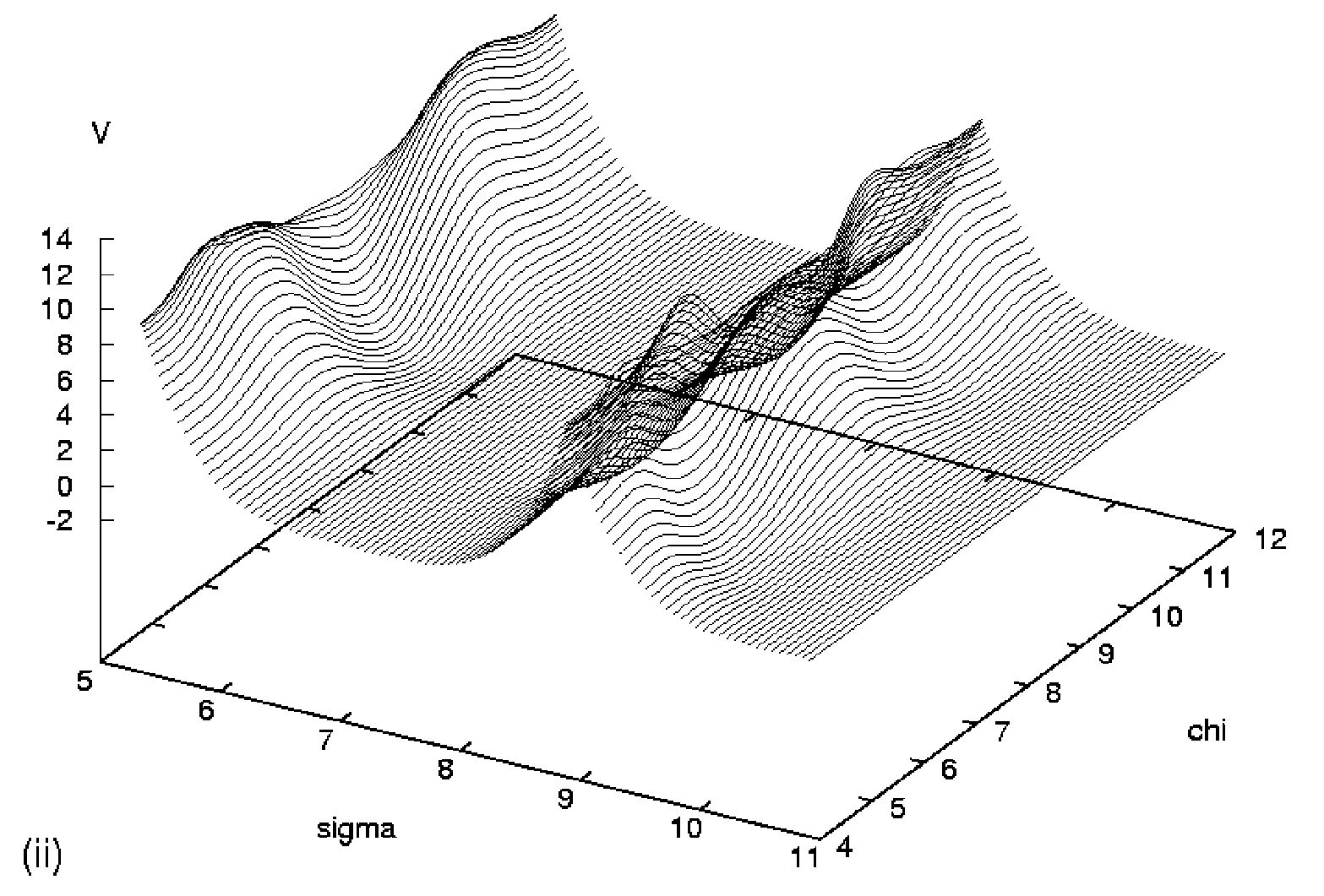} 
\end{tabular}}
\caption{\scriptsize {\textsl{Unfolded and rescaled shape of the potential $V$ as a function of $\sigma(t)$ and $\chi$ (see comments in the text).}
} \normalsize}\label{V}
\end{figure}

\begin{figure}
\centerline{%
\begin{tabular}{c@{\hspace{1mm}}c@{\hspace{1mm}}c}
\includegraphics[scale=0.25,angle=0]{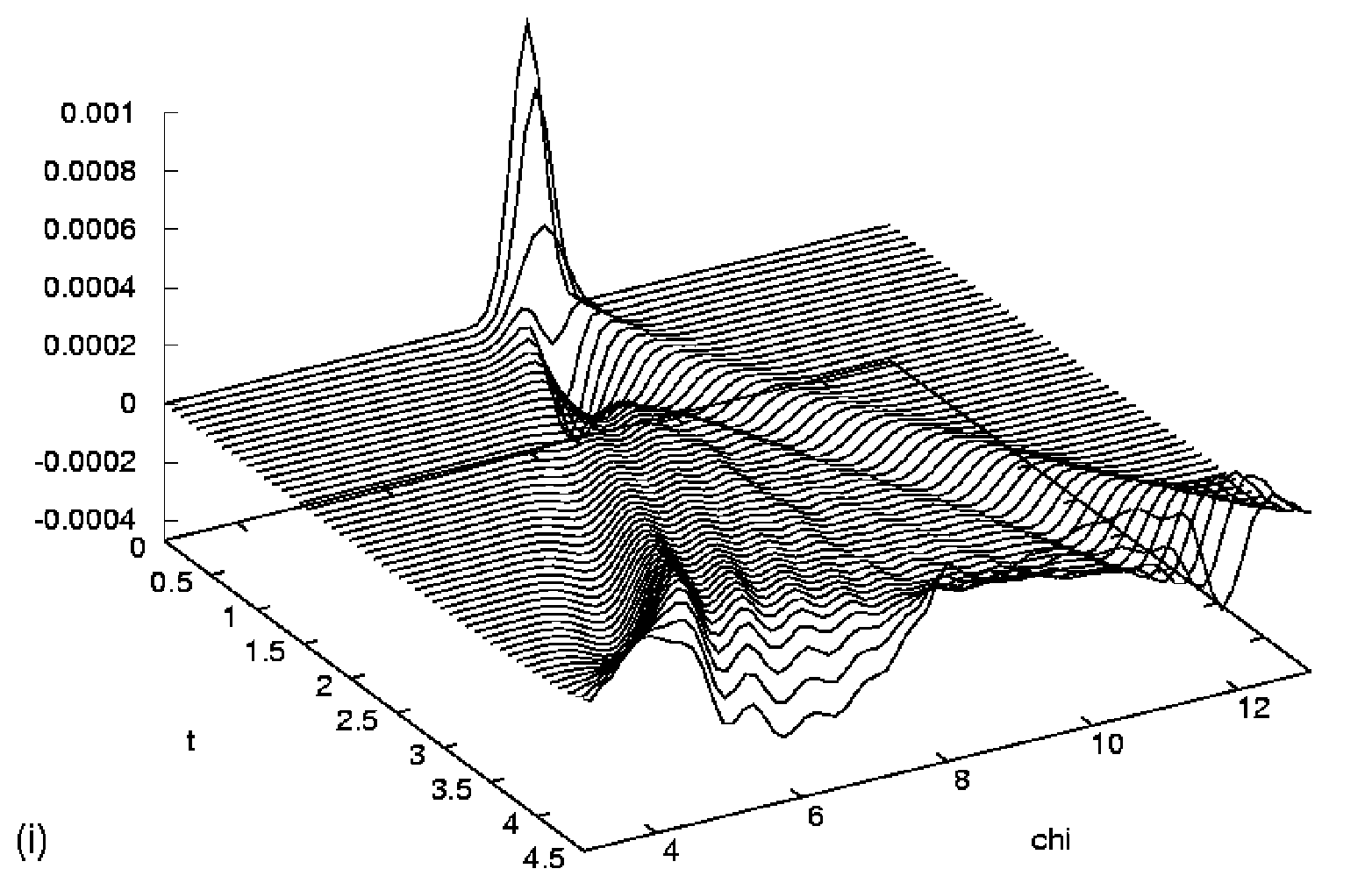} & 
\includegraphics[scale=0.25,angle=0]{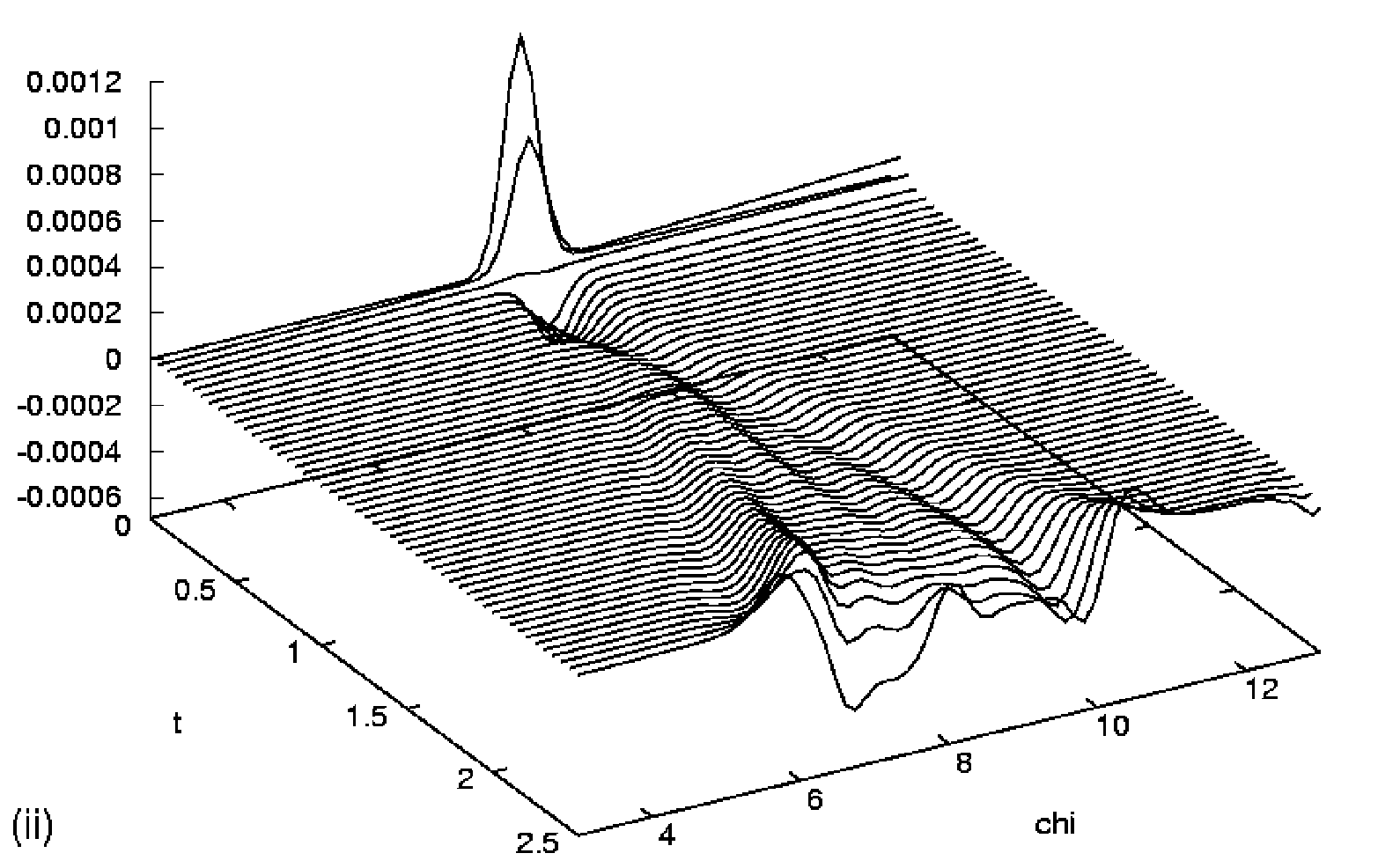} &
\includegraphics[scale=0.25,angle=0]{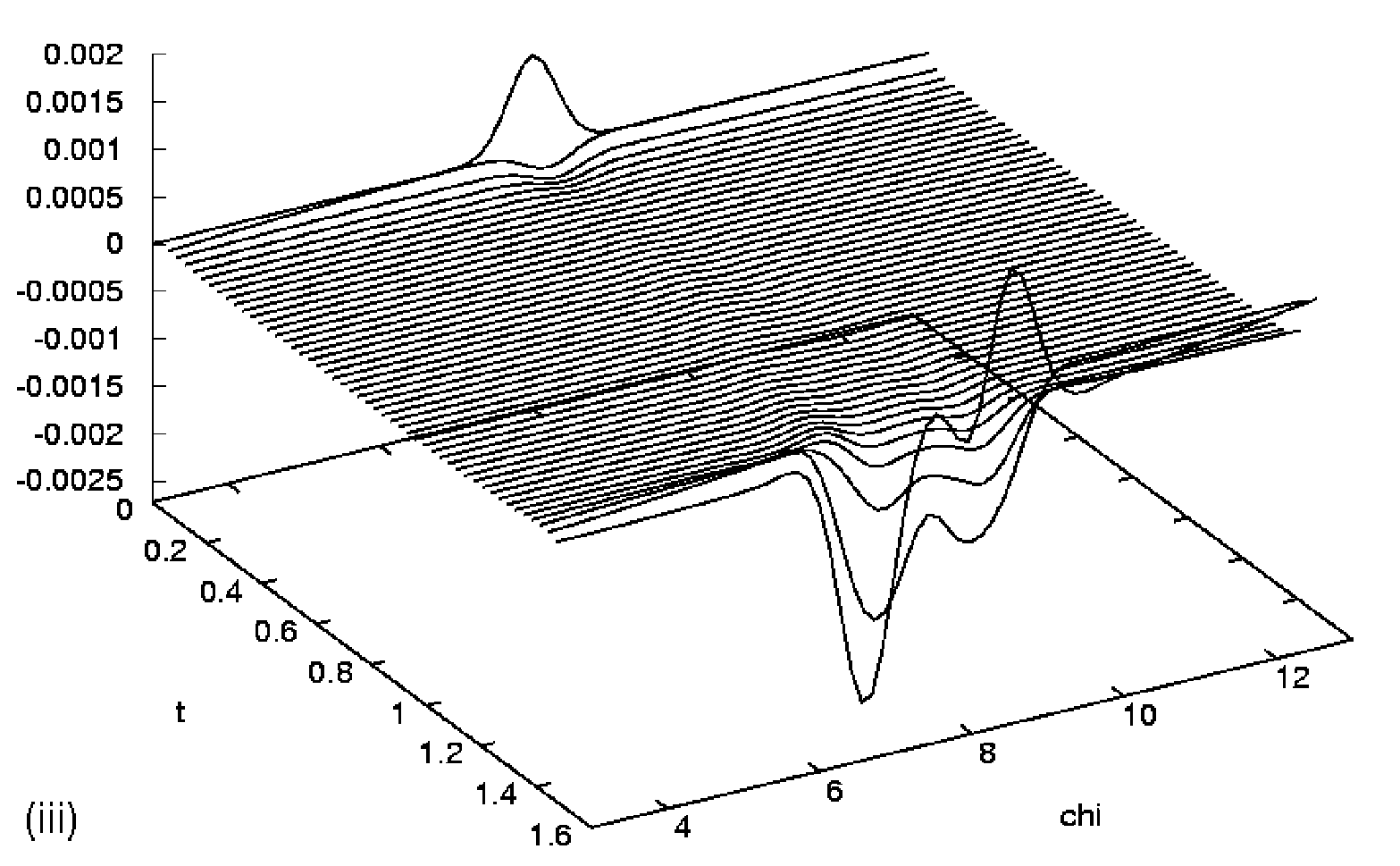} 
\end{tabular}}
\centerline{%
\begin{tabular}{c@{\hspace{1mm}}c@{\hspace{1mm}}c}
\includegraphics[scale=0.25,angle=0]{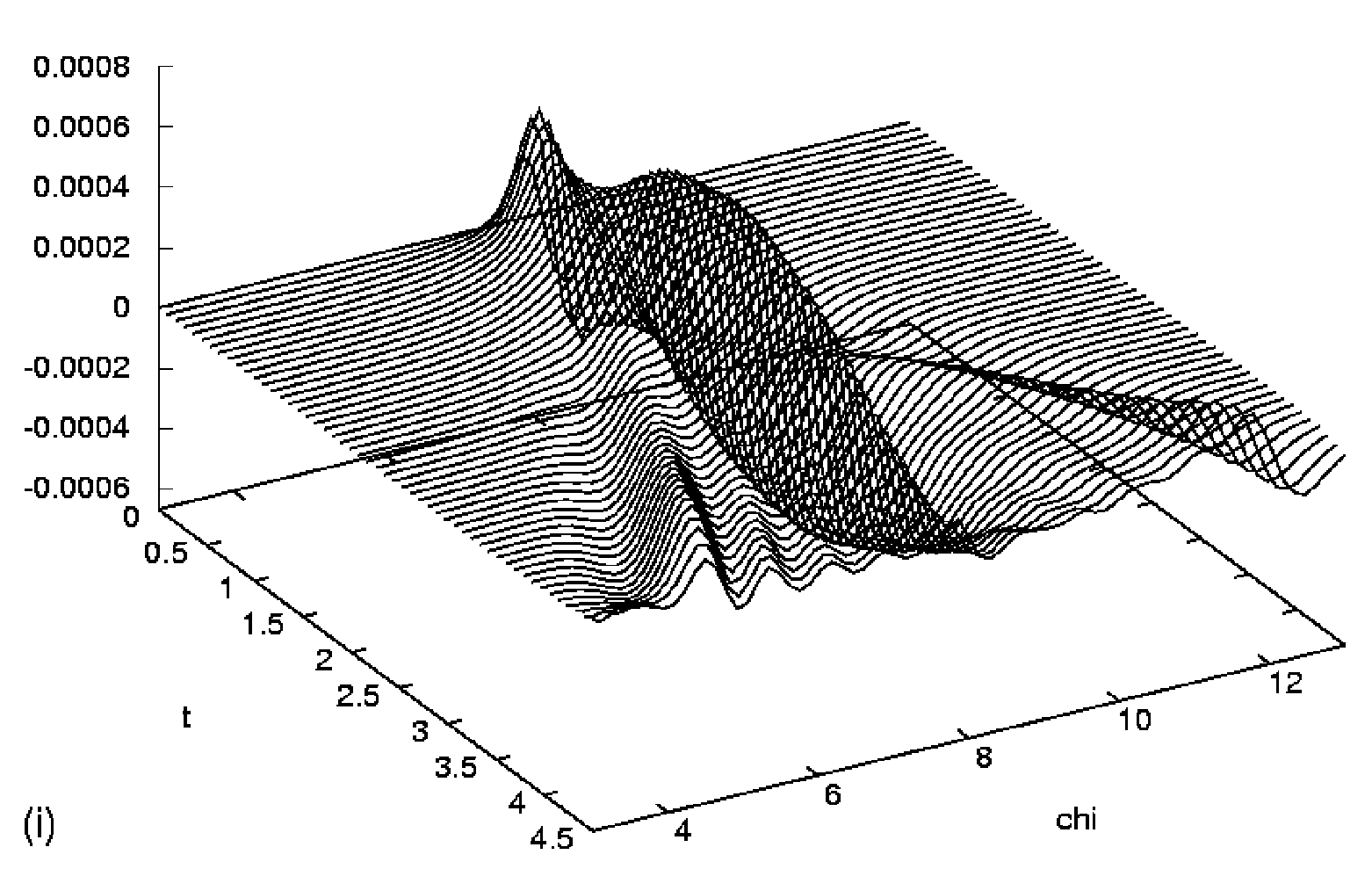} & 
\includegraphics[scale=0.25,angle=0]{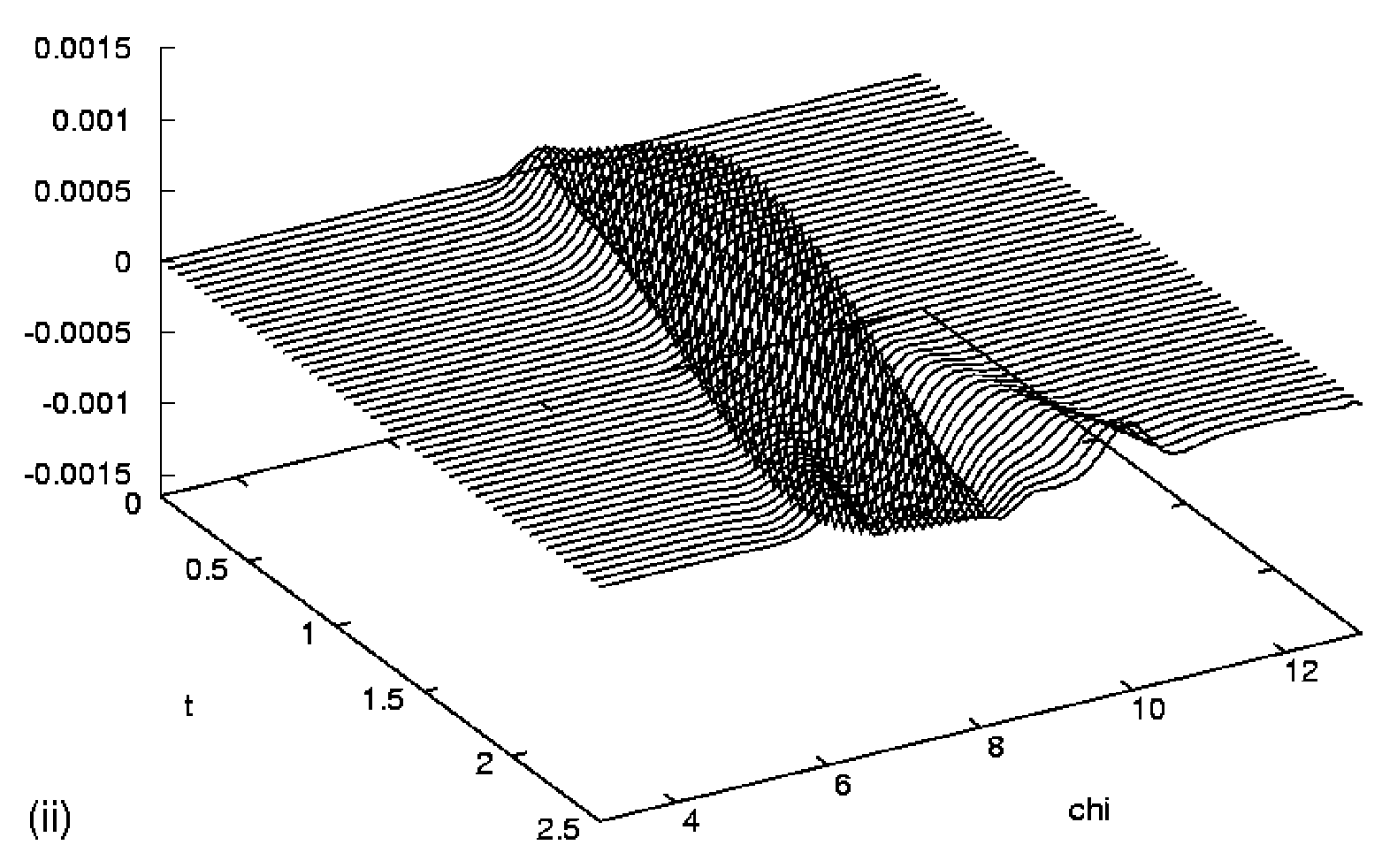} &
\includegraphics[scale=0.25,angle=0]{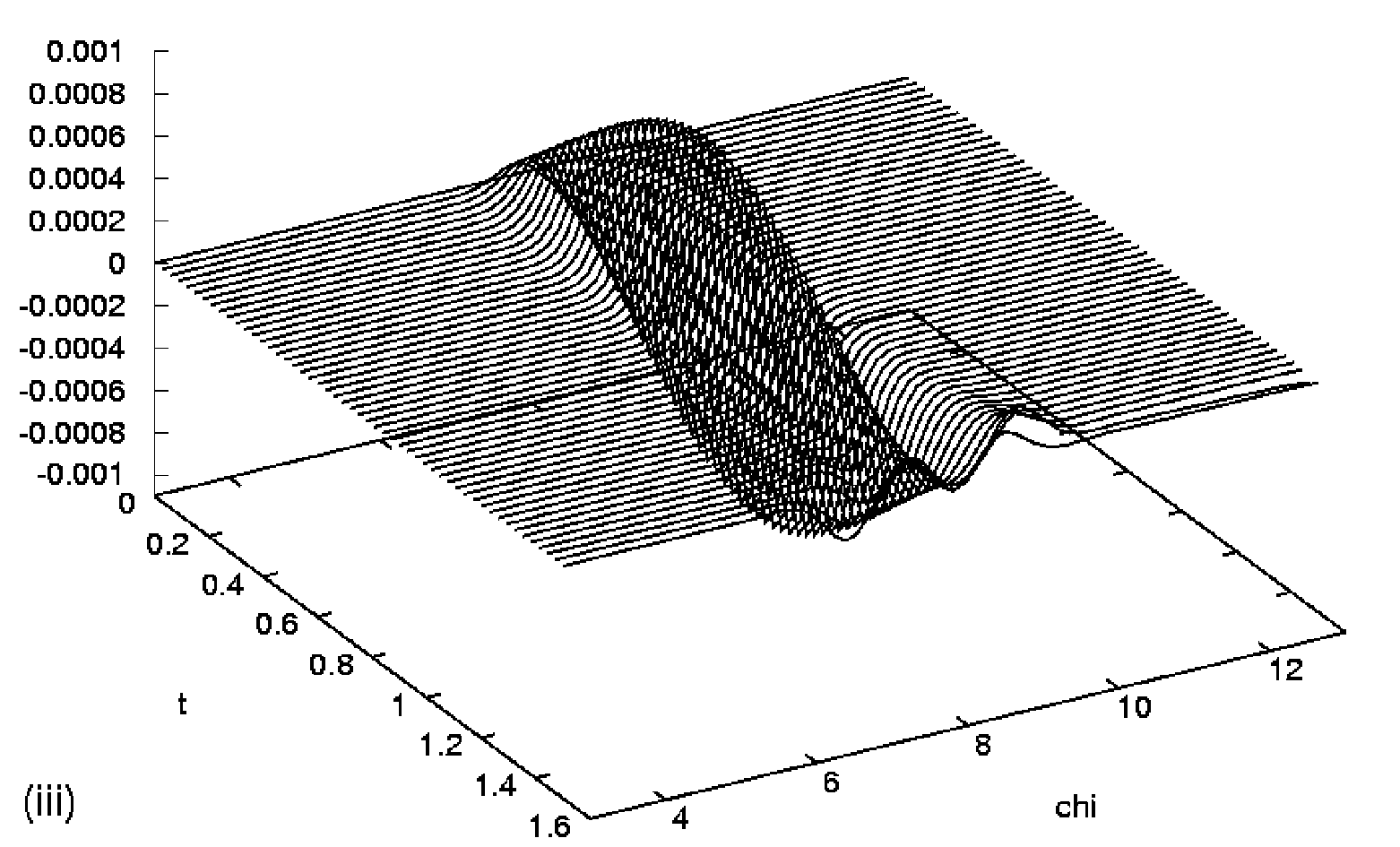} 
\end{tabular}}
\centerline{%
\begin{tabular}{c@{\hspace{1mm}}c@{\hspace{1mm}}c}
\includegraphics[scale=0.25,angle=0]{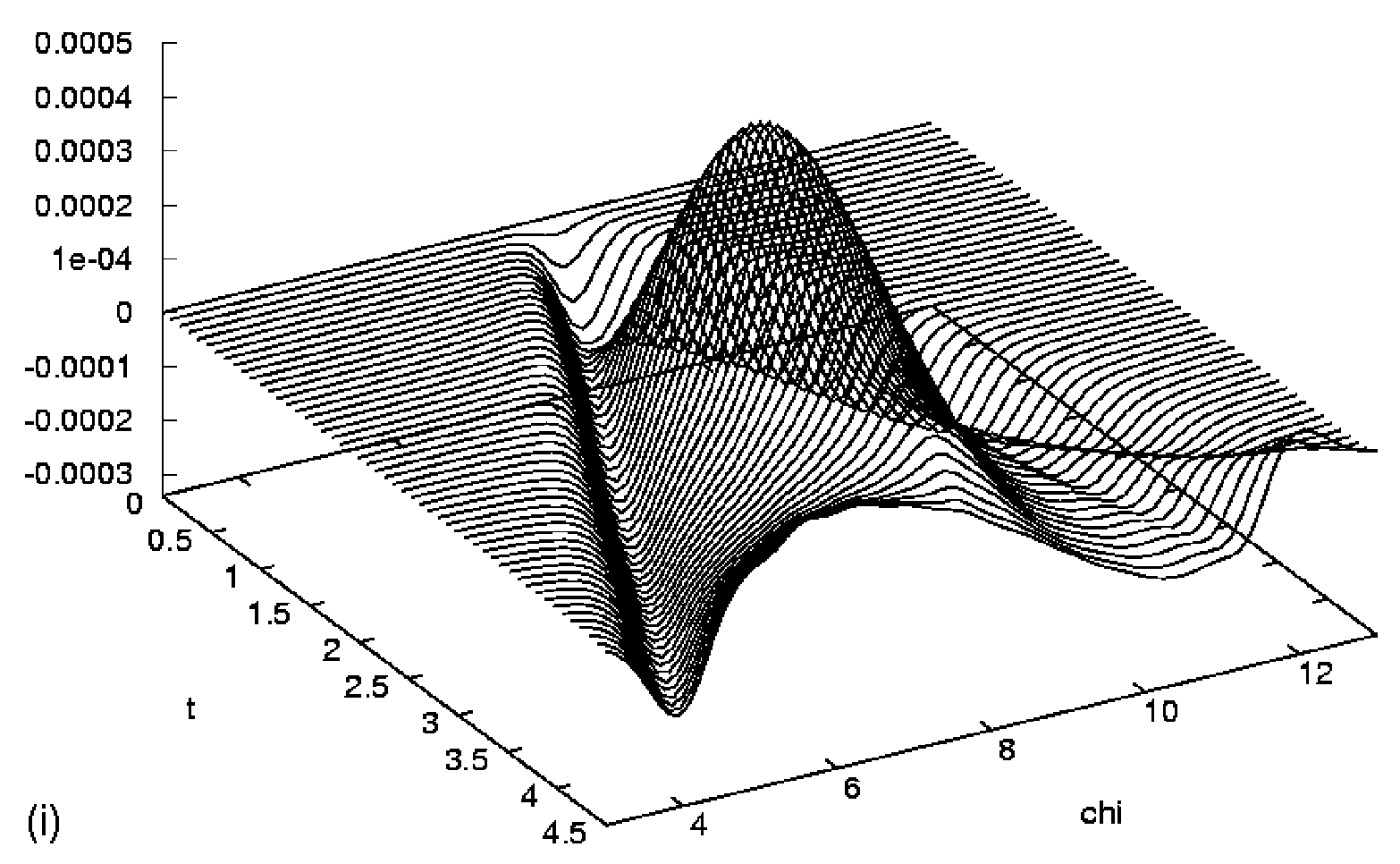} & 
\includegraphics[scale=0.25,angle=0]{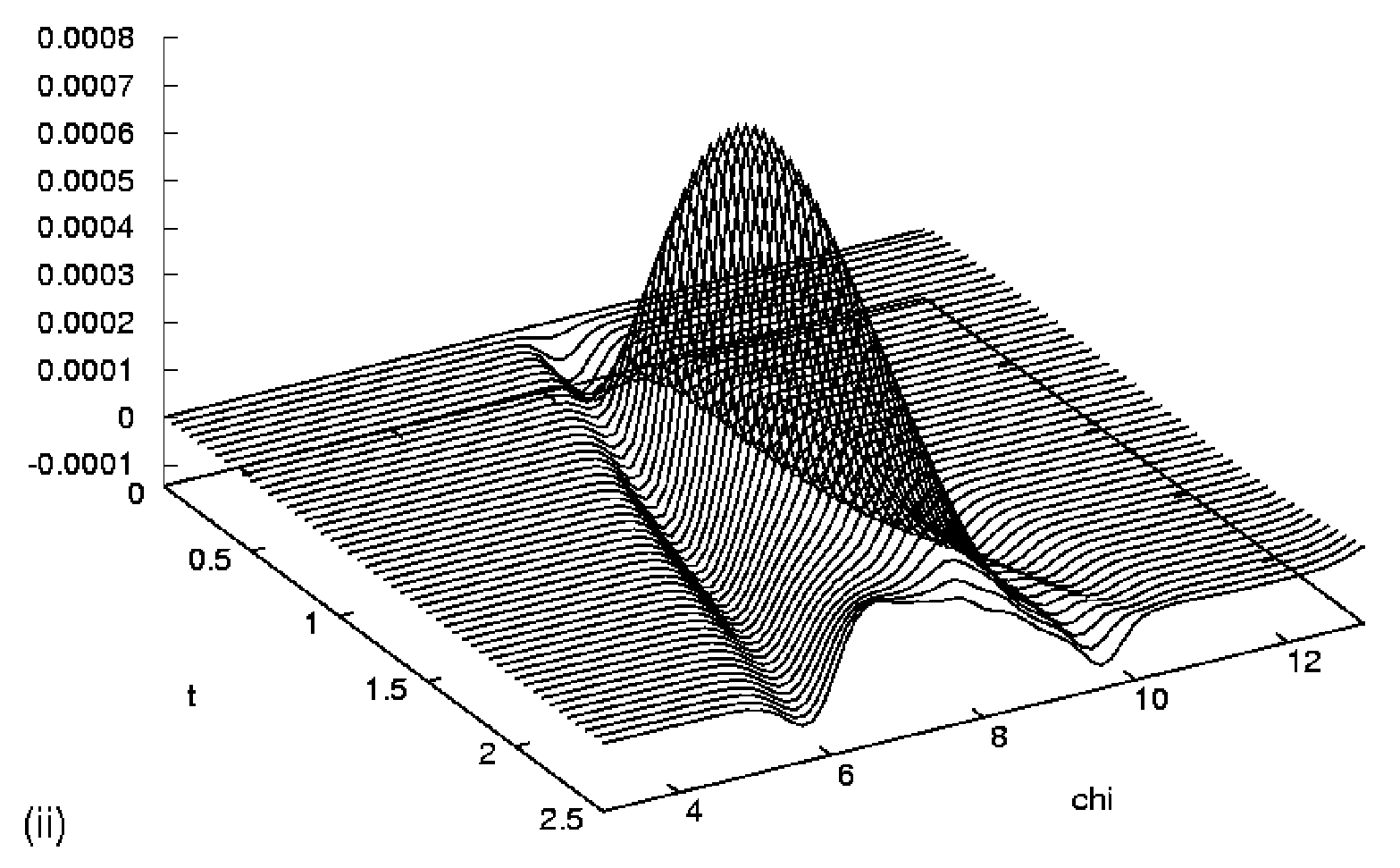} &
\includegraphics[scale=0.25,angle=0]{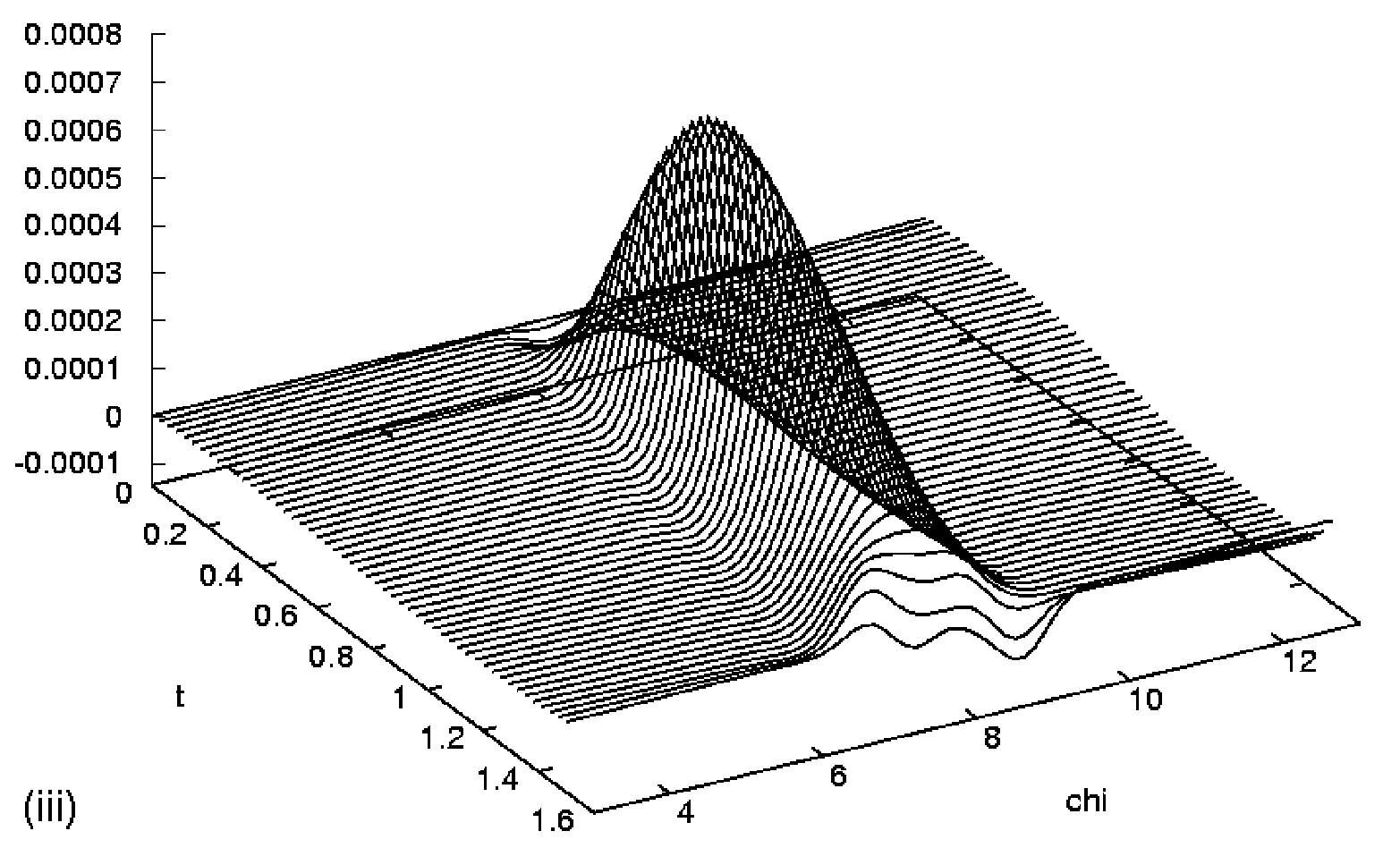} 
\end{tabular}}
\caption{\scriptsize {\textsl{Perturbations of the electric $a$ (above) and magnetic $b$ (middle) and
$c$ (below) potentials for different values of the
initial expansion rate. (i) left:  $\rho_0^B=0.15$ ($L_H^0\approx 0.9$) (ii) center: $\rho_0^B=1.5$ ($L_H^0\approx 0.3$) (iii) right: $\rho_0^B=15$
($L_H^0\approx 0.09$) ($\epsilon_a =10^{-3}$, $w=0.1$, $\sigma_0=0$).}
} \normalsize}\label{dif}
\end{figure}

\begin{figure}
\centerline{%
\begin{tabular}{c@{\hspace{5mm}}c}
\includegraphics[scale=0.3,angle=0]{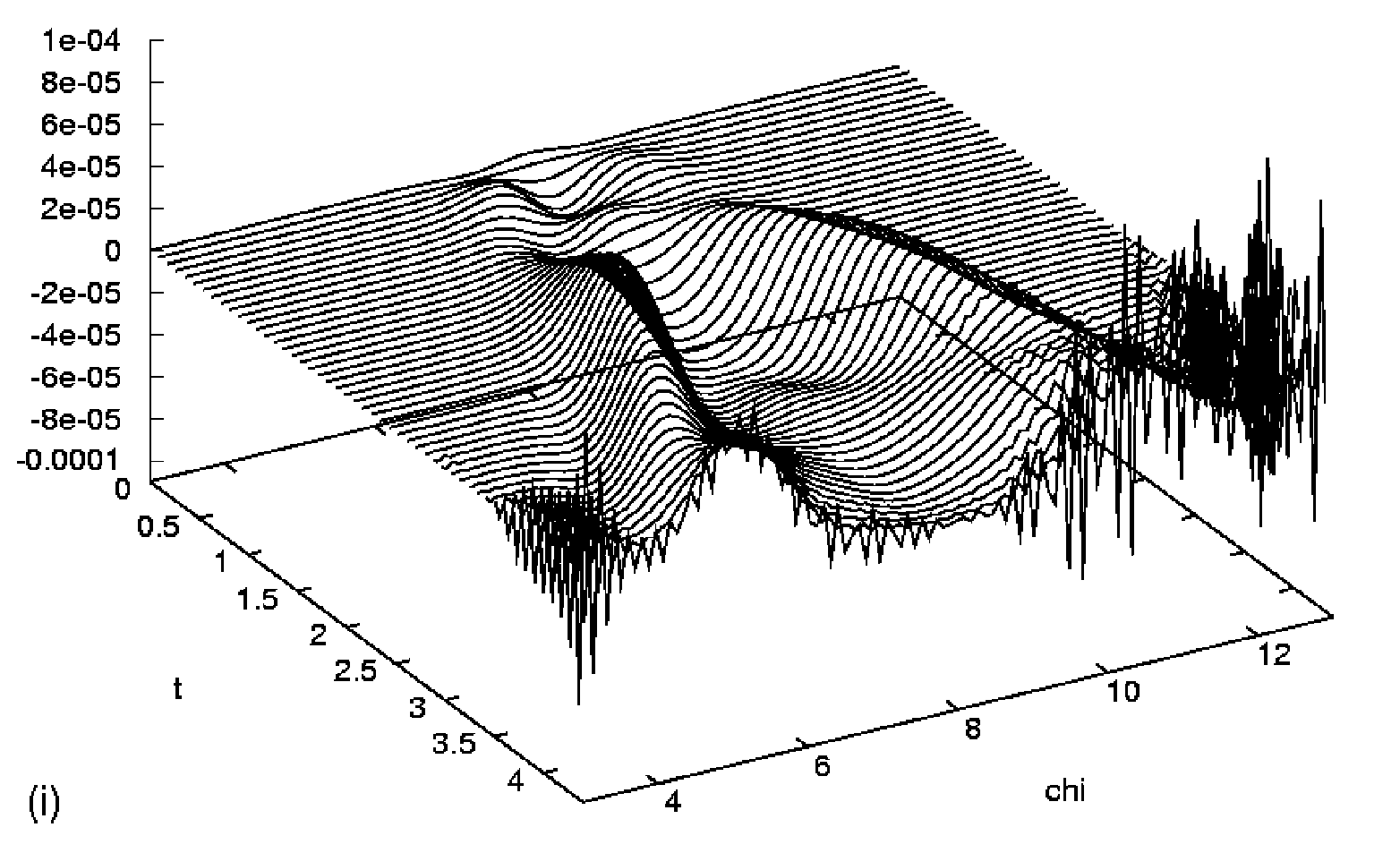} & 
\includegraphics[scale=0.3,angle=0]{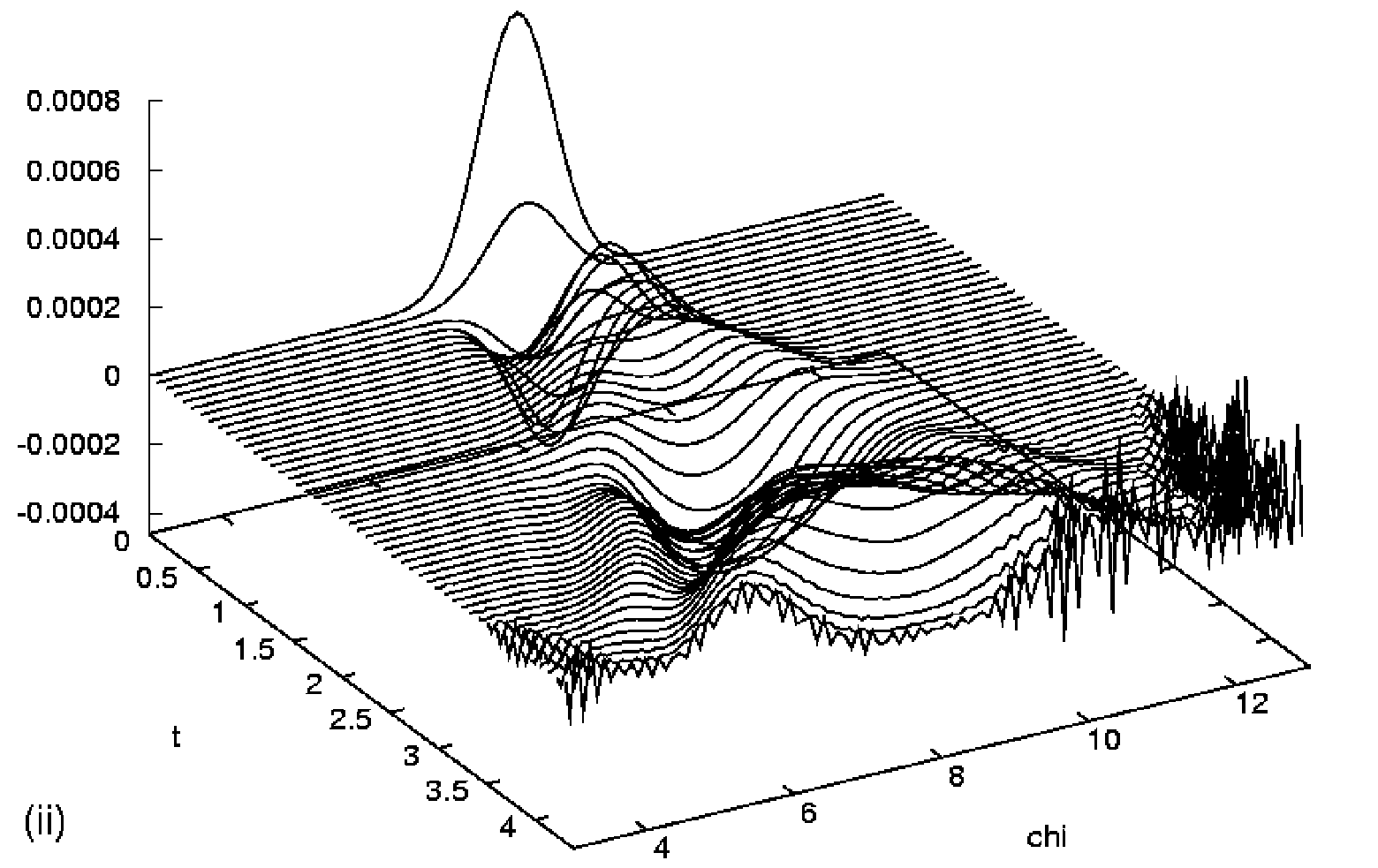} 
\end{tabular}}
\centerline{%
\begin{tabular}{c@{\hspace{5mm}}c}
\includegraphics[scale=0.3,angle=0]{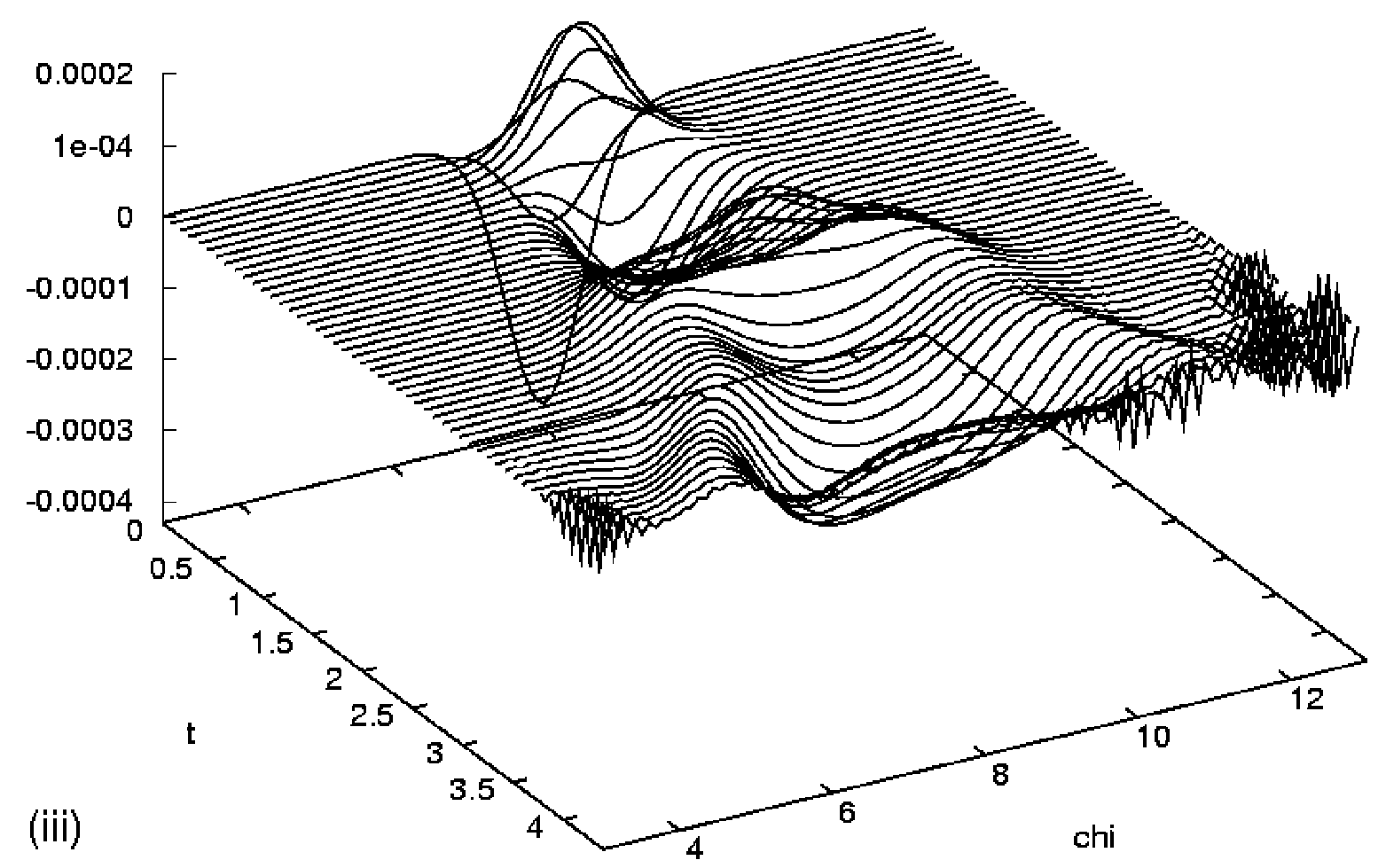} & 
\includegraphics[scale=0.3,angle=0]{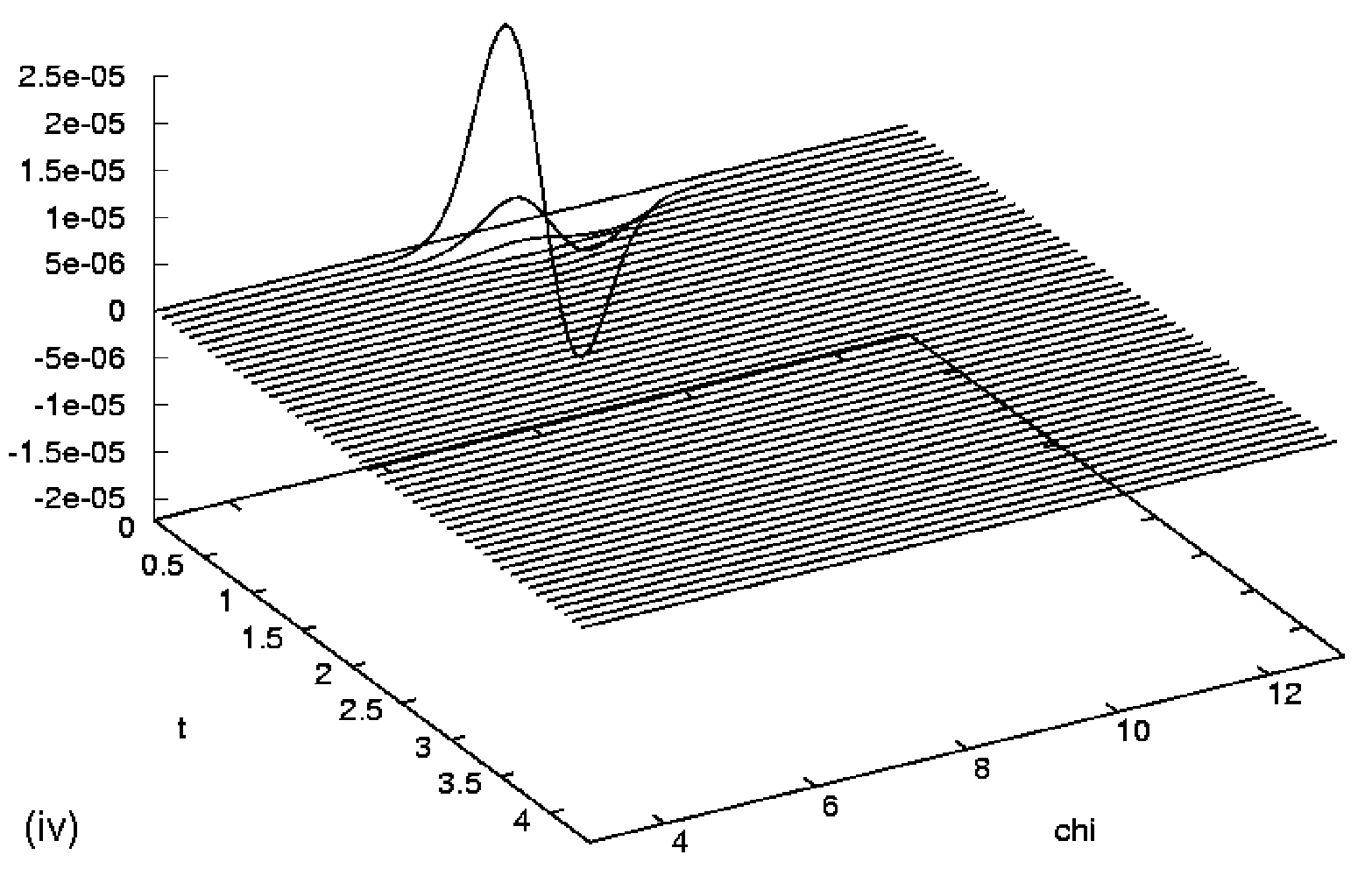} 
\end{tabular}}
\caption{\scriptsize {\textsl{Density and pressures contrasts with non-diagonal
component of the stress-energy tensor. (i) $\frac{T^0_0}{\rho^B}-1$ (ii) $-\frac{3T^1_1}{\rho^B}-1$ (iii) $-\frac{3T^2_2}{\rho^B}-1$ (iv) $T_0^1$
(same parameters as in Figure \ref{rsig}).}
} \normalsize}\label{contrast1}
\end{figure}

\begin{figure}
\centerline{%
\begin{tabular}{c@{\hspace{1mm}}c@{\hspace{1mm}}c}
\includegraphics[scale=0.25,angle=0]{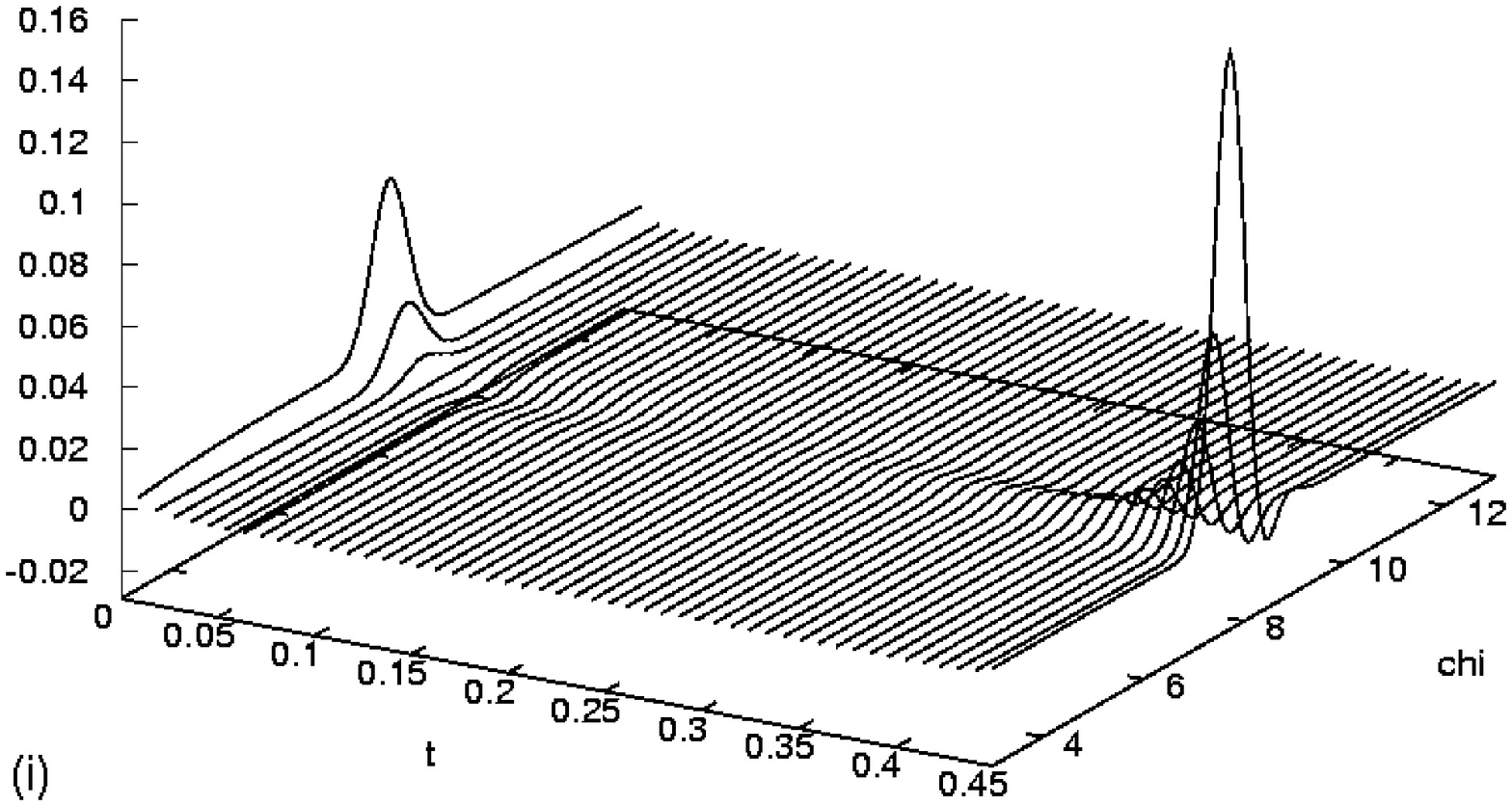} & 
\includegraphics[scale=0.25,angle=0]{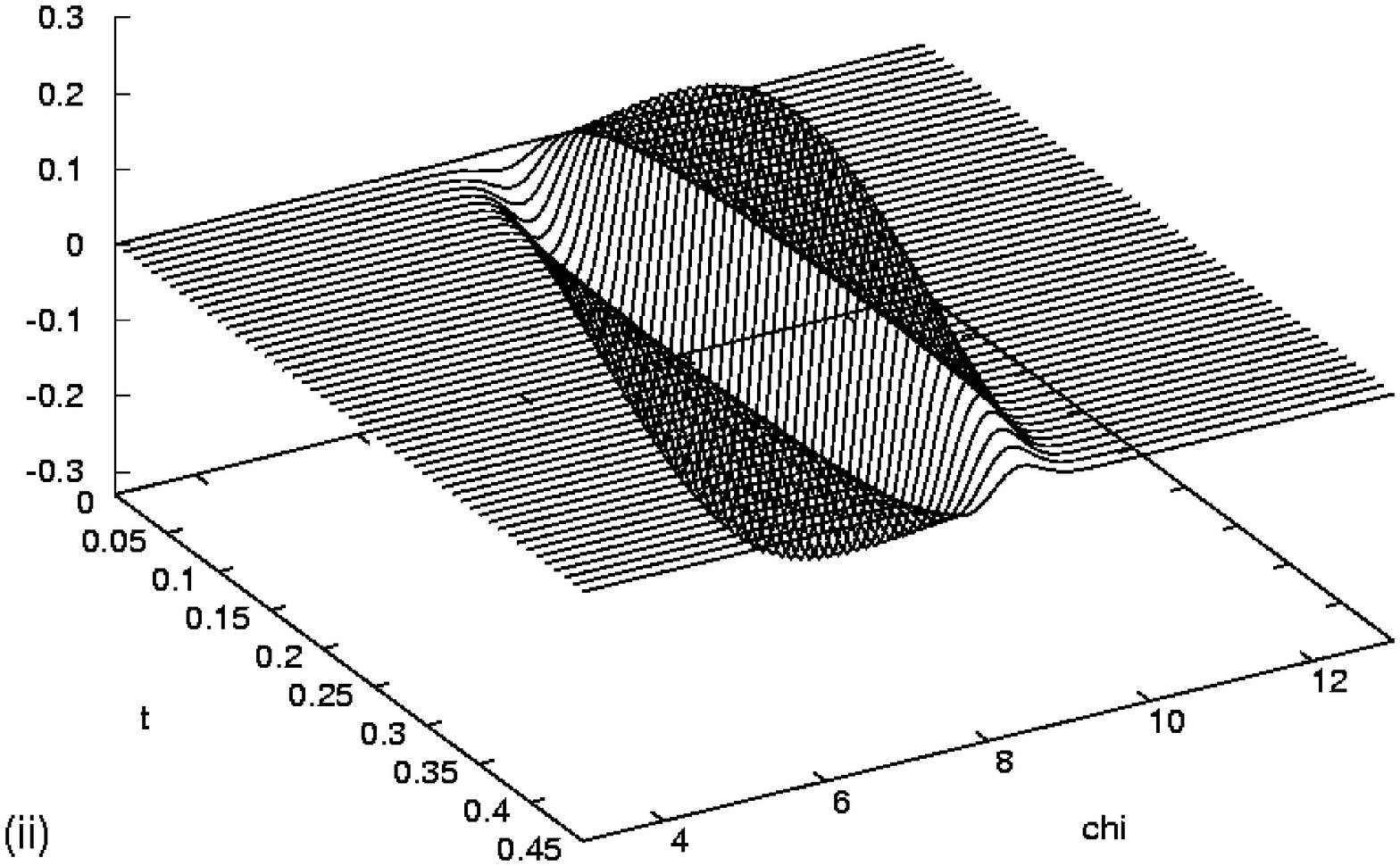} &
\includegraphics[scale=0.25,angle=0]{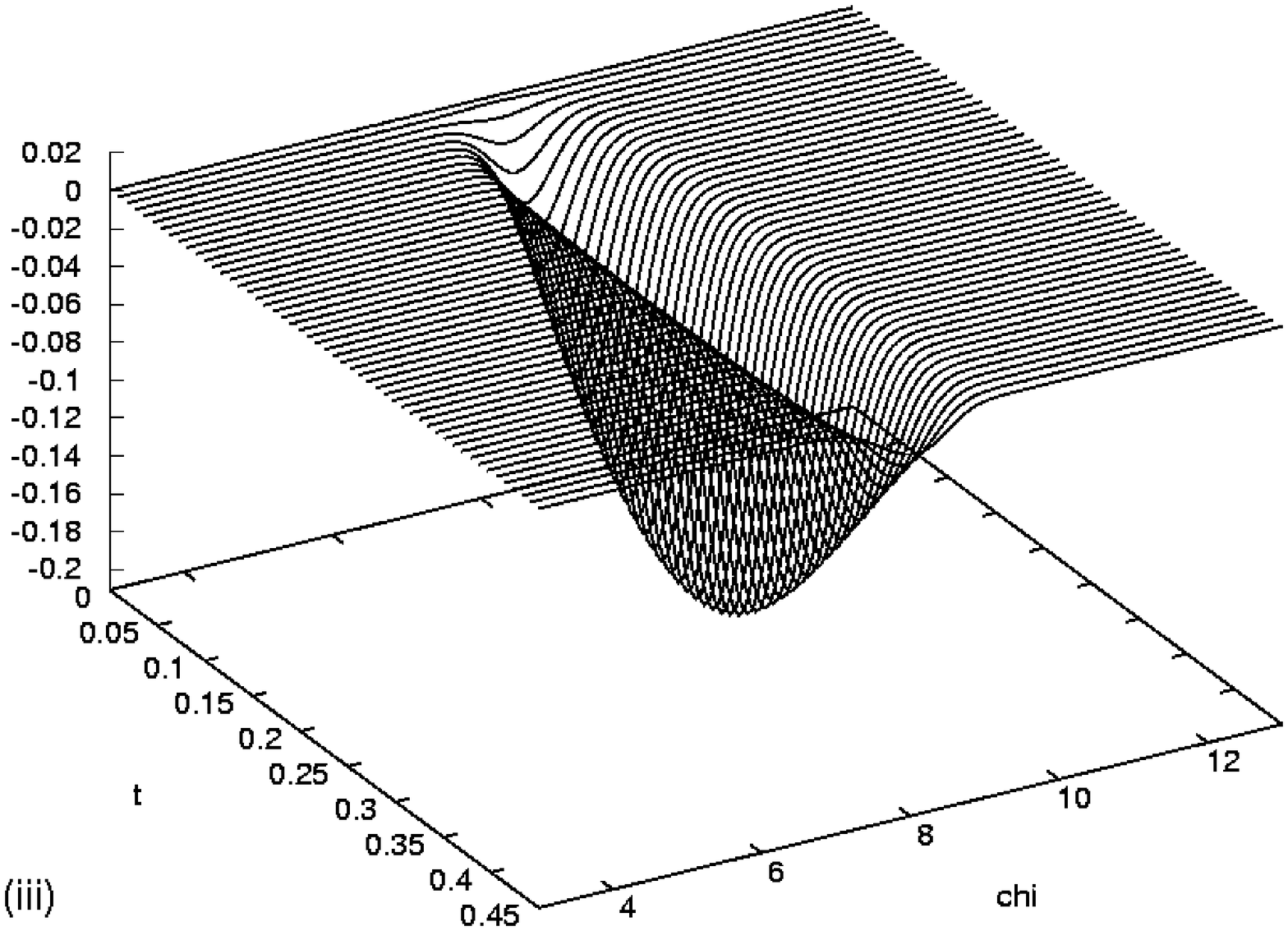} 
\end{tabular}}
\centerline{%
\begin{tabular}{c@{\hspace{1mm}}c@{\hspace{1mm}}c}
\includegraphics[scale=0.25,angle=0]{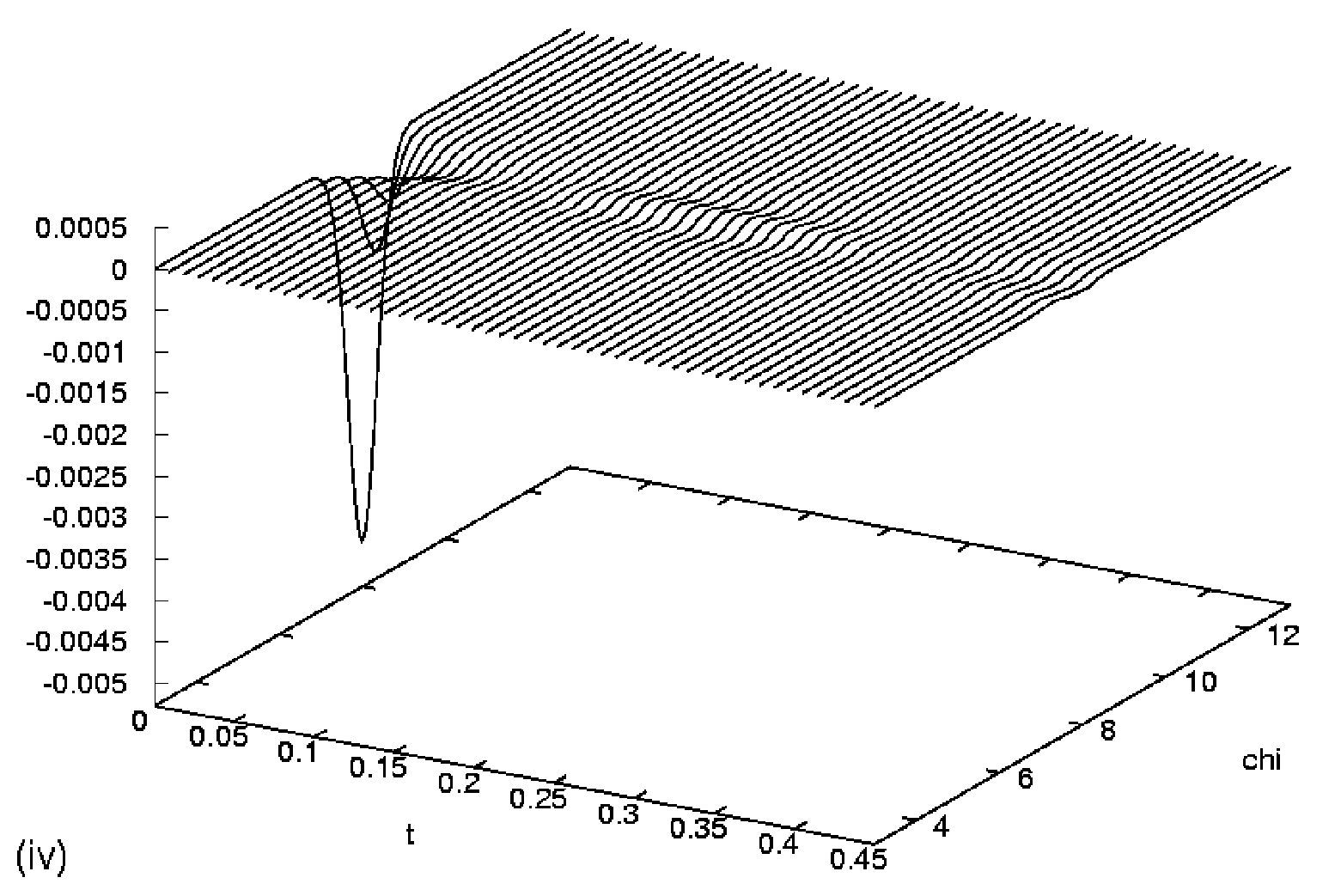} & 
\includegraphics[scale=0.25,angle=0]{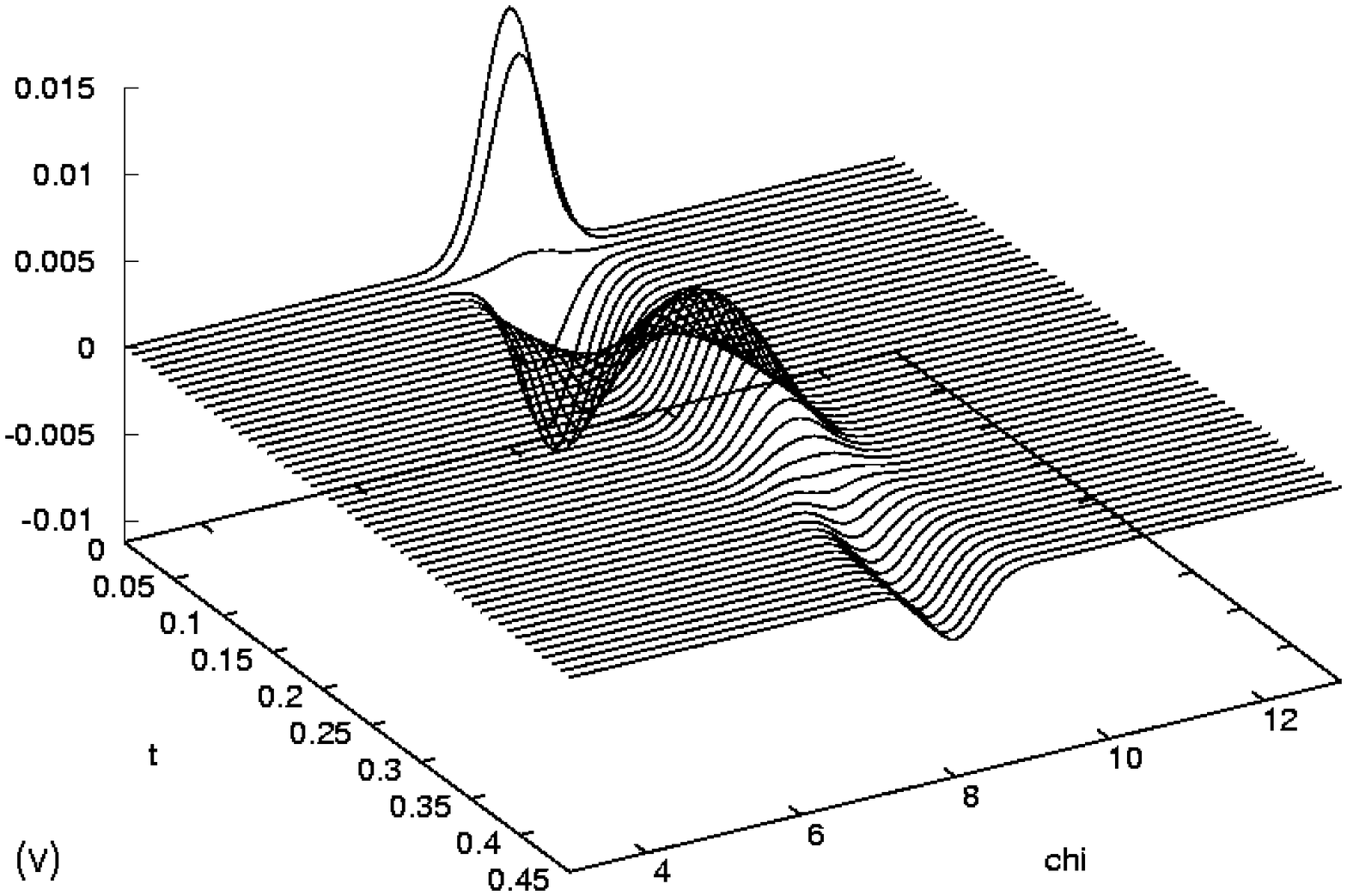} &
\includegraphics[scale=0.25,angle=0]{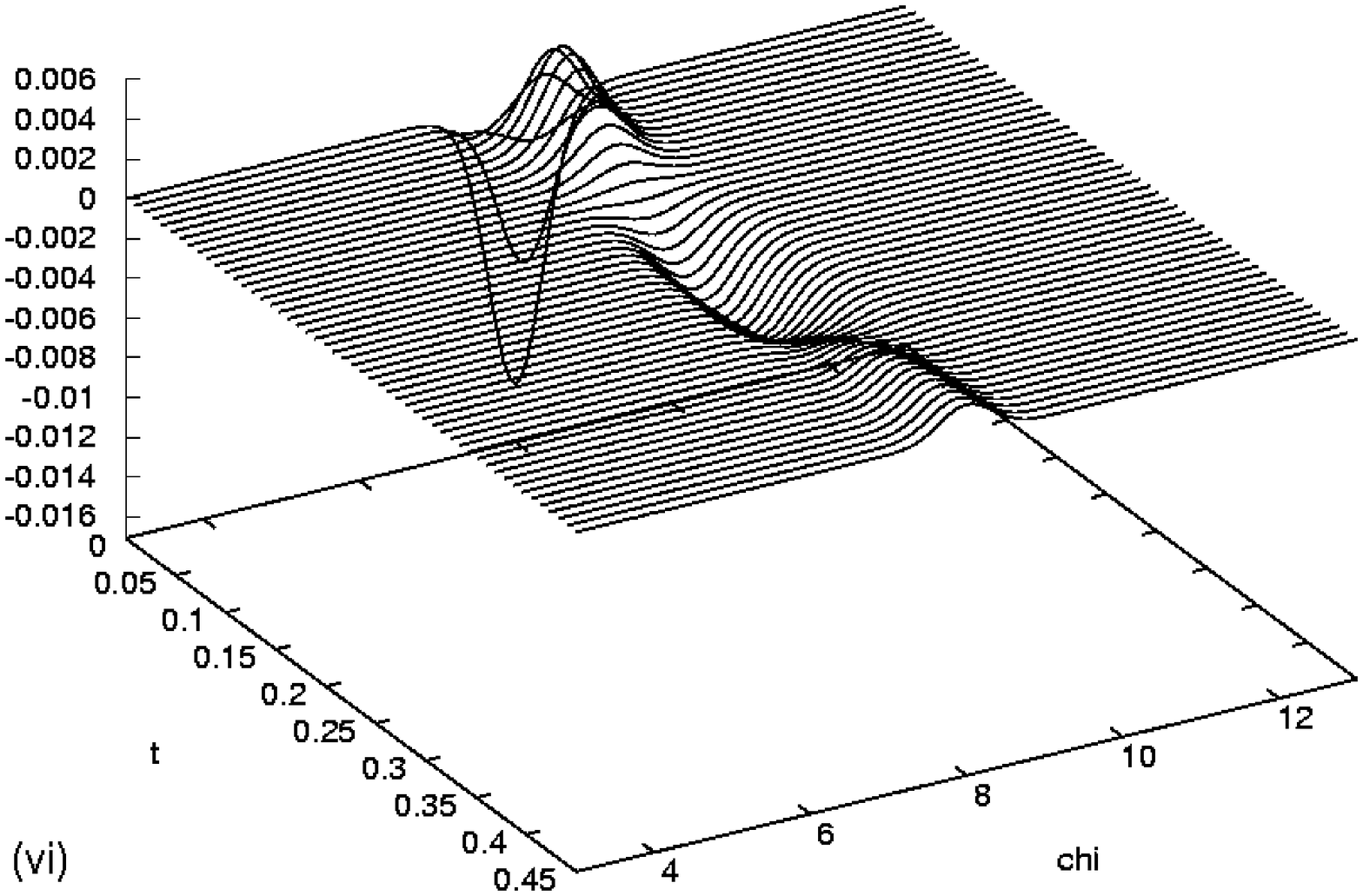} 
\end{tabular}
}
\caption{\scriptsize {\textsl{Example of decaying modes (early stages of the evolution). (i) $(a-a^B)$ (ii) $(b-b^B)$ (iii) $(c-c^B)$ (iv) $\left(\frac{T^0_0}{\rho^B}-1\right)$ 
(v) $\left(-\frac{3T^1_1}{\rho^B}-1\right)$ (vi) $\left(-\frac{3T^2_2}{\rho^B}-1\right)$
($\rho_0^B=1.5\times 10^3$, $L_H^0\approx 0.009$, $\epsilon_b=10^{-3}$, $w=0.25$, $\sigma_0=0$, $\Delta\chi=0.15$, $\Delta t=10^{-6}$).}
} \normalsize}\label{cosm}
\end{figure}

\begin{figure}
\centerline{%
\begin{tabular}{c@{\hspace{1mm}}c@{\hspace{1mm}}c}
\includegraphics[scale=0.25,angle=0]{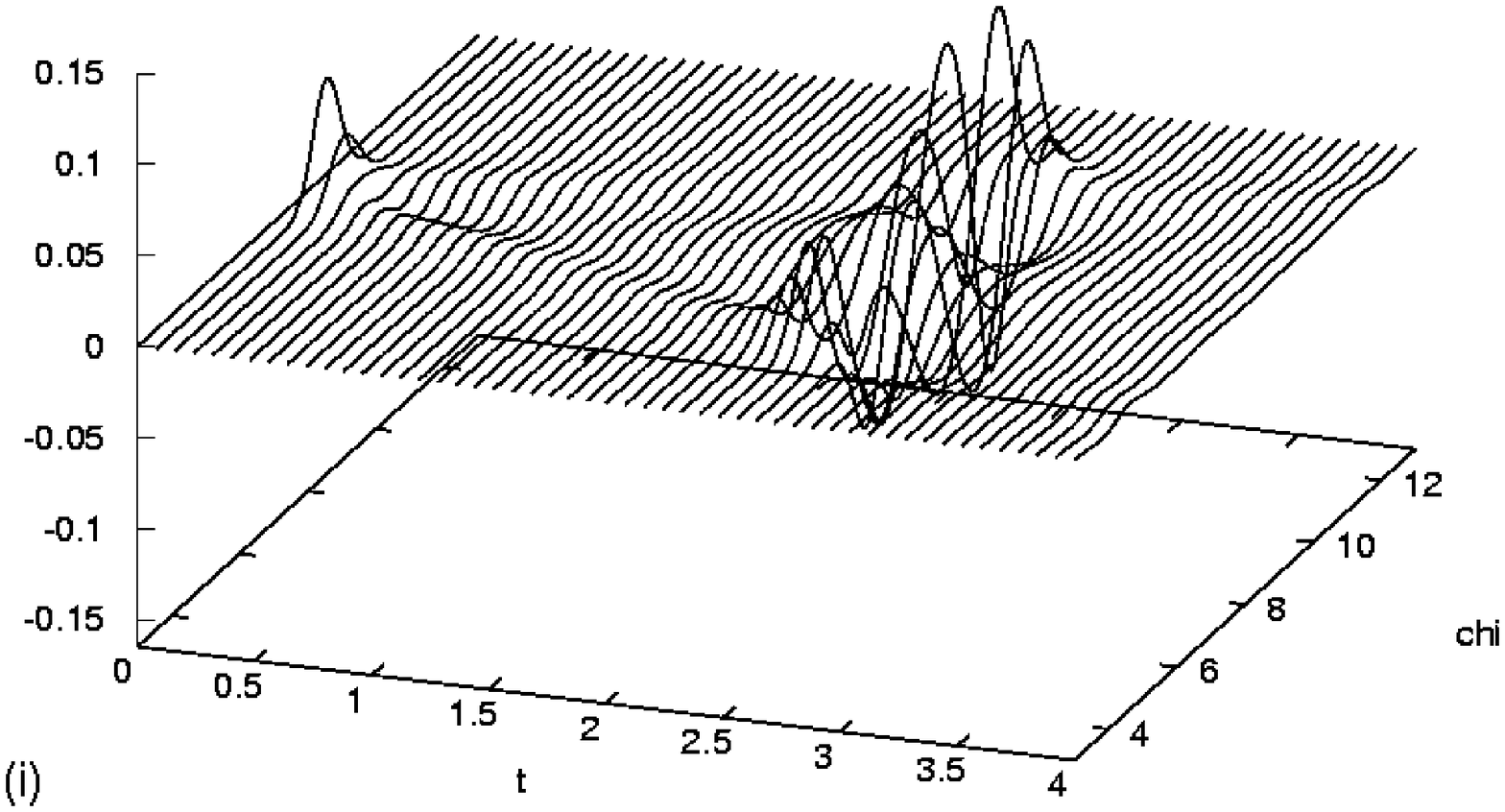} & 
\includegraphics[scale=0.25,angle=0]{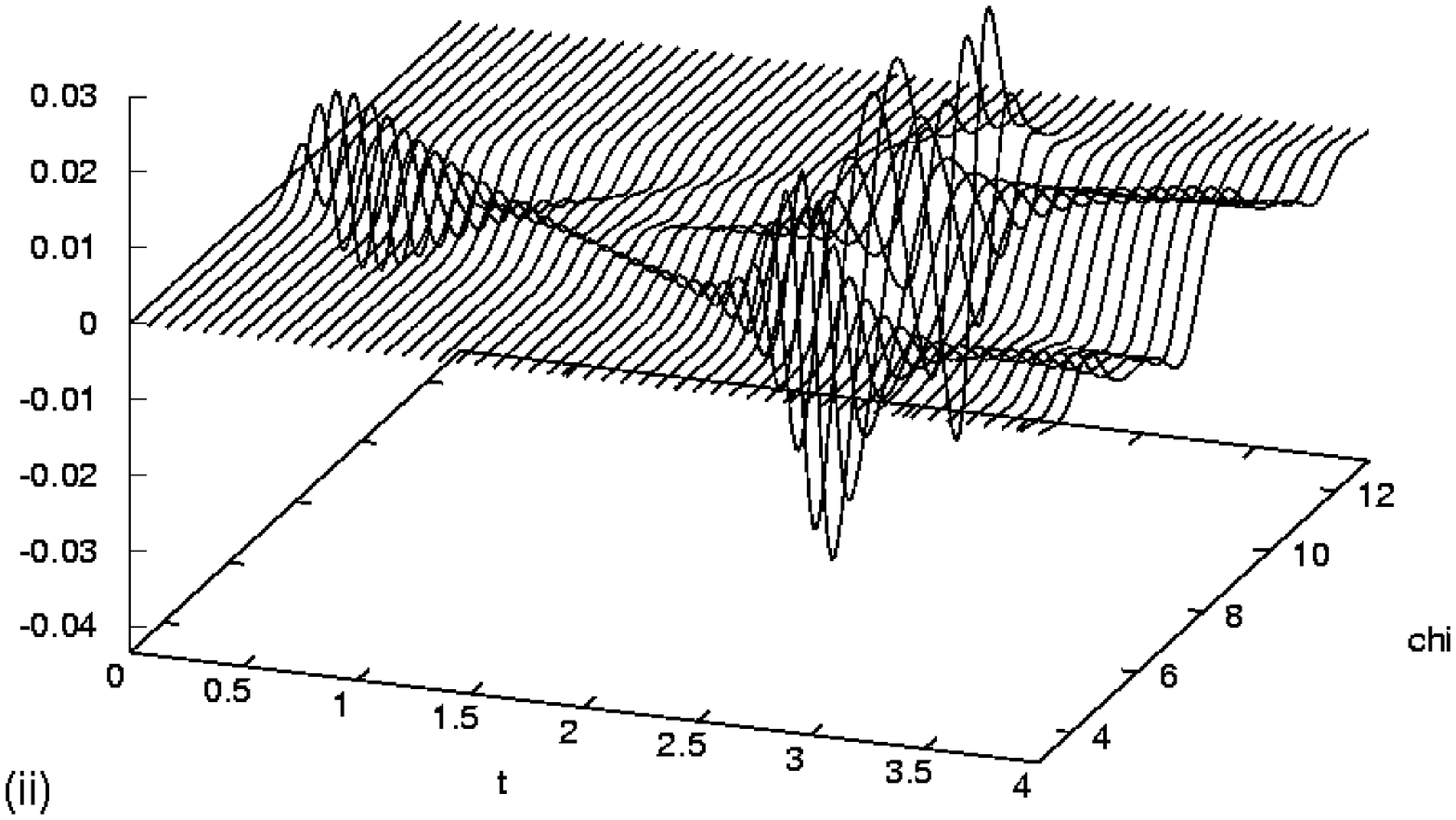} &
\includegraphics[scale=0.25,angle=0]{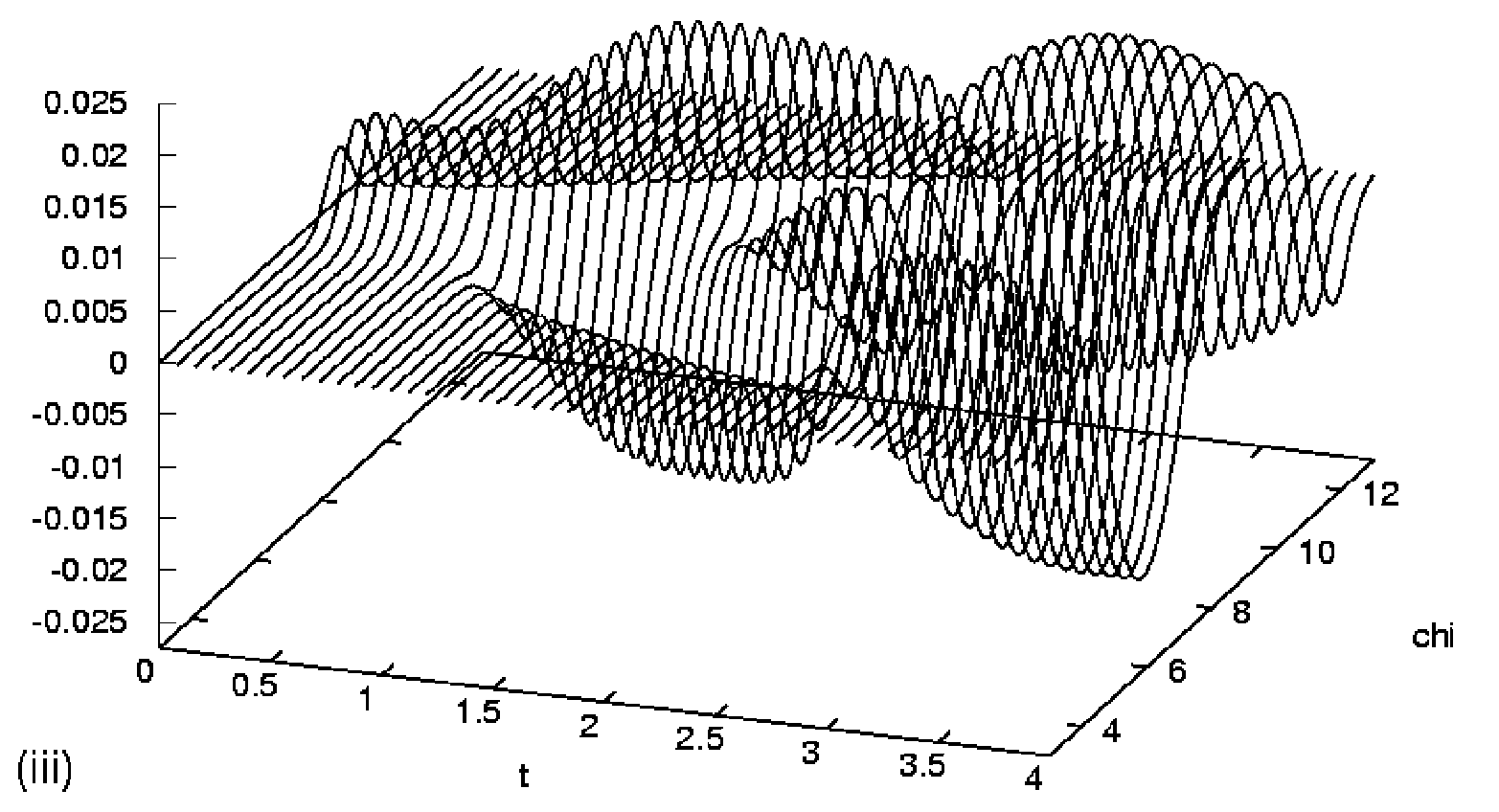} 
\end{tabular}}
\centerline{%
\begin{tabular}{c@{\hspace{1mm}}c@{\hspace{1mm}}c}
\includegraphics[scale=0.25,angle=0]{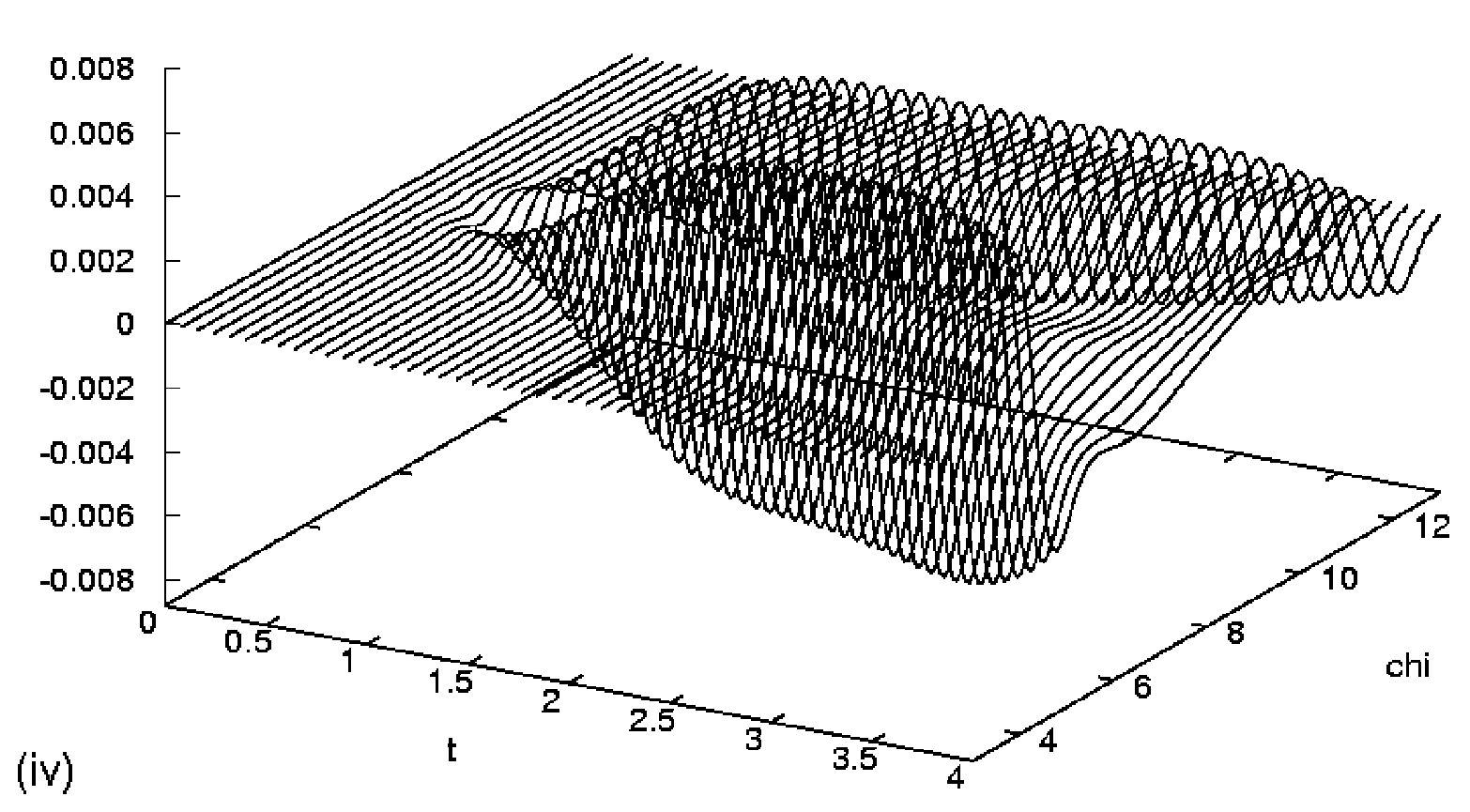} & 
\includegraphics[scale=0.25,angle=0]{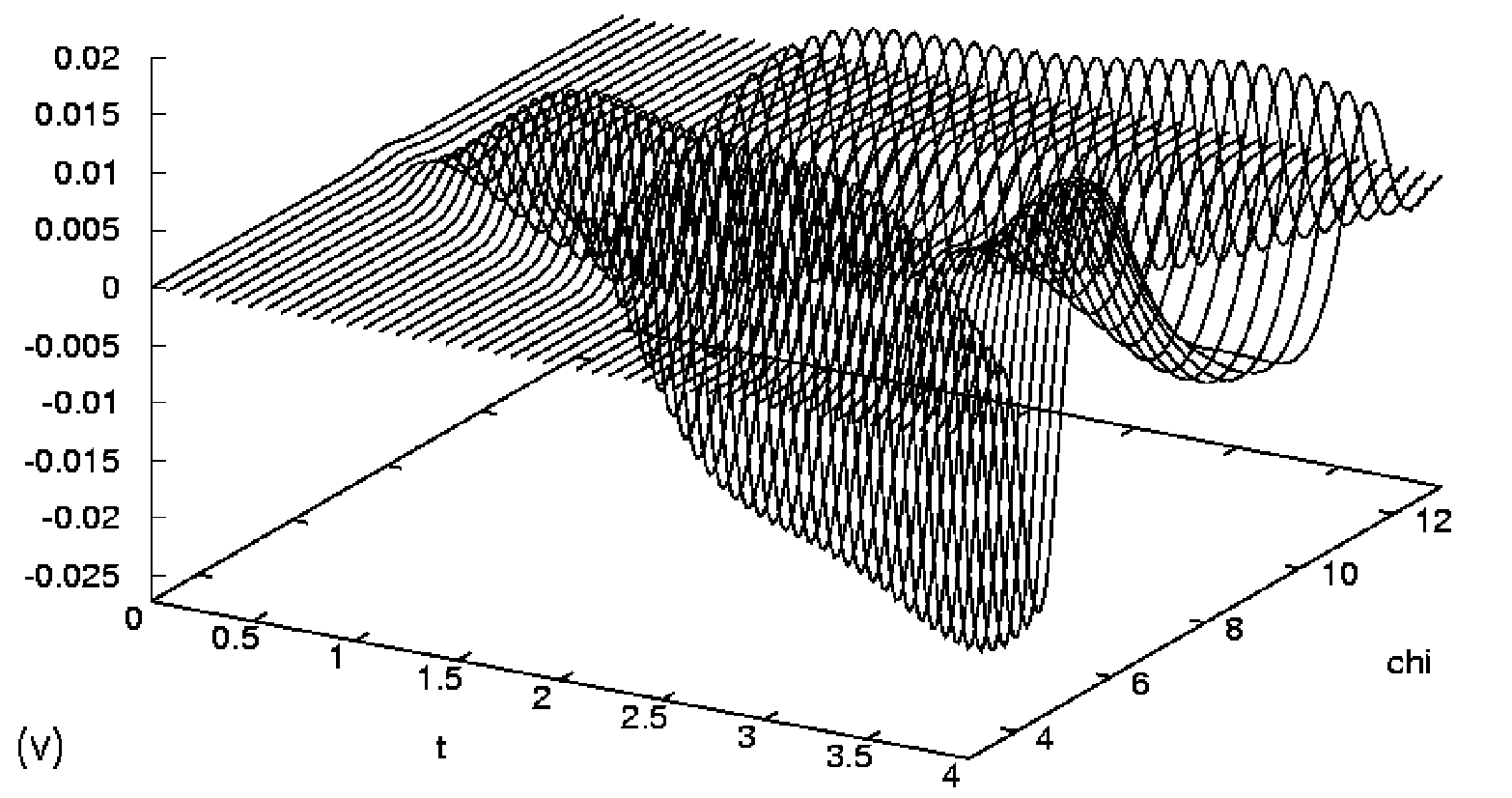} &
\includegraphics[scale=0.25,angle=0]{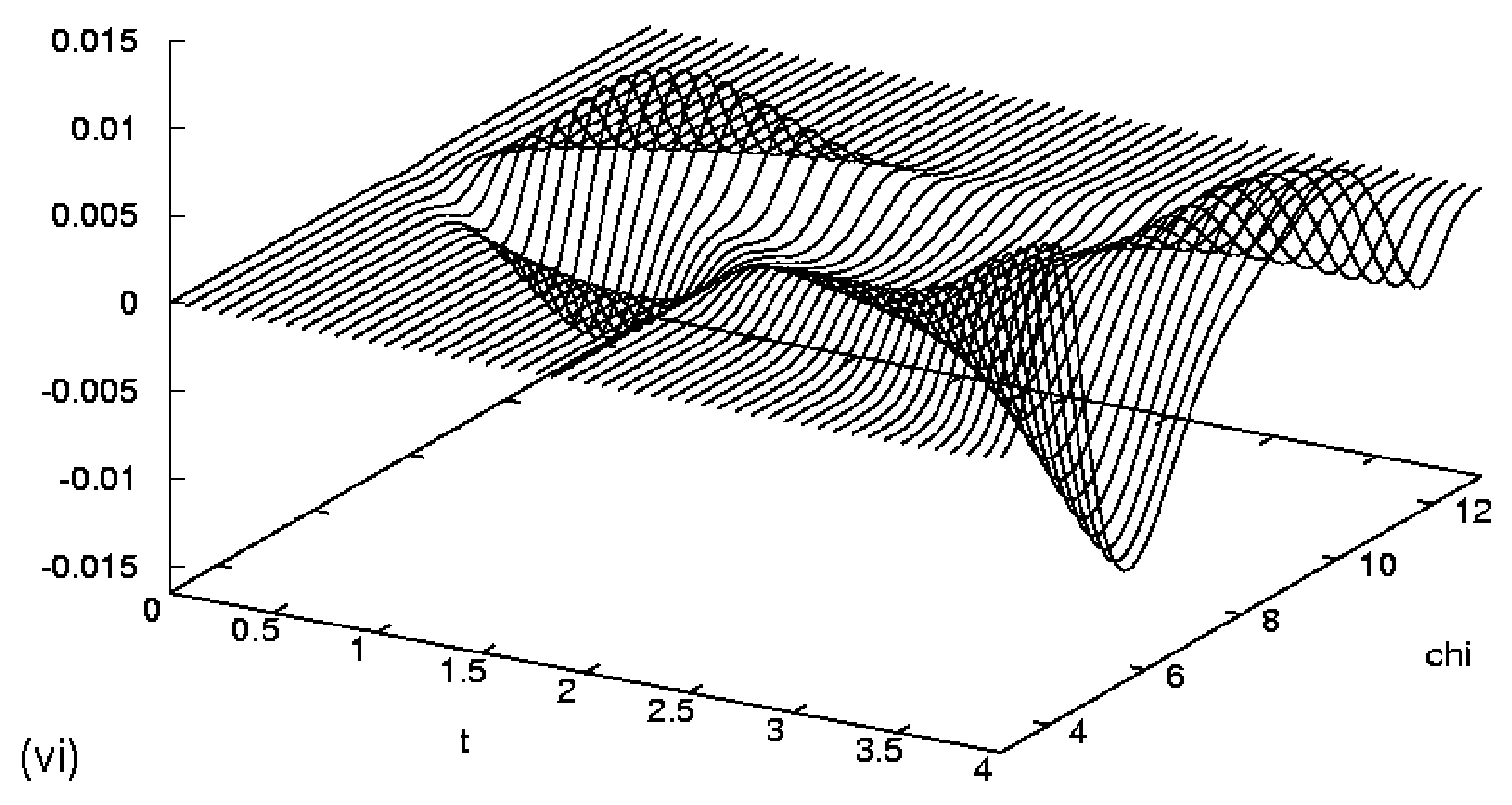} 
\end{tabular}
}
\caption{\scriptsize {\textsl{Example of growing modes (long time scale). (i) $(a-a^B)$ (ii) $(b-b^B)$ (iii) $(c-c^B)$ (iv) $\left(\frac{T^0_0}{\rho^B}-1\right)$ 
(v) $\left(-\frac{3T^1_1}{\rho^B}-1\right)$ (vi) $\left(-\frac{3T^2_2}{\rho^B}-1\right)$
($\rho_0^B=1.5$, $L_H^0\approx 0.3$, $\epsilon_{\dot{c}}=10^{-1}$, $w=0.3$, $\sigma_0=0$, $\Delta\chi=7\times 10^{-2}$, $\Delta t=10^{-7}$)}}
} \normalsize\label{cd}
\end{figure}

\begin{figure}
\centerline{%
\begin{tabular}{c@{\hspace{1mm}}c@{\hspace{1mm}}c}
\includegraphics[scale=0.25,angle=0]{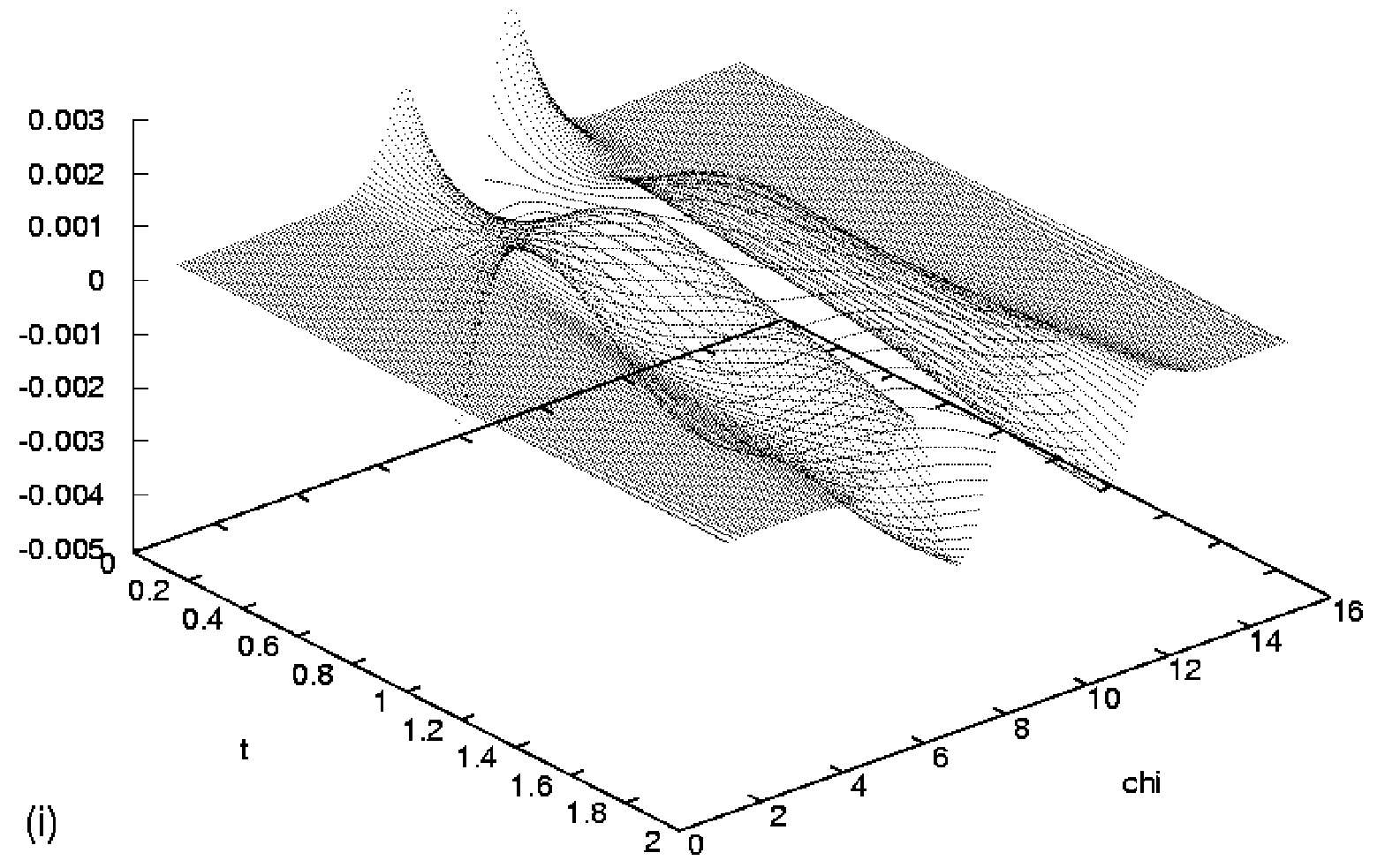} & 
\includegraphics[scale=0.25,angle=0]{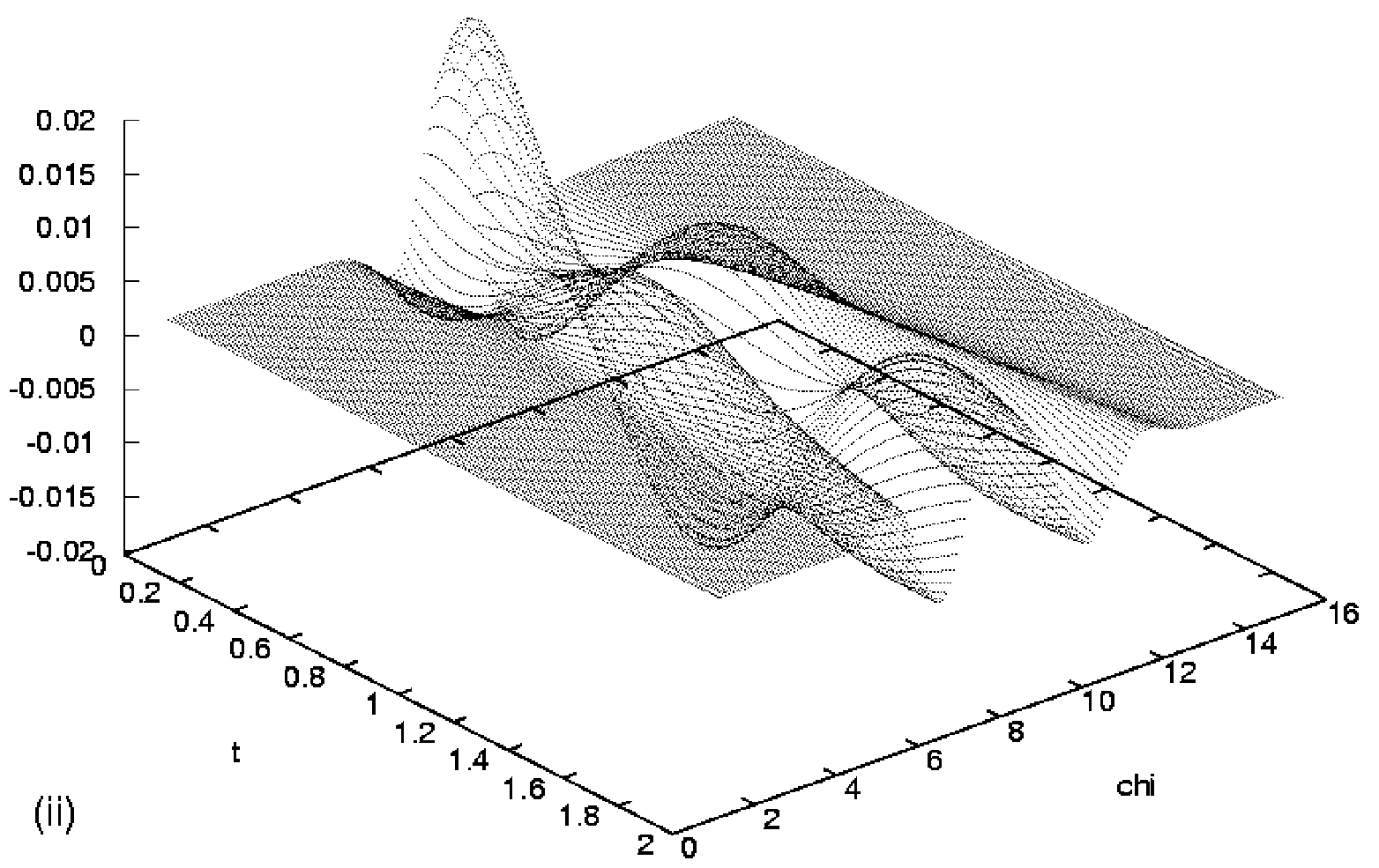} &
\includegraphics[scale=0.25,angle=0]{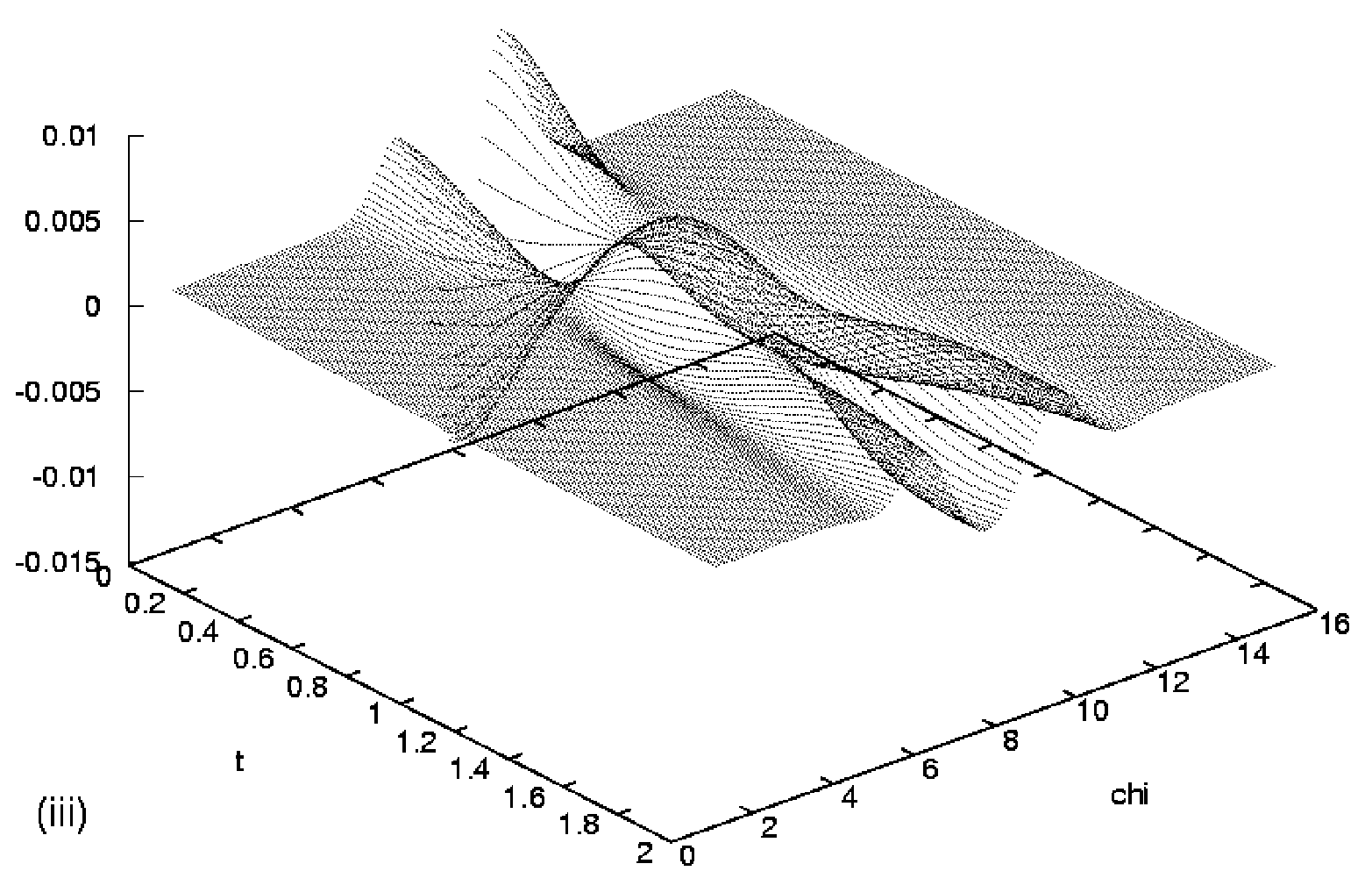} 
\end{tabular}}
\centerline{%
\begin{tabular}{c@{\hspace{1mm}}c@{\hspace{1mm}}c}
\includegraphics[scale=0.25,angle=0]{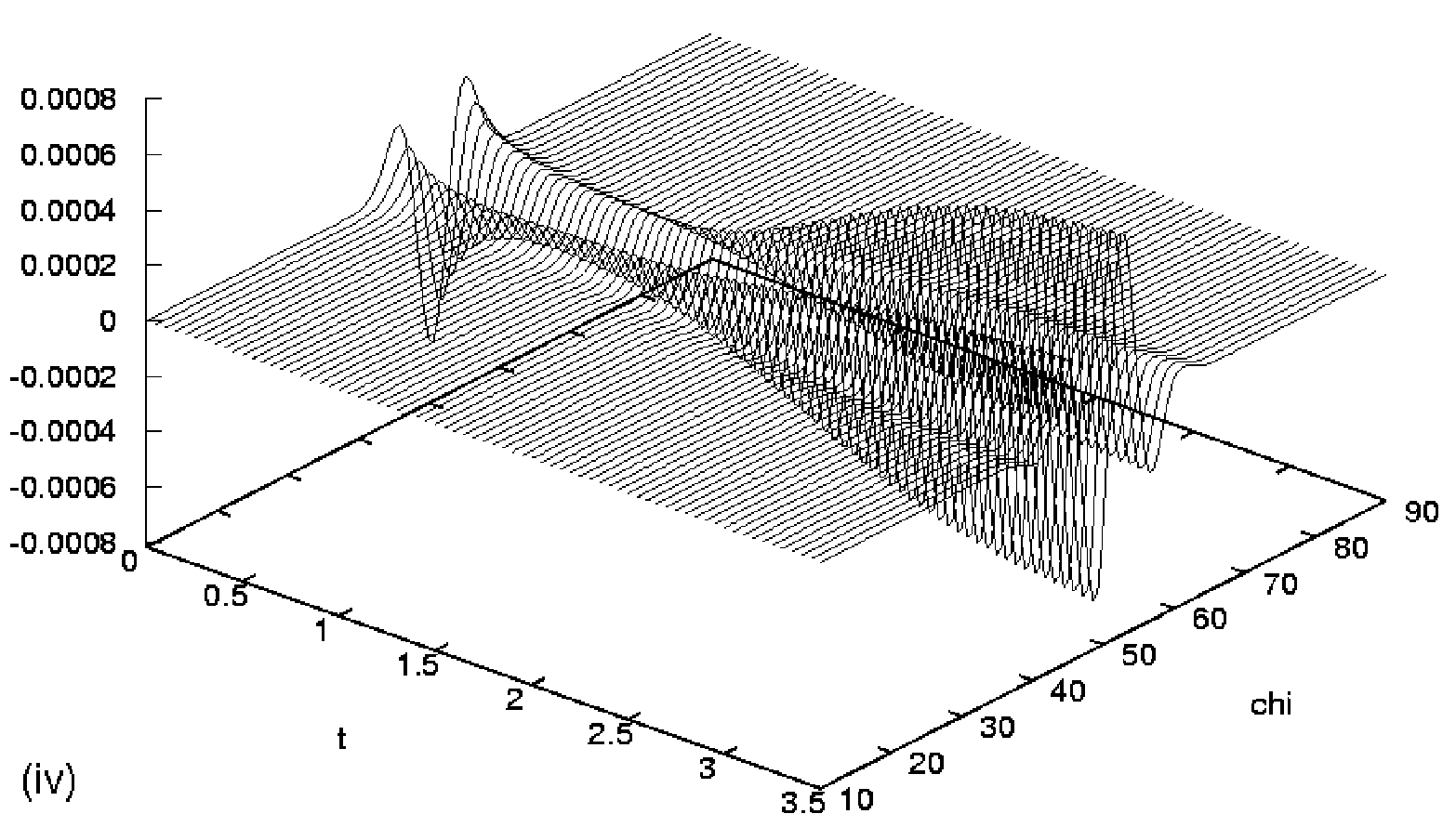} & 
\includegraphics[scale=0.25,angle=0]{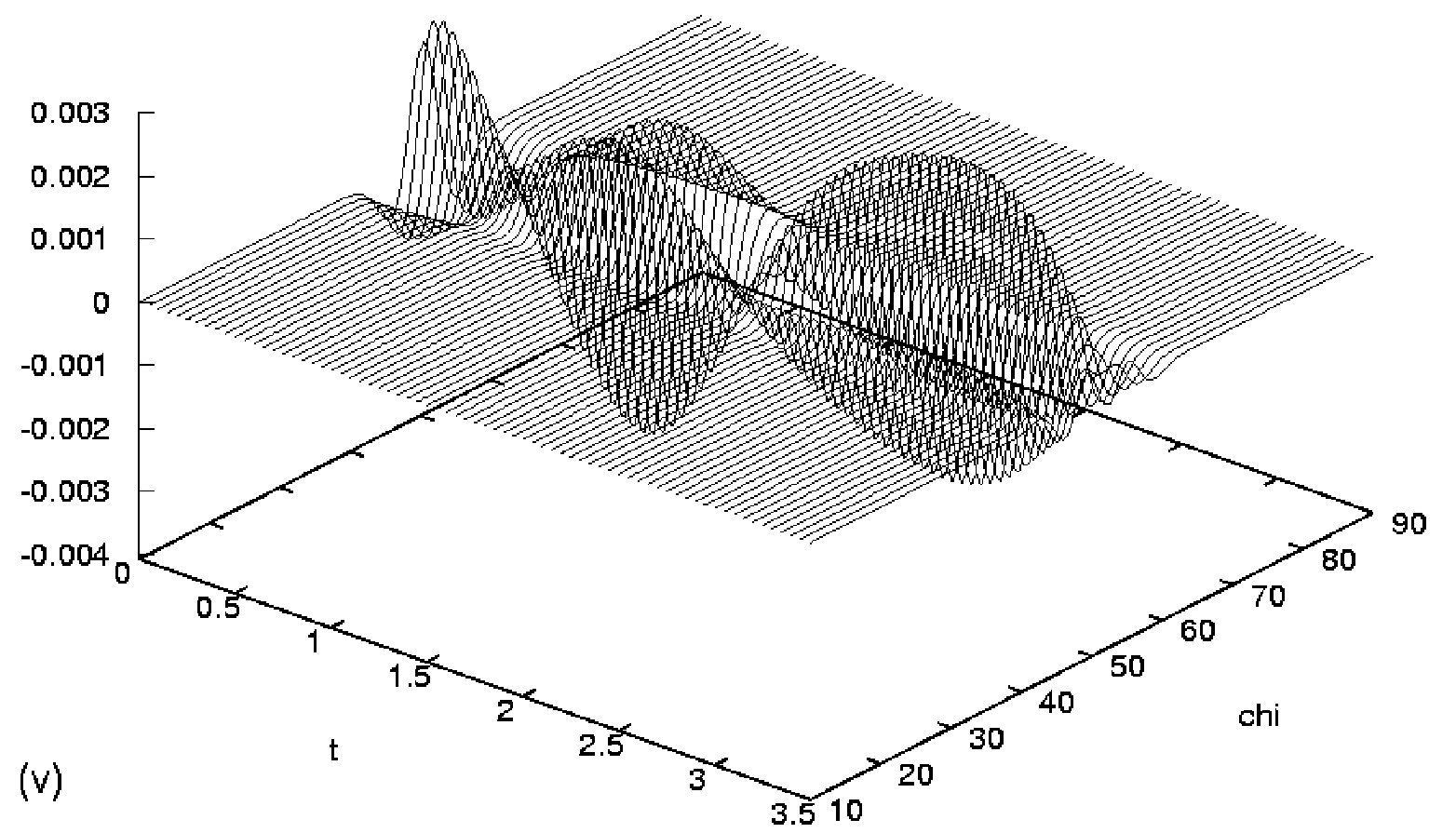} &
\includegraphics[scale=0.25,angle=0]{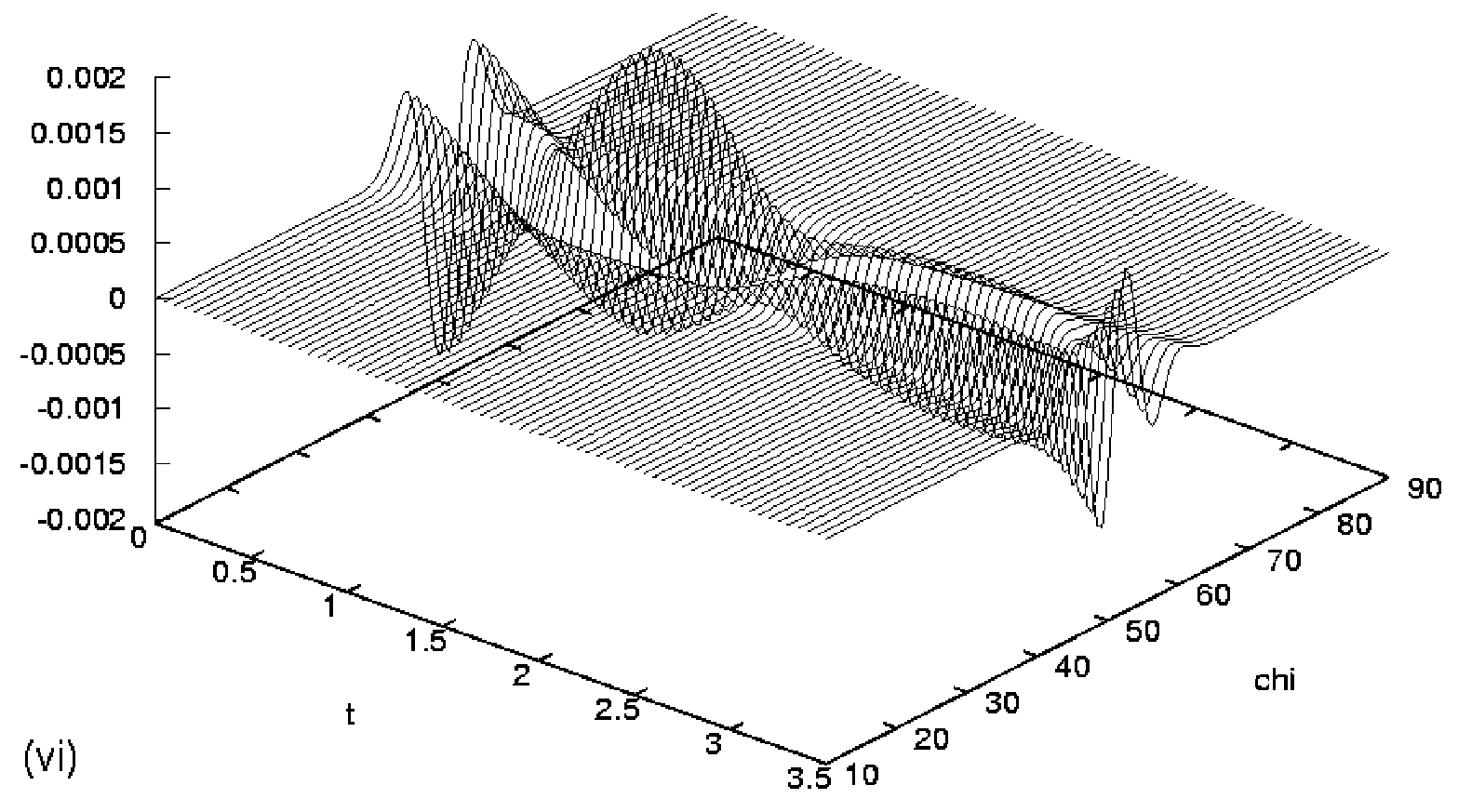} 
\end{tabular}
}
\caption{\scriptsize {\textsl{Evolution of fluctuations for two different coherent lengths: $\lambda_P\approx 6$ (i, ii, iii)
and $\lambda_p\approx 20$ (iv, v, vi). (i), (iv) $\left(\frac{T^0_0}{\rho^B}-1\right)$ ; (ii), (v) $\left(-\frac{3T^1_1}{\rho^B}-1\right)$ ; 
(iii), (vi) $\left(\frac{-3T^2_2}{\rho^B}-1\right)$ 
($\rho_0^B=1.5$, $L_H^0\approx 0.3$, $\epsilon_{c}=10^{-3}$, $\sigma_0=0.1$)}}
} \normalsize\label{hubble}
\end{figure}

\begin{figure}
\centerline{%
\begin{tabular}{c@{\hspace{5mm}}c}
\includegraphics[scale=0.3,angle=0]{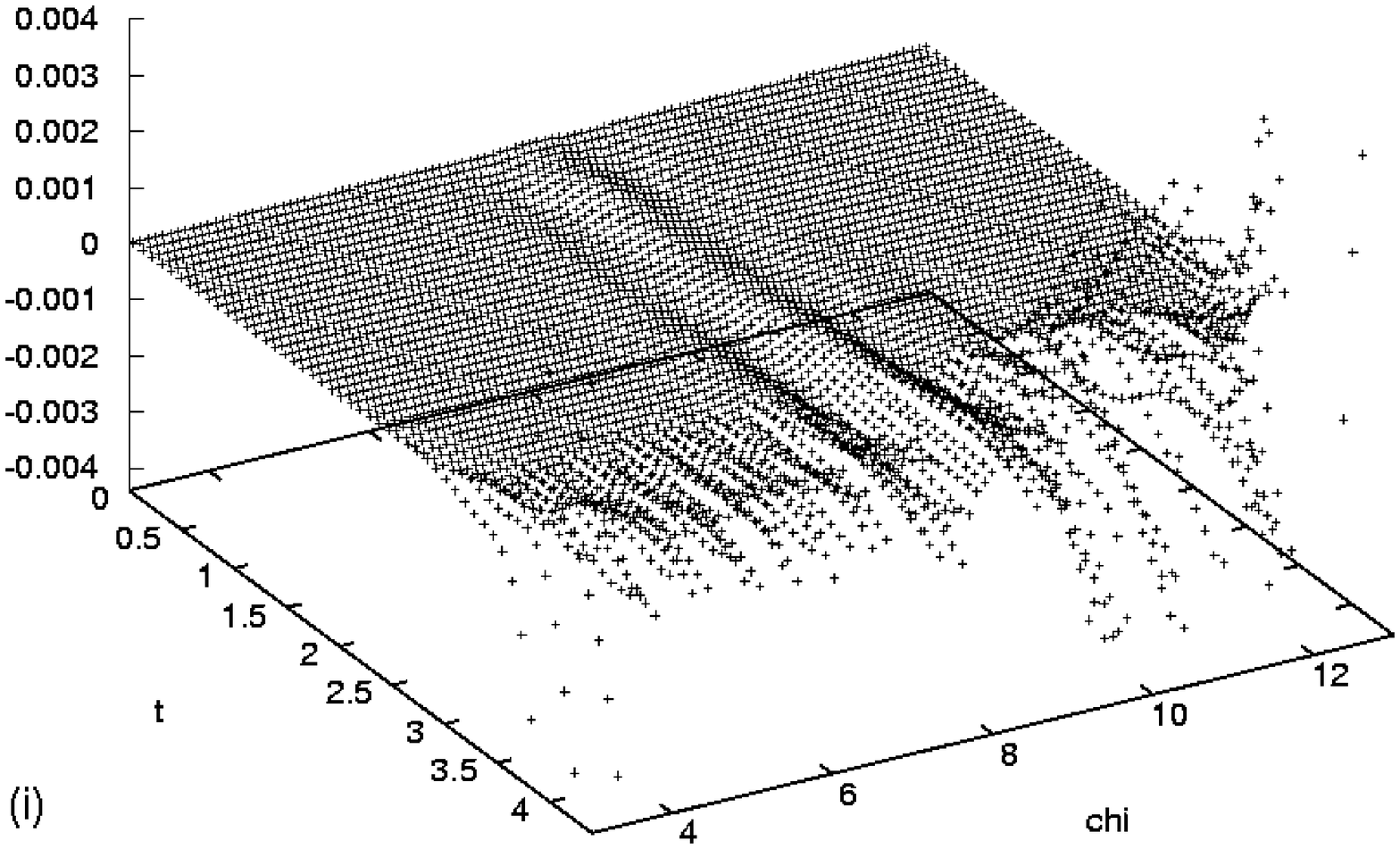} & 
\includegraphics[scale=0.3,angle=0]{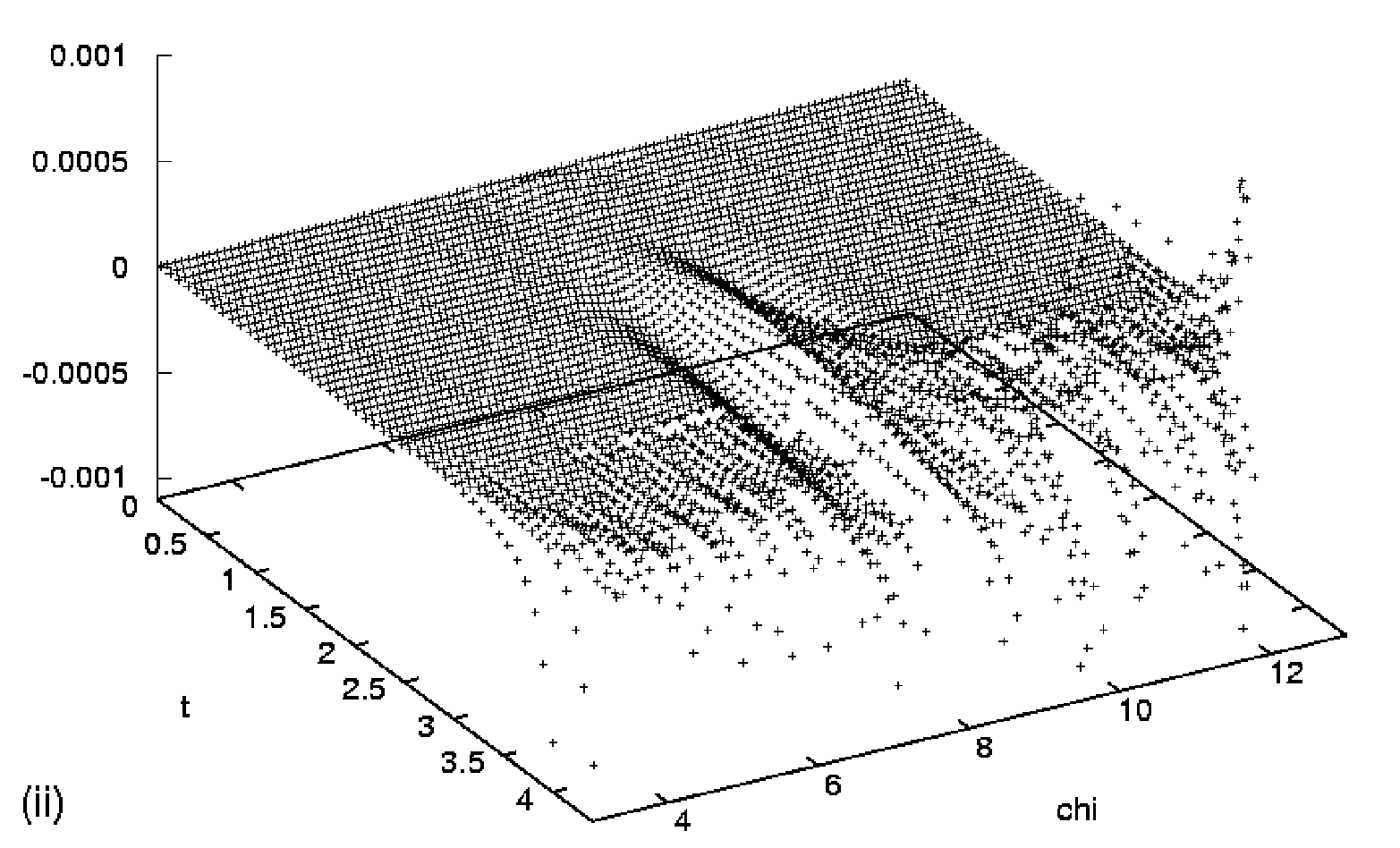}
\end{tabular}}
\begin{center}
\includegraphics[scale=0.3,angle=0]{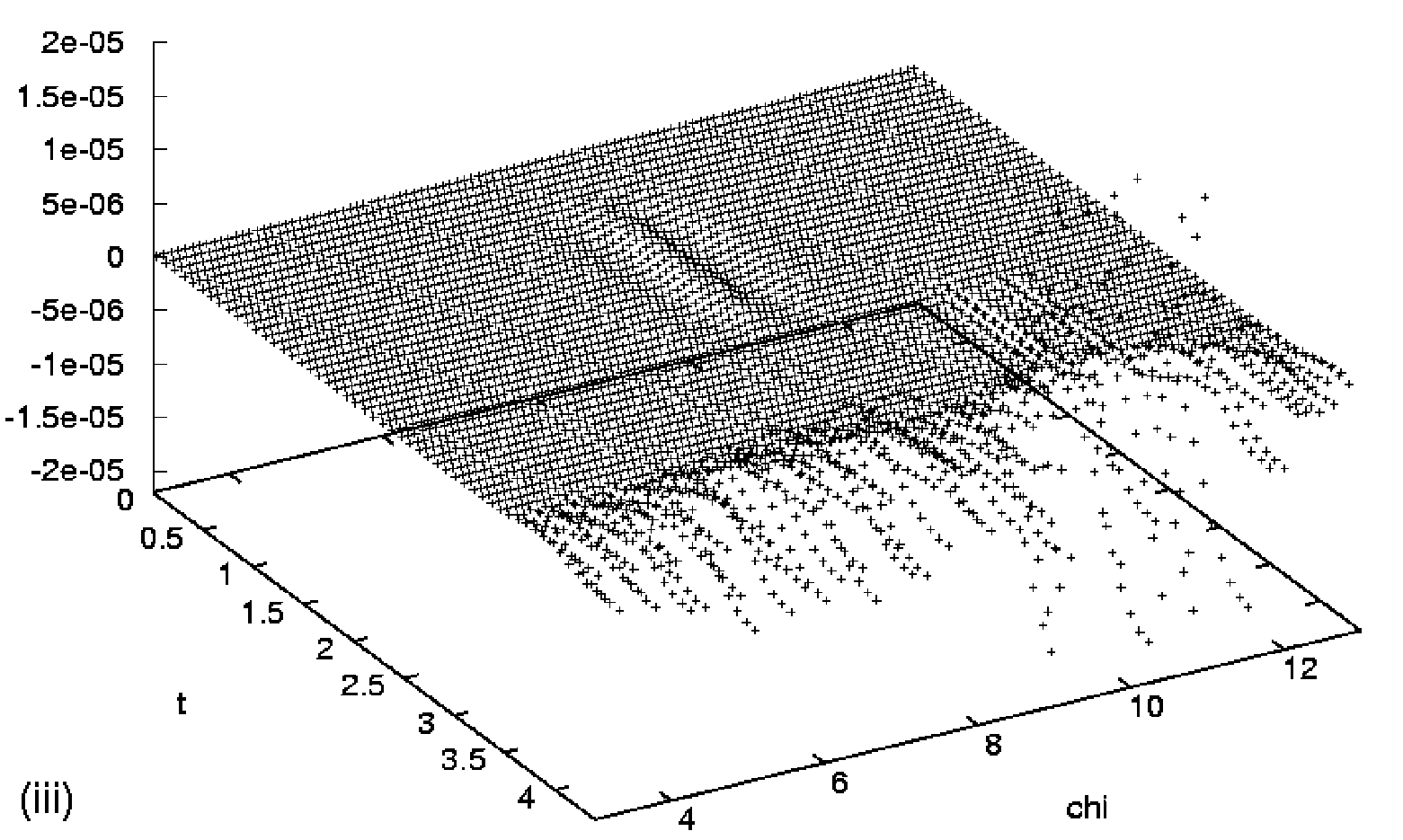}
\end{center}
\caption{\scriptsize {\textsl{Evolution of the constraints. (i) $\mathcal{H}$ (ii) $\mathcal{H}_{1}$ (iii) $\mathcal{G}^{\bf{3}}$ (same parameters as in Figure \ref{rsig})}
} \normalsize}\label{cons}
\end{figure}

In this section, we use the numerical results obtained from the method introduced before in order to present some interesting features
of the gravitational instability of YM cosmologies. For the sake of simplicity, we used for the initial pertubations 
a gaussian function of the form:
$$
\epsilon_i(\chi)=\epsilon_i\;e^{-\frac{\left(\chi-\chi_F\right)^2}{w}}
$$
where $\epsilon_i$, $\chi_F$ and $w$ are parameters indicating respectively the amplitude, position and width of the fluctuation. The last
is related to the $coherence$ $length$ $\lambda_P$ of the fluctuation, the length over which the perturbation changes significantly. The
figures were computed with only one field initially perturbed amongst the set $(a_0,\;b_0,\;c_0,\;\dot{c}_0,\;\dot{\lambda}_0)$.
Perturbing several fields will change the amplitude of density and pressures fluctuations on the initial space-like slice
specified by the IVP and will affect the developments of the resulting instabilities in the long term. Indeed,
as we will see further, the first stages of the evolution are characterized by a conformally-invariant regime maintained by
the too less inhomogeneous expansion. The instabilities arise only after some time, depending
on the initial conditions, when the fields reach a sufficient amount of inhomogeneity.\\ 
Figures \ref{ml}, \ref{a}, \ref{b}, \ref{c}, \ref{contrast1} and \ref{cons} show a typical evolution for the metric, the gauge potentials and the YM stress-energy tensor components as
well as the corresponding values of the three constraints under an initial perturbation of the gauge field $c$. It is useful
to notice that only $50$ iterations have been plotted in each 3D plot.\\
\\
For the gravitational sector, the perturbations of the spatial coefficients of the metric grow up as a power of conformal time (see Figure \ref{ml}). This is typical
of a radiation dominated universe and has been studied for a long time by several authors such as Lemaître, Tolman, Lifshitz and Bonnor 
(for a review, see \cite{peebles}).
From there, one should expect that the conformal invariance of YM lagrangian should result in a rather independent dynamics of the gauge
fields regards to the expansion. The last should thus dilute the energy densities of the YM initial fluctuations. In fact, this
is true when space-time is not significantly inhomogeneous i.e. when the perturbations of metric coefficients are negligible.
In this case, the slightly inhomogeneous expansion mimics a conformal transformation of the metric. However, when the
metric perturbations become important, we observe that the gauge fields start diffusing and the pressures and density contrasts start growing.\\
\\
Before going further, let us examine more precisely the evolution of the gauge potentials.
One can see in Figures \ref{a} to \ref{c} that the gauge fields seem to \textit{roll on} the potential well $V=\sigma^4$
of the homogeneous solution but this potential is now distorded by the perturbations. 
For example, the electric component $a$, related to the speed of the oscillation, reaches its maxima at the middle of
the phase space orbit $\sigma^B=0$ (see Figure \ref{rsig}), while magnetic components $b$ and $c$ oscillate at almost the same frequency
of the background solution $\sigma^B\cdot$ Let us give an idea of the distortion of the potential. Discarding the kinetic terms
into Eq.(\ref{t00}),
this potential can be written, in terms of the variables introduced in section 3.2:
\begin{equation}
V=\frac{2}{3}\left(\frac{\gamma^2}{2\;l^2}+\frac{\left(\gamma+\chi\frac{\gamma'}{2}\right)^2+\beta^2\chi^2}{m^2l\left(1+\chi^2\gamma\right)}\right)\cdot
\end{equation}
In Figure \ref{V}, we have represented the potential $V$ associated to the results in Figures \ref{ml}, \ref{a}, \ref{b}, \ref{c}, \ref{contrast1} and \ref{cons}.
The oscillation has been unfolded along the $\sigma-$axis and the resulting plot has been split in two
for sake of readiness. The distortion has also been scaled up to be visible (the function plotted is $\sigma^4+6\times 10^{3}\left(V-\sigma^4\right)$, the other parameters are same as in Figure \ref{rsig}). The potential of inhomogeneous YM cosmologies appears to be an implicit function
of time.\\
\\
Furthermore, the gauge fields diffuse along the $\chi-$axis, the perturbations spreading out with time.
This diffusion is due to the increasing importance of spatial gradients of the gauge fields
and the inhomogeneity of the metric in the YM equations. Another way
to see this is to draw the evolution of the gauge fields during half a period of $\sigma^B$ for the same amplitude of the initial perturbation
but for different values of the background initial expansion rate (parametrized here by $\rho^B_0$). This is done in Figure \ref{dif} for an initial
perturbation in the electric component $a$. For small values of the expansion rate (Figure \ref{dif} $(i)$),
the diffusion is important but it rapidly disappears for stronger expansion (Figure \ref{dif} $(ii)$ and $(iii)$). 
A much clearer way to express this is to compare the size of the fluctuations, given roughly by their coherence length $\lambda_P$, to
the Hubble length $L_H=1/H=t+\sqrt{\frac{3}{\kappa \mathcal{C}}}$ ($c=1$). At the beginning of the evolution ($t=0$), the last is given by 
$$
L_H^0=\frac{R_0}{\dot{R}_0}=\sqrt{\frac{3}{\kappa \rho^B_0}}\cdot
$$
Therefore, by varying the parameter $\rho^B_0$ and keeping the same coherence length through the same parameter $w$, we just modify
the ratio between both quantities. The corresponding initial Hubble lengths have been specified in the caption of Figure \ref{dif} with a
coherence length $\lambda_P\approx 1$. 
We can thus say that the diffusion is initially important for the short wavelengths $\lambda_P\approx L_H$ and, for wavelengths
greater than the Hubble length, becomes important only when the spatial inhomogeneity is strong enough, as in 
Figures \ref{ml}, \ref{a}, \ref{b}, \ref{c} and \ref{contrast1}.\\
\\
The observable quantities are represented in Figure \ref{contrast1} under the form of
the density contrast:
$$
\frac{T_0^0}{\rho^B}-1
$$ and similar relations for the radial ($T_1^1$) and tangential ($T_2^2$) pressures contrasts. 
It should also be noted that, according to the conformal invariance
of YM fields, the scalar curvature $\mathcal{R}$ vanishes exactly and therefore the components of the stress-energy tensor are
directly proportional to those of the Ricci tensor through Einstein equations and we have therefore a direct information on the spatial
inhomogeneities. \\
These contrasts oscillates in general, reflecting the underlying oscillations of gauge potentials, even though
the amplitude of the oscillation is damped in the first stages of the evolution and then grow after some time, with a
different regime for density and pressures contrasts. 
For example, in the case illustrated in Figure \ref{contrast1}, the density constrast first decays, then enter
a growing regime while the amplitude of the (radial and tangential) pressures contrasts slowly decreases. \\
\\
According to the radiation nature of YM, one should have expected that the evolution of the density contrast would have been a
superposition of a growing mode in squared power of the conformal time\footnote{The density contrast is proportional to the world time in
the synchronous gauge $N=1$.} 
and a decaying one in inverse squared power of conformal time. 
This is well-known from a simple first order perturbation computation (see \cite{peebles}).
In fact, when the perturbations of the geometry are small and their gradients negligible, the expansion can be considered as
roughly homogeneous and it dilutes the energy of the conformally invariant YM fields. This case is a pretty good approximation
of the early universe and is represented in Figure \ref{cosm}: the amplitudes of the density and pressures contrasts
decay with conformal time while the gauge potentials still oscillate in their distorded interaction potential, insensitive
to an expansion that is too close to a conformal transformation. The growing modes of the linear theory simply do not appear
immediately
due the scale invariance of YM fields. This means also that the YM cosmology can be much more inhomogeneous than
it looks at the level of the observable quantities as the non observable ones (the metric components and the gauge potentials)
do not decay in general, as we have seen before. \\
\\
By the way, when space inhomogeneity is too strong, growing modes for the contrasts 
and the perturbations of the gauge potentials (see Figure \ref{cd} for illustration) appear,
along with the diffusion already mentionned above.  In fact, the details of this regime seem to be quite complex 
as it strongly depends on the interactions
between the gauge components and on the field(s) initially perturbed. More work is under way to explore that regime and to
characterize it in terms of the two dimensional abelian Higgs model of the theory.\\
\\
To summarize the evolution of Figures \ref{ml}, \ref{a}, \ref{b}, \ref{c}, \ref{contrast1}, if one starts with some fluctuations 
in a universe dominated by YM fields, the density and pressures contrasts will
enter an oscillating regime, firstly damped by the quasi-homogeneous expansion. Then, when the expansion becomes significantly inhomogeneous,
the constrasts grow, at a time depending on the nature, the size and the amplitude of the initial perturbation. The transition
to growing modes (and diffusion) seems indeed to
occur when the perturbation of the metric coefficients $\delta_m=(m-1)R$ or $\delta_l=(l-1)R^2\chi^2$ reach the order of magnitude
of the scale factor (here a few percents, see
Figure \ref{ml}). Therefore, the transition to a non conformal expansion regime is to be linked directly with the emergence of
the instabilities.\\
\\
If the existence of growing modes is due here to the fact that the inhomogeneous expansion looks no longer
like a conformal transformation of the metric and therefore yields to non trivial interactions with the inhomogeneous gravitational
field, it would be interesting to see what happens to the growing modes with other gauges than the conformal one ($N=R(t)$)
used here, for example, in the case of $maximal$ $slicing$ (cf. \cite{seidel} and references therein). Indeed, this will lead to the study of a true \textit{gravitational collapse}
of YM fields instead of a shiny \textit{gravitational instability}.\\
\\
Now, let us be more precise about how the parameters of the simulation affect the evolution of the observable quantities. 
A run is specified by the following parameters: $\rho^B_0$, the initial background density ; $w$, related to
the coherence length $\lambda_P$ of the fluctuations ; $\epsilon$, the relative amplitude of the perturbation ; $\sigma_0$, the
initial position of the Gal'tsov-Volkov particle in the interaction potential, and the nature of the perturbation (what
fields were initially perturbed). The first two define the ratio between the coherence and Hubble lengths $\frac{\lambda_P}{L_H}$.
Here, we mostly focus on perturbations larger than the Hubble length: in Figure \ref{contrast1}, $\lambda_P\gg L_H$ ($\lambda_P\approx 3$,
$L_H^0\approx 0.09$);
in Figure \ref{cosm}, $\lambda_P>>> L_H$ ($\lambda_P\approx 2$, $L_H^0\approx 0.009$) and finally in Figure \ref{cd}, $\lambda_P> L_H$
($\lambda_P\approx 2$, $L_H^0\approx 0.3$). In the first case, we see that the density contrast
has grown after half a period of $\sigma$ ; in the second case and for the same period, 
we do not see yet any instability arising and finally, in Figure \ref{cd}, the contrasts have grown quite significantly during half a period of
$\sigma$. A more complete analysis shows that, all other parameters being equal, the large fluctuations $\lambda_P\gg L_H^0$ 
evolves more slowly than the small $\lambda_P\approx L_H^0$. This can be intuitively deduced from the previous comments on
Figures \ref{contrast1}-\ref{cd}
and is clearer by examining Figure \ref{hubble} where the evolutions of contrasts for two different coherence lengths are
represented, all other
parameters being equal. The evolution of the contrasts is qualitatively the same, although it is slower for large fluctuations 
(Figure \ref{hubble} $(iv)$, $(v)$, $(vi)$). Please note the different duration and amplitudes of the two simulations.
Therefore, the details of the evolution, e.g.
the growing time, the period and/or the amplitudes of the contrasts, depend on the coherence length, the absolute amplitude
of the perturbations (defined by $\frac{f_0-f^B_0}{f_0^B}$ for a certain field $f$) which is related to $\epsilon$ and $\sigma_0$,
and the nature of the perturbation. However, the scheme damped-then-growing oscillations seems to be general. We even performed
some computations at short wavelengths $\lambda_P\ll L_H$ but we did not recover 
the usual Jeans criterion for radiation (cf. \cite{peebles}). We think that this criterion could be modified somehow due the locally
anisotropic nature of YM pressures\footnote{Locally anisotropic fluids are characterized by $T_2^2=T_3^3\ne T_1^1$ in
spherical symmetry and therefore have an equation of state different in the inhomogeneous case than the simple barotropic
equation $p=K\rho$ with constant $K$.} 
(cf. \cite{anisotropy} and references therein for some
properties of locally anisotropic fluids). More work is under way to fix that point.
\\
\\
In Figure \ref{contrast1} $(iv)$, we see that the non-diagonal component of the stress-energy tensor $T_0^1$ becomes more and more homogeneous with expansion. This was
observed in all simulations and can be deduced from Eqs.(\ref{t01} - \ref{tmunu}) where one can see that this quantity evolves as the
inverse fourth power of the scale factor $R(t)$.\\
\\
For hyperbolic and closed cosmologies, we have performed some computations that lead essentially to analogous features as
those presented here in the flat case, for fluctuations of the order of the Hubble length.
\\
\\
Finally, let us check out the violation of the three constraints $\mathcal{H}$, $\mathcal{H}_1$ and $\mathcal{G}^{\bf{3}}$
(see Figure \ref{cons}).
The central bump in each constraint is related to the truncation errors when evaluating the spatial gradients, this
can be improved by increasing the number of discretization points in the variable $\chi$.
The interesting information is the time-evolution of the constraints: we see that the violation increases almost linearly during the computation
and accelerates at the end. This is typical of the very simple explicit scheme we used here. However, it is important
to note that the constraints remain several orders of magnitude below the perturbations of the related fields thanks to the
very strong Courant condition we used. Another crucial point is the boundary effect 
as the careful reader will have already noticed from  previous graphs. This can partly be improved by refining the accuracy of
the IVP and partly by using more localized fluctuations. 
Some compactification of the $\chi$ domain could help in verifying more exactly
the boundary conditions. In order to produce a more accurate method, one should also focus on improving
the stability by including implicit features in the scheme as well as a better control of boundary conditions.
\section{Conclusion}
By using a numerical method based on the hamiltonian formulation of the EYM system with $SU(2)$ gauge group, we have
checked that the primeval excitations of YM fields have been diluted by strong expansion in
the very early universe, due to the scale invariance of YM equations. The gauge fields
oscillate in a distorded potential well inherited from the homogeneous solution while the metric suffers an ever increasing departure from 
homogeneity. Meanwhile, the corresponding density and pressures contrasts undergo firstly damped oscillations.\\
\\
However, the YM fields are gravitationally unstable in the long term. Indeed, the metric perturbations
grow monotonically with time and, when the inhomogeneity of the metric and the gauge potentials cannot be neglected anymore, the 
last start diffusing non-linearly. This yields growing oscillation modes for the density and pressures contrasts, 
with a transition from damping depending on the interactions between the perturbed gauge fields,
the nature, size and amplitude of the primeval fluctuations. For example, the larger the fluctuation w.r.t. the Hubble length,
the slower it will grow. A fruitful comparison can be made
with the gravitational instability of the inflaton which lead us to the preliminary conclusion that the growth rate
of the YM fluctuations should depend also on the strength of the gauge coupling constant 
and should be more important for strong interaction.\\
\\
Nevertheless, in the real universe, we can conjecture that the phase transitions that made YM fields short-ranged
occur probably long before
their inhomogeneities reach a sufficient size. Therefore such fields could be of minor importance for cosmology. 
There are, however, two main objections to that affirmation.
First, a more physical and quantitative analysis, including precise 
considerations about the amplitude and size of the primeval excitations of the gauge fields,
other components of matter, precise cosmological parameters as well as quantum considerations on the status of the gauge fields after
the phase transitions should be performed before concluding.
In second, considering a Born-Infeld type modification of the YM lagrangian, deriving from string theory, which breaks the conformal invariance
can be of interest. In particular, an interesting work would be to study the gravitational instability of non-abelian Born-Infeld cosmology
which was examined in details in \cite{galtsov2}.\\
\\
Furthermore, another interesting point for cosmology is that the conformally invariant YM fields can hide for a while
the deep inhomogeneity of
space-time. Indeed, in a universe dominated by such fields, the observable quantities tend first to mimic a homogeneous universe while
the metric is getting more and more inhomogeneous and the perturbations of the gauge fields oscillate.
The question is to know which role such inhomogeneities could have played
after the phase transitions
that left the YM fields short-ranged as other types of matter fields feel differently the riddles of space-time hidden during the gauge
dominating era than the conformally invariant gauge fields did. \\
\\
For all these reasons, we think that our results are of interest 
at least for the investigation of mathematical properties of the pure EYM theory,
if not 
for a tiny step in the understanding of the very early universe .
\subsubsection*{Acknowledgements}
We would like to thank warmly Professor M. VOLKOV for useful discussions and his interest in this research as well as Professors
D. LAMBERT and A. LEMAITRE for their unconditional support and unshakeable enthusiasm.
This work was performed under the auspices of the Belgian National Fund for Scientific Research (F.N.R.S.).

\pagebreak

\end{document}